\documentclass{iopjournal}

\usepackage{physics}
\usepackage{bm}
\usepackage{cleveref}
\usepackage{amssymb}
\usepackage{mathdots}
\usepackage{cite}
\usepackage{yfonts}

\newcommand{\gradpar}[2][]{\nabla_{\parallel}^{#1} #2}
\newcommand{\htorloc}{h_{\mathrm{tor}}^{\mathrm{loc}}}
\newcommand{\specprefac}{\sum_{s} \frac{Z_s^2 n_s T_{\mathrm{e}}}{n_{\mathrm{e}} T_s}}
\newcommand{\resfacspec}[1][]{\left(\omega-\omega_{ds}\right)^{#1}}
\newcommand{\gradB}[1][]{\gradpar[#1]{\ln B}}
\newcommand{\gradFLR}[1][]{\gradpar[#1]{\ln(\frac{k_\perp^2}{B})}}
\newcommand{\graddriftmixterm}[1][]{\left[\frac{\omega_{\grad{B},s}}{2} \gradpar[#1]{\ln( \bm{k_\perp}\vdot\bm{\mathcal{L}})}-\omega_{\kappa,s}\gradpar[#1]{\ln B}\right]}
\newcommand{\gradcurv}[1][]{\omega_{\kappa,s} \gradpar[#1]{\ln(\frac{\bm{k_\perp}\vdot\bm{\mathcal{K}}}{B})}}

\newcommand{\gradtwodriftmixgradBterm}{\frac{\omega_{\grad{B},s}}{2}\left(\gradpar[2]{\ln(\bm{k_\perp}\vdot\bm{\mathcal{L}})} +\left[\gradpar{\ln(\bm{k_\perp}\vdot\bm{\mathcal{L}}})\right]^2\right)}
\newcommand{\gradtwodriftmixcurvterm}{\omega_{\kappa,s} \left(\gradB[2]-\left[\gradB\right]^2 + 2 \gradB \gradpar{\ln(\bm{k_\perp}\vdot\bm{\mathcal{K}})}\right)}
\newcommand{\gradtwocurv}{\omega_{\kappa,s} \left(\gradpar[2]{\ln(\bm{k_\perp}\vdot\bm{\mathcal{K}})}-\gradB[2]+\left[\gradpar{\ln(\bm{k_\perp}\vdot\bm{\mathcal{K}})}-\gradB\right]^2\right)}

\newcommand{\FLRdiff}[1][]{\left(\Gamma_{1}(b_{#1})-\Gamma_0(b_{#1})\right)}
\newcommand{\diamagterm}{\left(1-\frac{\omega_{\grad{p_s}}}{\omega}\right)}

\newcommand{\acond}{\abs{\frac{k_\parallel(\bm{x}) v_T}{\tilde{\omega}_d(\bm{x})}}^2}
\newcommand{\bcond}{\Re\left[\frac{\omega}{\tilde{\omega}_d(\bm{x})}\right]}

\DeclareMathOperator{\erfc}{erfc}
\DeclareMathOperator{\sgn}{sgn}

\let\bmog\bm 
\renewcommand{\bm}[1]{{\mathbf{#1}}}

\begin{document}

\articletype{Paper} 

\title{Resonant theory with asymptotic-transit effects of toroidal ion-temperature-gradient instability in general geometry}

\author{M.C.L. Morren$^{1,*}$\orcid{0009-0004-4565-083X}, M.J. Pueschel$^{2,1,3}$\orcid{0000-0003-3895-1920}, J.H.E. Proll$^{4,1}$\orcid{0000-0001-8865-9448}, G.G. Plunk$^{4}$\orcid{0000-0002-4012-4038}, Ö.D. Gürcan$^{5}$\orcid{0000-0002-2278-1544} and A. Manole$^{1}$}

\affil{$^1$ Department of Applied Physics and Science Education, Eindhoven University of Technology, 5600 MB Eindhoven, The Netherlands}

\affil{$^2$ Dutch Institute for Fundamental Energy Research, 5612 AJ Eindhoven, The Netherlands}

\affil{$^3$ Department of Physics \& Astronomy, Ruhr-Universität Bochum, 44780 Bochum, Germany}

\affil{$^4$ Max-Planck-Institut für Plasmaphysik, 17491 Greifswald, Germany}

\affil{$^5$ Laboratoire de Physique des Plasmas, CNRS, Ecole Polytechnique, Sorbonne Université, Université Paris-Saclay, Observatoire de Paris, F-91120 Palaiseau, France}

\affil{$^*$Author to whom any correspondence should be addressed.}

\email{m.c.l.morren@tue.nl}


\begin{abstract}
Based on an asymptotic expansion for weak parallel-streaming effects, we derive a simplified dispersion relation for the curvature-driven ion-temperature gradient (ITG) instability in the adiabatic-electron limit. The theory is valid for arbitrary toroidal geometry, does not impose an ansatz on the eigenmode structure, includes kinetic resonances with the magnetic drift, orders the variation of the magnetic geometry along the field to be comparable to the length scale of variation of the electrostatic potential, and can account for impurities. We validate the model's applicability against high-fidelity gyrokinetic simulations in a circular tokamak, and the Large Helical Device (LHD) and the Wendelstein 7-X (W7-X) stellarators. Additionally, the model's sensitivity to  geometric changes is probed by performing a shear scan in the circular-tokamak case. \par
These tests show the model is capable of providing quantitative predictions for ITG eigenmodes and eigenfrequencies in the transport-relevant regime beyond conventional tokamaks down to global shear of $\hat{s}=-0.3$, although the agreement at shorter wavelengths and/or larger negative shear deteriorates due to breakdown of the assumption of weak transit effects. The latter effect is exacerbated in stellarator geometries -- in particular for W7-X -- where the model predictions follow the trends observed in high-fidelity simulations qualitatively albeit with greater discrepancies than in the axisymmetric case. The introduction of additional corrections to the dispersion relation associated with the variation of the magnetic geometry along the field line is found to generally result in an upshift in the growth-rate spectrum, with the effect greatest in many-field-period stellarator geometries and low-negative-shear tokamaks, and is observed to further deteriorate the agreement in eigenfunctions.
\end{abstract}

\section{Introduction} \label{sec:intro}
In magnetic confinement fusion, anomalous heat losses due to microturbulence-induced transport represents the dominant energy-loss mechanism in tokamaks and stellarators optimised for neoclassical transport. The source of this microturbulence can be traced back to pressure-gradient-driven Larmor-radius-scale instabilities, of which one of the most commonly observed is the ion-temperature-gradient (ITG) mode. Both the anomalous transport and the underlying microinstabilities are known to be sensitive to the magnetic configuration, allowing one to optimise for reduced microturbulence through geometry tuning. Including linear\cite{Jorge2024DirectDevices} and nonlinear\cite{Kim2024OptimizationStellarators} gyrokinetic simulations directly in the stellarator design optimisation loop has only recently been considered in the case of adiabatic-electron ITG, however, is generally considered prohibitively expensive for routine use. \par
Hence, alternative proxy models for the predictions of microstability and microturbulence are desired. For tokamaks, TGLF\cite{Staebler2007APhysics,Staebler2010ElectronModel}, QuaLiKiz\cite{Bourdelle2007APlasmas,Citrin2017}, and the recently developed PQLS \cite{Wilms2026PQLS:Closure} may be used as reduced models for fast transport predictions, however, their predictions have been found to suffer from reduced accuracy at e.g. low magnetic shear\cite{Citrin2012,Najlaoui2025VerifyingPlasmas} or strong plasma shaping in the edge regions\cite{Snoep2025CharacterizationJET}, associated with the difficulty for the enforced ansatz of the mode structure to represent the eigenfunctions in those parameter regimes. 
The ITG microinstability has previously been analysed with simplified dispersion relation models, however to gain analytical leeway, these studies have been limited to ansatz-based methods for the mode structure in simplified geometries such as a quadratic drift well \cite{Rodriguez2025TheLocalisation}, or by invoking non-resonant limits in a circular tokamak\cite{Plunk2014,Romanelli1989IonTokamaks,Horton1981ToroidalGradients,Guo1989IonTransport,Coppi1990CandidatePlasmas}. These latter studies often neglect the role of the variation of the magnetic geometry, with noticeable exceptions being Refs.~\cite{Guo1989IonTransport,Coppi1990CandidatePlasmas}, though still limited to a non-resonant expansion in a circular tokamak geometry. Additionally, the ITG has also been extensively studied in the local limit\cite{Biglari1998b,Zocco2018ThresholdPlasmas,Gurcan2014NumericalCurvature,Ivanov2023AnField,Romanelli1989IonTokamaks,Kuroda1998InitialMode,Kim1994KineticMode,Kim1991TransitionModes,Calvo2026ModificationPlasmas}, where, assuming the eigenmodes to be strongly ballooned, the variation of the eigenmode and the magnetic geometry may be neglected altogether. \par 
In this work, we develop a semi-analytical dispersion relation for ITG modes under the assumption of adiabatic electrons based on an asymptotic expansion of the Vlasov-Poisson system in the limit of weak parallel streaming. This dispersion relation retains kinetic resonances with the magnetic drift, though resonances with the transit motion of particles are neglected, and is therefore only suited to describe the toroidal branch of the ITG mode. Additionally, the dispersion relation retains all terms relating to inhomogeneities in the magnetic geometry along the field line, and does not \textit{a priori} impose any assumptions on the shape of the eigenmode solutions. \par 
In addition to this self-consistent asymptotic dispersion relation, we also consider two heuristic alternative dispersion relations which rely on a guess for the mode structure to yield a prediction for the eigenfrequency. The eigenfrequencies from all three models are subsequently compared to high-fidelity simulation with the \textsc{Gene} code\cite{Jenko2000} -- examining the geometries of an $\hat{s}$-$\alpha$ tokamak\cite{Connor1978}, the inward-shifted configuration of the Large Helical Device (LHD) stellarator \cite{Iiyoshi1990DesignDevice}, and the high-mirror configuration of the Wendelstein 7-X stellarator \cite{Beidler1990} -- where the high-fidelity eigenmodes are used as mode structure guesses for the heuristic models. \par
The remainder of the paper is organised as follows; in \Cref{sec:Vlasov-Poiss} we introduce the linearised electrostatic collisionless gyrokinetic framework which is used in \Cref{sec:disp_deriv} to derive the adiabatic-electron ITG dispersion relation under asymptotic expansion for weak particle streaming. Subsequently, \Cref{sec:altmod} introduces the two heuristic reduced models considered for comparison against the asymptotic dispersion relation. \Cref{sec:numerics} then provides a brief overview of the magnetic configurations, \textsc{Gene} simulation parameters and numerical technique used to solve the asymptotic dispersion relation. Thereafter, in \Cref{sec:results}, we compare the ITG eigenfrequency and (where applicable) eigenmode predictions from the reduced models with those from \textsc{Gene}. Lastly, \Cref{sec:discussion} provides a summary and discussion for future improvements of the model.

\section{The gyrokinetic Vlasov-Poisson system} \label{sec:Vlasov-Poiss}
The ITG mode may in be characterised as an unstable electrostatic drift wave, hence we consider the $\beta\rightarrow 0$ limit of the gyrokinetic framework\cite{Catto1981,Brizard2007}, such that magnetic fluctuations are assumed negligible. Here $\beta = 2\mu_0 p/B^2$ is the ratio of kinetic pressure $p$ to magnetic pressure, with $\mu_0$ the vacuum permeability and $B$ is the equilibrium magnetic field strength. Additionally, since we are concerned with describing the behaviour of fluctuations in core plasmas, we assume the plasma to be collisionless. These assumptions are fairly modest as finite-$\beta$ and finite collisionality tend to have a moderately stabilising effect on the ITG without significantly affecting its fundamental properties\cite{Dong1987FiniteModes,Hirose2000OnMode,Pueschel2008GyrokineticBeta,Pueschel2010TransportMicroturbulence,Belli2017ImplicationsSimulation,Kumar2025ITGPlasmas,Mulholland2025Finite-Modes}. \par 
We decompose the distribution function of each species $s$ as
\begin{align}
        f_s(\bm{x},\bm{v},t) = F_{Ms}(\bm{x},\bm{v}) \left(1-\frac{q_s \phi(\bm{x},t)}{T_s(\bm{x})}\right) + g_s(\bm{R},\bm{v},t),
        \label{eq:fs_decomposition}
    \end{align}
with $q_{s},T_{s}, F_{Ms}$ the charge, temperature and equilibrium Maxwellian 
    \begin{align}
        F_{Ms}(\bm{x},\bm{v}) = \frac{n_s(\bm{x})}{\left(2\pi v_{Ts}(\bm{x})^2\right)^{3/2}} \exp(-\frac{v^2}{2v_{Ts}(\bm{x})^2})
    \end{align}
of species $s$, respectively, with $n_s(\bm{x})$ and $v_{Ts}(\bm{x})=\sqrt{T_s(\bm{x})/m_s}$ the equilibrium density and thermal velocity of species s, respectively.  The first term of \Cref{eq:fs_decomposition} includes both the equilibrium distribution and the adiabatic response to electrostatic potential fluctuations in the $q_s\phi/T_s\ll1$ limit, while the second term is the non-adiabatic gyrocenter distribution $g_s$ defined at guiding center positon $\bm{R} = \bm{x} - \bm{e_b}\cross\bm{v}/\Omega_s$, where $\bm{e_b}=\bm{B}/B$ and $\Omega_s = q_s B/m_s$ the cyclotron frequency of species $s$, with $\bm{B}$ being the (equilibrium) magnetic flux density, with $B=\norm{\bm{B}}$, and $m_s$ the mass of species $s$. \par
The non-adiabatic part of the distribution function evolves according to linear gyrokinetic Vlasov equation (GKVE) \cite{Rutherford1968,Catto1981}
    \begin{align}
         v_\parallel \nabla_{\parallel} \hat{g}_s - i(\omega-\omega_{ds}) \hat{g}_s = - i \frac{q_s}{T_s} F_{Ms} \left(\omega-\omega_{*s}^{T}\right) J_{0s} \hat{\phi}
         \label{eq:lin_GKVE}
    \end{align}
where we have adopted the following short-hand notations; $\gradpar{} = \bm{e_b}\vdot\grad{}$ as the directional derivative along the field line, $J_{0s} = J_0(k_\perp v_\perp/\Omega_s)$ where $J_0$ is the zeroth-order Bessel function of the first kind, $v_\parallel, v_\perp$ denote the velocity components with respect to their alignment of the magnetic field, and $k_\perp = \sqrt{\bm{k_\perp}\vdot\bm{k_\perp}}$ is the perpendicular wavenumber. Furthermore $\omega_{ds}=\bm{k_\perp}\vdot\bm{v_{ds}}$ with $\bm{v_{ds}} = \bm{e_b} \cross \left(v_\parallel^2 \bmog{\kappa} + (v_\perp^2/2) \grad{\ln B}\right)/\Omega_s$ the magnetic drift velocity, where $\bmog{\kappa}=\left(\bm{e_b}\vdot\grad\right){\bm{e_b}}$ is the magnetic curvature vector, and $\omega_{*s}^T=\omega_{\grad{n_s}}+\omega_{\grad{T_s}} \times$ $\left(v^2/(2v_{Ts}^2)-3/2\right)$ is the temperature-dependent diamagnetic drift frequency, with $\omega_{\grad{n_s}} = T_s/(q_s B) \ (\bm{k_\perp}\cross\bm{e_b})\vdot\grad \ln{n_s}$ known as the diamagnetic drift frequency, with an analogous definition for $\omega_{\grad{T_s}}$.
In writing \Cref{eq:lin_GKVE} it has been assumed that all fluctuations are characterised as small-amplitude low-frequency drift waves with $\omega/\Omega_s \ll 1$ and a strong anisotropy with respect to magnetic field alignment $k_\parallel/k_\perp \ll 1$, admitting the use of an WKB-like representation for the fluctuating quantities $\xi(\bm{x},t)=\hat{\xi}(\bm{x}) \exp(iS(\bm{x}) - i\omega t)$ where $\hat{\xi}(\bm{x})$ describes the slow variation along the magnetic field, while the phase factor -- subject to $\bm{B}\vdot\grad{S}=0$ -- describes the rapid cross-field variation on length scales comparable to the thermal gyroradius $k_\perp \rho_{Ts} \sim \order{1}$. Additionally, the GKVE has been considered under the ballooning transform\cite{Connor1978,Connor1998,Dewar1983,Candy2004SmoothnessSurface}, with $\grad{S}=\bm{k_\perp}$ as a perpendicular wavenumber, reducing \Cref{eq:lin_GKVE} to a one-dimensional differential equation in $\hat{g}_s(\bm{x})$ along the field line, while accommodating for the lack of physical cross-field periodicity in the presence of finite shear. \par
The electrostatic potential fluctuations need to be self-consistently determined from Poisson's law, which for the length scales under consideration for typical core fusion plasma conditions may be reduced to the quasi-neutrality condition for the density fluctuations 
    \begin{align}
        \sum_{s} \frac{Z_s^2 n_s T_{\mathrm{e}}}{n_{\mathrm{e}} T_s} \hat{\phi} = \sum_{s} \frac{Z_s T_{\mathrm{e}}}{en_{\mathrm{e}}} \int J_{0s} \hat{g}_s \dd^3{\bm{v_s}},
        \label{eq:QN_cond}
    \end{align}
as $k_\perp \lambda_{\mathrm{De}} \ll 1$, where $\lambda_{\mathrm{De}}=\sqrt{\epsilon_0 T_{\mathrm{e}}/(n_{\mathrm{e}} e^2)}$ is the electron Debye length.

\section{ITG dispersion relation in asymptotically-small streaming limit} \label{sec:disp_deriv}
As a result of the coupling between the GKVE [\Cref{eq:lin_GKVE}] and the quasi-neutrality condition [\Cref{eq:QN_cond}], the admissible eigenmodes and eigenfrequencies are dictated by a dispersion relation. While a general solution $\hat{g}_s$ to \Cref{eq:lin_GKVE} may be obtained, this involves an arduous integral equation for the dispersion which obfuscates the underlying salient physics causing the instability, see e.g. Refs.~\cite{Connor1980StabilityIII} for such an approach. \par
Rather, exploiting the fact that for fusion-relevant applications typically the toroidal branch of the ITG instability as driven unstable by magnetic curvature is of primary interest, the resulting dispersion relation for toroidal ITGs may be considerably simplified by considering the influence of parallel streaming to be weak with respect to the magnetic drift frequency in \Cref{eq:lin_GKVE}. Here, instead of completely neglecting the influence parallel streaming  altogether, we consider it a small effect and obtain a solution to \Cref{eq:lin_GKVE} under an asymptotic expansion for $\omega_{\mathrm{ti}}/\omega \ll 1$, where $\omega_{\mathrm{ti}}=v_{T\mathrm{i}}/L_\parallel$ is the ion transit frequency, with the connection length $L_\parallel$, a proxy for the norm of the $v_\parallel \nabla_\parallel \sim \omega_{\mathrm{t}s}$ operator. Note that as a consequence of the assumption on the smallness of the transit frequency, the model is incapable of describing trapped-ion modes, which would require frequencies comparable the the ion bounce frequency, being considerably lower than permitted by our ordering \cite{Romanelli1989IonTokamaks}, nor can the model describe the slab-branch of the ITG instability \cite{Coppi1967InstabilitiesConfigurations,Kadomtsev1970TurbulenceSystems}. Additionally, for simplicity, we assume the electrons to be adiabatic ($\hat{g}_e=0$) to focus on the main instability drive from ions, though kinetic-electron effects are known to provide additional destabilisation to ITG modes, see e.g. Refs.~\cite{C.J.2020HowSize, Dominski2015HowTransport, Dominski2012IdentifyingSimulations, Hardman2022ExtendedElectrons}. 
As shown in \Cref{app:disp_rel_deriv_details}, to lowest order in $\omega_{\mathrm{ti}}/\omega \ll 1$, the dispersion relation for adiabatic-electron ITG under this expansion follows as
    \begin{align}
        \sum_{s'} \frac{Z_{s'}^2 n_{s'} T_{\mathrm{e}}}{n_{\mathrm{e}} T_{s'}} \hat{\phi}(l) = \specprefac \int J_{0s} \left[ 1 -\frac{v_\parallel}{\omega-\omega_{ds}}\gradpar{\frac{v_\parallel}{\omega-\omega_{ds}}\gradpar{}} \right] \mathcal{F}_{Ms}  \frac{\omega-\omega_{*s}^{T}}{\omega-\omega_{ds}} J_{0s} \hat{\phi}(l) \dd^3{\bm{v_s}}, 
        \label{eq:asymp_disp_rel}
    \end{align}
where the sum over $s'$ on the left-hand side includes electrons, while the sum over $s$ on the right-hand side only runs over ions, and we have introduced $\mathcal{F}_{Ms}=F_{Ms}/n_s$. In the limiting case of a single hydrogenic-ion species, \Cref{eq:asymp_disp_rel} matches with forms obtained in e.g. Equations (14) and (10) of Refs.~\cite{Plunk2014,Romanelli1989IonTokamaks}, respectively, although the latter specifically only applies in tokamaks. Note that while $\omega_{\mathrm{ti}}/\omega \ll 1$ is a commonly applied expansion parameter in literature for its convenience, a more formal requirement for \Cref{eq:asymp_disp_rel} to be valid is that $v_\parallel \gradpar{\ln \hat{g}_s}/(\omega-\omega_{ds}) \ll 1$, which for an ordering of $\omega_{ds}/\omega \sim \order{1}$ where resonant effects are relevant can be more restrictive, though it reduces to the latter when $\omega_{ds}/\omega \ll 1$. We examine the restrictions this imposes in detail in \Cref{app:asymp_breakdown}, observing that this condition may be violated in some regions of phase-space depending on the eigenfrequency and parallel eigenmode profile. \par
The $\gradpar{}\rightarrow 0$ limit of \Cref{eq:asymp_disp_rel} constitutes the usual local toroidal ITG dispersion relation, while the second term in square brackets forms the finite transit correction to the toroidal ITG dispersion relation. Retaining such corrections, despite being formally only $\order{(\omega_{\mathrm{ti}}/\omega)^2}$, is strictly necessary in order to provide regularisation to the eigenmode problem. If the weak-transit limit is taken too egregiously by neglecting the effect of parallel streaming altogether, then formally the eigenmodes of \Cref{eq:asymp_disp_rel} would correspond to an eigenbasis of $\hat{\phi}(l)=\delta(l-l_0)$ in the local limit \cite{Connor1980StabilityIII}, with $\delta(x)$ the Dirac-delta function, which violates the assumptions of the WKB approximation used to represent fluctuations in the gyrokinetic framework. Therefore, despite being ostensibly small in our ordering, the correction terms due to finite-transit effects need to be retained as they determine the admissible eigenmodes to $\order{1}$.
Alternatively, regularisation of the eigenmodes could also be obtained by including the effects of trapped electrons, for which the parallel streaming term in \Cref{eq:lin_GKVE} is formally the dominant contribution, and in an equivalent asymptotic expansion for $\omega/\omega_{be} \ll 1$, with $\omega_{\mathrm{be}}$ the electron bounce frequency, the resulting bounce-average would introduce another source of non-local effects along the field line \cite{Dominguez1992,Morren2026FastApproach}, but this route is not further pursued here as we adopt an adiabatic-electron approximation. \par 
By expanding the nested parallel streaming operator in \Cref{eq:asymp_disp_rel} and grouping terms with equal derivative of the eigenfunction, the dispersion relation can be rearranged to yield an advection-diffusion-like equation along the field line determining the eigenfrequency-eigenmode pairs
    \begin{align}
        h_{\parallel}^{(2)}(\omega,l) \gradpar[2]{\hat{\phi}}(l) + h_{\parallel}^{(1)}(\omega,l) \gradpar[]{\hat{\phi}}(l)  + \left(h_{\mathrm{tor}}^{\mathrm{loc}}(\omega,l)-h_{\parallel}^{(0)}(\omega,l)\right) \hat{\phi}(l) = 0,
        \label{eq:adv-diff}
    \end{align}
where we have introduced the resonant integrals
    \begin{align}
        h_{\parallel}^{(2)} =& \specprefac \int \mathcal{F}_{Ms} v_\parallel^2 \frac{\omega-\omega_{*s}^T}{\left(\omega-\omega_{ds}\right)^3} J_{0s}^2 \dd^3{\bm{v_s}} \nonumber \\
        h_{\parallel}^{(1)} =& \specprefac \int \mathcal{F}_{Ms}  \frac{\omega-\omega_{*s}^T}{\omega-\omega_{ds}} J_{0s} v_\parallel \left(\frac{v_\parallel}{\omega-\omega_{ds}}\gradpar{\frac{J_{0s}}{\omega-\omega_{ds}}}+\gradpar{\frac{v_\parallel J_{0s}}{\left(\omega-\omega_{ds}\right)^2}}\right) \dd^3{\bm{v_s}} \nonumber \\
        h_{\mathrm{tor}}^{\mathrm{loc}} =& 1 + \specprefac \left(1-\int \mathcal{F}_{Ms} \frac{\omega-\omega_{*s}^T}{\omega-\omega_{ds}} J_{0s}^2 \dd^3{\bm{v_s}}\right) \nonumber \\
        h_{\parallel}^{(0)} =& \specprefac \int \mathcal{F}_{Ms} \frac{\omega-\omega_{*s}^T}{\omega-\omega_{ds}} J_{0s} v_\parallel \gradpar{\left(\frac{v_\parallel}{\omega-\omega_{ds}}\right)\gradpar{\frac{J_0}{\omega-\omega_{ds}}}} \dd^3{\bm{v_s}},
        \label{eq:ODE_ints}
    \end{align}
with the local dispersion relation simply given by $h_{\mathrm{tor}}^{\mathrm{loc}}(\omega,l)=0$. 
\Cref{eq:adv-diff} is commonly considered in the simplified form of $h_{\parallel}^{(2)}\gradpar[2]{\hat{\phi}}+\htorloc\hat{\phi}=0$, as may be justified by imposing the assumption of strong eigenmode localisation such that $\gradpar{\ln \mathrm{geo}} /\gradpar{\ln \hat{\phi}}\ll 1$, where $\gradpar{\ln \mathrm{geo}}$ is the characteristic length-scale of variation in the geometric quantities, which is typically on the order of the connection length. In a tokamak, one can estimate $\gradpar{\ln \mathrm{geo}}\sim 1/(qR)$, where $q$ is the local safety factor and $R$ the major radius, but it may be considerably larger in a stellarator; if the stellarator magnetic field possesses a discrete toroidal symmetry due to $N_{fp}$ field periods, then one may instead estimate $\gradpar{\ln \mathrm{geo}}\sim$ $N_{fp}/(qR)$ \cite{Mackenbach2025OnEnergy}. By contrast, $\gradpar{\ln \hat{\phi}}$ is an a priori unknown length scale, thus potentially making parallel dynamics more important in stellarators as passing particles may experience a stronger variation of the geometry while traversing a coherent structure of the electrostatic potential. Here we explicitly scrutinise this simplifying assumption by considering both this ``minimal model" and the full asymptotic dispersion relation retaining the effects of geometric variation along the field line. \par
The various kinetic integrals in \Cref{eq:ODE_ints} include resonances with the magnetic drift, and may be solved analytically in terms of the two-pole generalisation of the plasma dispersion function \cite{Gurcan2014NumericalCurvature}
    \begin{align}
        G_{n}(z_1,z_2) = \frac{1}{\sqrt{\pi}} \int_{-\infty}^{\infty} \frac{x^n \exp(-x^2)}{(x-z_1)(x-z_2)}\dd{x}
        \label{eq:Gn_int},
    \end{align}
which, as shown in \Cref{app:Gn_props}, can be related to the usual plasma dispersion integral 
    \begin{align}
        Z_0(z) = \frac{1}{\sqrt{\pi}} \int_{-\infty}^{\infty} \frac{\exp(-x^2)}{x-z}\dd{x}
        \label{eq:Z0int}
    \end{align}
through\footnote{In \Cref{eq:Gn_sequence}, the factor of $(1+(-1)^{n-k-1})/2$ is a correction with respect to Equations (4)--(6) of Ref.~\cite{Gurcan2014NumericalCurvature} to account for the vanishing of odd moments of the Gaussian, affecting the calculation of $G_{3}$ and upwards, as required for $h_{\parallel}^{(2)}, h_{\parallel}^{(1)},h_{\parallel}^{(0)}$ (see \Cref{eq:semi-ana-integrals_expand}). Note that this correction, however, does not affect the previous findings of Ref.~\cite{Gurcan2014NumericalCurvature}, which focussed on using \Cref{eq:Gn_int} to study the local limit of the ITG dispersion relation ($\htorloc=0$), under the substitution of $\omega-\omega_{ds}\rightarrow \omega-\omega_{ds}-k_\parallel v_\parallel$ with $k_\parallel \in \mathbb{R}$ rather than a streaming operator, thus only necessitating the use of $G_0,G_2$ (see \Cref{eq:semi-ana-integrals_expand}).}
    \begin{align}
        G_n(z_1,z_2) = \frac{1}{\sqrt{\pi}} \frac{1}{z_1-z_2} \sum_{k=0}^{n-1} \left(z_1^{k}-z_2^{k}\right) \Gamma\left(\frac{n-k}{2}\right) \frac{1+(-1)^{n-k-1}}{2} + \frac{z_1^{n}Z_0(z_1)-z_2^nZ_0(z_2)}{z_1-z_2},
        \label{eq:Gn_sequence}
    \end{align}
where it has been assumed that $z_1 \neq z_2$, see \Cref{app:Gn_props} for a treatment of the degenerate case. \par
Note that in accordance with Ref.~\cite{Gurcan2014NumericalCurvature}, the path of integration in \Cref{eq:Z0int} is taken to be strictly along $\mathbb{R}$, thus retaining the essential singularity at $z\in\mathbb{R}$, rather than being the plasma dispersion function $Z(\zeta)=i\sqrt{\pi}\mathcal{W}(\zeta)$ where $\mathcal{W}(\zeta)=\exp(-\zeta^2)\erfc[-i\zeta]$ is the Faddeeva function. Consequently, only in the upper half plane do these definitions of $Z(z)$ and $Z_0(z)$ agree. In the context of Laplace-Fourier transforms to describe the plasma fluctuations, this agreement of $Z_0(z)$ with the plasma dispersion function is typically further extended by analytically continuing \Cref{eq:Z0int} into the lower half-plane by deforming the path of integration \cite{Kuroda1998InitialMode,Ivanov2023AnField}. However, we do not pursue this approach here for three reasons; first, the analytical continuation of \Cref{eq:Z0int} only makes sense to consider for modes characterised by $\Im[\omega]<0$, while the poles of \Cref{eq:Gn_int} may lie in the lower-half plane even when the eigenmodes satisfy $\Im[\omega]>0$ (see below \Cref{eq:Inmpq_main_def}), thus necessitating a way to solve \Cref{eq:Z0int} when $\Im[z]<0$. Second, we restrict our attention to unstable eigenmodes in the $\Im[\omega]>0$ region. Lastly, even in the event of damped modes, the fluctuations described by the subsequent Laplace-Fourier analysis of $\Im[\omega]<0$ are not eigenmodes of the collisionless gyrokinetic system, thus violating our normal ansatz $\sim \exp(-i\omega t)$ for all fluctuation quantities. Taking into account the above considerations, we make use of the fact that for $\Im[z]\neq0$ away from the essential singularity in \Cref{eq:Z0int}, $Z_0(z)$ may be solved analytically as \cite{Gurcan2014NumericalCurvature,Morren2026FastApproach}
    \begin{align}
        Z_0(z) = \sigma_z i\sqrt{\pi} \exp(-z^2)\erfc[-i\sigma_z z] = \sigma_z Z(\sigma_z z) \quad \Im[z]\neq0
        \label{eq:Z0-Faddeeva_rel}
    \end{align}
where $\sigma_z = \sgn{\Im[z]}$, therewith making the agreement with the plasma dispersion function in the upper half-plane and the discontinuity along the real axis apparent. Thus, throughout this work, \Cref{eq:Z0-Faddeeva_rel} is the expression for $Z_0(z)$ that is used in \Cref{eq:Gn_sequence} to evaluate $G_n(z_1,z_2)$. \par
After some tedious algebra which is relegated to \Cref{app:geom_derivs,app:res_int_calculations}, the integrals determining the coefficients of our advection-diffusion-like dispersion relation [\Cref{eq:adv-diff,eq:ODE_ints}] may then be written in terms of \Cref{eq:Gn_int} as
    \begin{align}
        h_{\parallel}^{(2)} =&  \specprefac  v_{Ts}^2 \{\mathbb{I}_{2,1}^{3,0}\} \nonumber \\
        h_{\parallel}^{(1)} =& -\specprefac v_{Ts}^2  \left(\gradFLR   \sqrt{b_s} \{\mathbb{I}_{2,2}^{3,1}\}  + \gradB \{\mathbb{I}_{0,3}^{3,0}\} \right. \nonumber \\
        &\left. -3 \left( \graddriftmixterm  \{\mathbb{I}_{2,3}^{4,0}\} + \gradcurv \{\mathbb{I}_{4,1}^{4,0}\}\right)\right)\nonumber \\
        \htorloc =& 1 + \specprefac \left(1-\{\mathbb{I}_{0,1}^{1,0}\}\right) \nonumber
    \end{align}
    \begin{align}
        h_{\parallel}^{(0)} =& \specprefac v_{Ts}^2 \Biggl(\frac{\sqrt{b_s}}{4} \gradB \gradFLR \{\mathbb{I}_{0,4}^{3,1}\}  - \frac{1}{8} b_s \left(\gradFLR\right)^2 \left(\{\mathbb{I}_{2,3}^{3,0}\}-\{\mathbb{I}_{2,3}^{3,2}\}\right) \nonumber \\
        & -\frac{\sqrt{b_s}}{2} \left[\gradFLR[2] + \frac{1}{2} \left(\gradFLR\right)^2\right] \{\mathbb{I}_{2,2}^{3,1}\}  - \frac{\gradB}{2} \biggl( \gradcurv \{\mathbb{I}_{2,3}^{4,0}\} +  \nonumber \\
        & \graddriftmixterm \{\mathbb{I}_{0,5}^{4,0}\} \biggr) - \frac{3}{2} \sqrt{b_s} \gradFLR \biggl(\gradcurv \{\mathbb{I}_{4,2}^{4,1}\}  \nonumber \\
        & + \graddriftmixterm \{\mathbb{I}_{2,4}^{4,1}\}\biggr) + \Bigl\{\gradtwodriftmixgradBterm  \nonumber  \\
        & - \gradtwodriftmixcurvterm \Bigr\} \{\mathbb{I}_{2,3}^{4,0}\} + \nonumber \\
        & \gradtwocurv \{\mathbb{I}_{4,1}^{4,0}\}
        \nonumber \\
        & + 3 \graddriftmixterm^{2} \{\mathbb{I}_{2,5}^{5,0}\} + 3 \left(\gradcurv\right)^{2} \{\mathbb{I}_{6,1}^{5,0}\}  \nonumber \\
        & + 6 \gradcurv \graddriftmixterm \{\mathbb{I}_{4,3}^{5,0}\}\Biggr) 
        \label{eq:semi-ana-integrals_expand}
    \end{align}
having introduced the convenient short-hand notations $b_s = (k_\perp \rho_{Ts})^2$, $\bm{\mathcal{L}} = \bm{e_b}\cross\grad{B}$, and $\bm{\mathcal{K}}=\bm{e_b}\cross\bmog{\kappa}$, for the magnitude of the finite Larmor radius (FLR) effect, and geometric factors of the $\grad{B}$ and curvature components of the magnetic drift, respectively, along with the recurring combination
    \begin{align}
         \{\mathbb{I}_{n,m}^{p,q}\} = \left(\omega-\omega_{\grad{n},s}+\frac{3}{2}\omega_{\grad{T},s}\right) \mathbb{I}_{n,m}^{p,q}-\frac{\omega_{\grad{T},s}}{2} \left(\mathbb{I}_{n+2,m}^{p,q}-\mathbb{I}_{n,m+2}^{p,q}\right)
         \label{eq:Inmpq_triplet_shorthand}
    \end{align}
and the parametrised integral
    \begin{align}
        \mathbb{I}_{n,m}^{p,q} = \frac{\sqrt{2}^{n+m-1}}{\omega_{\kappa,s}} \frac{(-1)^{p}}{\left(p-1\right)!} \left(\pdv{}{\omega}\right)^{p-1}  \int_{0}^{\infty} \dd{u_\perp} u_\perp^m e^{-u_\perp^2} J_0(\sqrt{2b_s}u_\perp) J_q(\sqrt{2b_s} u_\perp) \mathsf{g}_n (z_s(u_\perp,\omega))
        \label{eq:Inmpq_integral}
    \end{align}
where $\mathsf{g}_n(z)  = G_n(z,-z)$ (see \Cref{eq:smollgn_z})-- note the distinction from the non-adiabatic distribution function $g_s(\bm{R},\bm{v},t)$ --  and $z_s(u_\perp,\omega) = \sqrt{\frac{\omega}{2\omega_{\kappa,s}}-\frac{\omega_{\grad{B},s}}{2\omega_{\kappa,s}}u_\perp^2}$. Here and throughout the manuscript, $\sqrt{\xi}$ or $\xi^{1/2}$ refer to the principal branch of the square root function with a branch cut along the negative real axis. In obtaining \Cref{eq:semi-ana-integrals_expand}, it has been assumed that the magnetic drift is non-vanishing\footnote{In particular $\omega_{\kappa,s}\neq 0$ is required in order for \Cref{eq:Inmpq_integral} to be valid, however, in the electrostatic limit of $\beta\rightarrow 0$ this also implies $\omega_{\grad{B},s}\neq 0$, and therefore is equivalent to $\omega_{ds}\neq0$.}, and that the kinetic energy and magnetic moment of each particle are fixed, necessitating to account for the orbit modulation by the magnetic field strength when evaluating the parallel derivative of velocity-dependent quantities. After evaluation of these parallel derivatives, the velocity variables are subsequently treated as local quantities in the dispersion relation for the evaluation of the resonant integrals, as \Cref{eq:asymp_disp_rel} must hold at each position along the field line. Note, however, that our choice of modelling the plasma dispersion integral [\Cref{eq:Z0int}] according to \Cref{eq:Z0-Faddeeva_rel} makes the model ill-suited to investigate the plasma behaviour near marginality, as this renders $z_s(u_\perp,\omega)\in \mathbb{R}$ through a fraction of phase space.  \par
Dispersion relations similar to \Cref{eq:adv-diff} -- including the effects of kinetic resonances with the magnetic drift -- have been obtained in Refs.~\cite{Cheng1981ElectrostaticTokamaks,Connor1980StabilityIII} (Eqns. (10), (35) therein, respectively), albeit in the case of wholly neglecting any of the geometric derivatives (i.e. $h_{\parallel}^{(1)}=h_{\parallel}^{(0)}=0$) for Ref.~\cite{Cheng1981ElectrostaticTokamaks} or while neglecting the effect of second-order geometry variations (i.e. $h_{\parallel}^{(0)}=0$) in case of Ref.\cite{Connor1980StabilityIII}. Note that this latter simplification may be made regardless of the ordering on length scales on $\gradpar{\ln \hat{\phi}},\gradpar{\ln \mathrm{geo}}$, as the contributions to the linear source term in \Cref{eq:adv-diff} scale as $h_\parallel^{0}/\htorloc \sim (v_{T\mathrm{i}} N_{fp} / (\omega qR))^2 \sim \order{(\omega_{\mathrm{ti}}/\omega)^2}$, provided that the eigenmodes are not too strongly localised to regions where $\htorloc(\omega,l)\approx 0$. Another subtle difference with respect to Ref.\cite{Connor1980StabilityIII}, concerns the expression for $h_\parallel^{(1)}$, which they implicitly define as $h_{\parallel}^{(1)}=\gradpar{h_{\parallel}^{(2)}}-h_\parallel^{(2)}\gradB$, which differs by a factor of $\gradB (\{\mathbb{I}_{2,1}^{3,0}\}+ \{\mathbb{I}_{0,3}^{3,0}\}/2 )$ from \Cref{eq:semi-ana-integrals_expand} obtained from explicitly expanding the nested parallel derivatives in \Cref{eq:asymp_disp_rel}. 
Lastly, if a subsidiary high-frequency expansion for $\omega_{ds}/\omega \ll 1$ is considered, the kinetic integrals $\mathbb{I}_{n,m}^{p,q}$ may be solved analytically, and, as shown in \Cref{app:nonres_binom_lim}, the resulting dispersion relation can be reduced to the various non-resonant forms mentioned in \Cref{sec:intro}, depending on the choice of further approximations. \par
Compared with the recently considered approach by Ref.~\cite{Rodriguez2025TheLocalisation} to develop a kinetic ITG model including the effects of mode localisation, their model permits retaining the effect of parallel streaming to leading order along with the drift resonance rather than through an asymptotic expansion, as considered here. However, this increase in fidelity of finite-transit effects comes at the cost of requiring several additional approximations in order to obtain a closed form dispersion relation. Our model does not require to limit all inhomogeneities in the magnetic geometry along the field line to that of a quadratic profile in the magnetic drift, does not impose a form of the eigenfunction solutions $\hat{\phi}$ restricted to a Gaussian envelope with a fluid-closure for its width, and retains toroidal resonance effects for both curvature and $\grad{B}$ components of the magnetic drift, rather than adopting a so-called curvature model for the drift where resonances in the latter component are neglected. Particularly, this last approximation has been shown to poorly reproduce the effects from drift resonances, resulting in underestimated growth rates of the toroidal ITG branch \cite{Terry1982KineticMode,Morren2026FastApproach}. All things considered, we thus expect our model to provide more accurate predictions for toroidally-driven ITG modes in realistic magnetic geometries -- especially in stellarator equilibria without quasi-symmetry where the curvature will not be periodic along the field line and could not simply be described by a single quadratic well \cite{Boozer1983TransportEquilibria,Nuhrenberg1988Quasi-helicallyStellarators} -- provided that the conditions for the validity of the asymptotic expansion (see \Cref{app:asymp_breakdown}) are satisfied.

\section{Alternative reduced dispersion models} \label{sec:altmod}
In addition to \Cref{eq:asymp_disp_rel}, we also consider two additional simplified dispersion relation models for toroidal ITG which can be applied to arbitrary geometry. Our goal in this endeavour is to provide alternative reduced models that the asymptotic dispersion relation can be compared to and tested against, in particular with regard to its treatment of non-local effects along the field. \par 
The first alternative stems from a recently developed dispersion model which assumes the influence of parallel streaming on ions to be negligible, but reincorporates some global effects along the field line by integrating the local dispersion relation along the field line as weighted by the eigenmode \cite{Morren2026FastApproach}
    \begin{align}
        D_{\mathrm{global}} = \int \htorloc(\omega,l) \hat{\phi}(l)^2 \frac{\dd{l}}{B}.
        \label{eq:adhoc_aeitg_global_disp}
    \end{align}
The form of \Cref{eq:adhoc_aeitg_global_disp} originates from a variational treatment in Ref.~\cite{Morren2026FastApproach} where the effects of trapped electrons are included; however, as also demonstrated therein, \Cref{eq:adhoc_aeitg_global_disp} is capable as an ad-hoc model to predict ITG eigenfrequencies in the adiabatic-electron limit. \par
Secondly we consider the so-called semi-local dispersion relation \cite{Xie2016GlobalGradient}
    \begin{align}
        D_{\mathrm{semi-loc}} = 1 + \specprefac \left(1-\int \mathcal{F}_{Ms} \frac{\omega-\omega_{*s}^T}{\omega-\langle{\omega_{ds}\rangle} - \sqrt{\langle{k_\parallel^2\rangle}} v_\parallel} J_{0}\left(\sqrt{\expval{b_s}} \frac{v_\perp}{v_{Ts}}\right)^2 \dd^3{\bm{v_s}}\right),
        \label{eq:semiloc_disp}
    \end{align}
which reinstates the effects of parallel streaming to $\order{1}$ by means of an effective parallel wavenumber, which is based on the operator norm of the streaming term in the GKVE [\Cref{eq:lin_GKVE}] \cite{Hirose1995IonTokamaks}
    \begin{align}
        \expval{k_\parallel^2} = - \int \hat{\phi} \nabla_\parallel^2 \hat{\phi} \dd{l} / \int \hat{\phi}^2 \dd{l} = \int \left(\gradpar{\hat{\phi}}\right)^2/\int \hat{\phi}^2 \dd{l}.
        \label{eq:kpar_SL_norm}
    \end{align}
Here used $\gradpar{} = \pdv*{}{l}$ and integration by parts are used. The norms of the perpendicular operators are defined analogously as
    \begin{align}
        \expval{\omega_{ds}} =& \int \omega_{ds}(l)\hat{\phi}(l)^2 / \int \hat{\phi}^2(l) \dd{l}, 
        \label{eq:omd_SL_norm} \\
        \expval{b_s} =& \int (k_\perp \rho_{Ts})^2 \hat{\phi}(l)^2 \dd{l} / \int \hat{\phi}(l)^2 \dd{l}.
        \label{eq:kperp_SL_norm}
    \end{align}
The heuristic inclusion of parallel dynamics into the semi-local model of \Cref{eq:semiloc_disp} modifies the ion Landau resonance compared with the toroidal limit, but the modification is essentially in the same form as the local limit of the ITG dispersion relation, and thus the resonance may be described with \cite{Gurcan2014NumericalCurvature}
    \begin{align*}
        \int \mathcal{F}_{Ms} \frac{\omega-\omega_{*s}^T}{\omega-\langle{\omega_{ds}\rangle} - \sqrt{\langle{k_\parallel^2\rangle}} v_\parallel} J_{0}\left(\sqrt{b_s} \frac{v_\perp}{v_{Ts}}\right)^2 \dd^3{\bm{v_s}} = \{\mathbb{I}_{0,1}^{1,0}\}
    \end{align*}
having made use of the notation of \Cref{eq:Inmpq_triplet_shorthand}, with the sole modification that we now consider the integrals $\mathbb{I}_{n,m}^{p,q}$ under the substitution of $\mathsf{g}_n(z_s(u_\perp,\omega))\rightarrow G_n(z_{s,+}'(u_\perp,\omega),z_{s,-}'(u_\perp,\omega))$ in \Cref{eq:Inmpq_integral}, where
    \begin{align}
        z_{s,\pm}'= \frac{1}{2} \left(-\frac{\sqrt{\expval{k_\parallel}^2}v_{Ts}}{\sqrt{2}\expval{\omega_{\kappa,s}}} \pm \sqrt{\frac{\expval{k_\parallel^2}v_{Ts}^2}{2 \expval{\omega_{\kappa,s}}^2} - 4 \left(\frac{\expval{\omega_{\grad{B},s}}}{2 \expval{\omega_{\kappa,s}}} u_\perp^2-\frac{\omega}{2\expval{\omega_{\kappa,s}}}\right)} \right)
    \end{align} 
are the roots of the shifted Landau resonance. \par
Note that in our definition of the operator norms in \Cref{eq:kpar_SL_norm,eq:kperp_SL_norm,eq:omd_SL_norm}, we choose to deviate from Refs.~\cite{Xie2016GlobalGradient,Hirose1995IonTokamaks} where Hermitian norms are employed (i.e. $\hat{\phi}^2 \rightarrow |\hat{\phi}|^2$ and likewise $(\gradpar{\hat{\phi}})^2 \rightarrow $ $|\gradpar{\hat{\phi}}|^2$). We motivate this deviation by the fact that the gyrokinetic Vlasov-Poisson system is fundamentally non-Hermitian \cite{Stephens2025AInstabilities}, and consequently the streaming term in particular may be poorly represented by a real-valued wavenumber. By employing non-Hermitian norms, the operator norms for the magnetic drift frequency and normalised perpendicular wavenumber formally also become complex-valued, even though the underlying operators are strictly real-valued. In \Cref{app:Herm_vs_nonHerm} we investigate the severity of this discrepancy for the cases studied in this manuscript, where it is shown that the (artificially) introduced imaginary component in the norm of both perpendicular operators is relatively small and the overall magnitude of these norms does not differ significantly from their Hermitian counterparts, particularly in the case of $\expval{b_s}$, with $\expval{\omega_{ds}}$ showing a somewhat greater sensitivity to the choice of the type of norm. By contrast, the effective parallel wavenumber is more substantially affected by the choice to adopt non-Hermitian norms, in accordance with our motivation to introduce this change. We further \textit{a posteriori} validate our use of these non-Hermitian norms in \Cref{sec:results} by comparing the obtained eigenfrequencies with predictions from high-fidelity gyrokinetic simulations. \par
Both alternative dispersion relation models rely on knowledge of an eigenmode guess for $\hat{\phi}$ in order to find the corresponding eigenfrequency satisfying $D(\omega)=0$, rather than self-consistently providing an eigenvalue-eigenmode pair like \Cref{eq:adv-diff}. Comparing the solutions obtained from the asymptotic dispersion relation to the global model may reveal whether retaining streaming effects also has an $\order{1}$ effect on the corresponding eigenfrequencies, or mostly determines the shape of the corresponding eigenmodes. Meanwhile, comparing against the semi-local model may reveal whether streaming effects ought to be retained resonantly and may not accurately be accounted for asymptotically.

\section{Numerical setup} \label{sec:numerics}
\subsection{Flux-tube geometries and normalisations}
Throughout this manuscript we adopt a flux-tube framework\cite{Beer1995Field-alignedTurbulence} to solve the various dispersion relations numerically. Thus, we consider a domain spanning on the order of a few poloidal turns along the magnetic field, while extending a handful of gyroradii across $\bm{B}$. To this end, we adopt field-aligned coordinates
\begin{align}
    x = L_{\mathrm{ref}} \sqrt{s}, \quad y= L_\mathrm{ref} \sqrt{s_0} \alpha/q_0, \quad z = \theta, 
    \label{eq:FTcoo}
\end{align}
with $L_\mathrm{ref}$ being a macroscopic normalisation length scale, $s=\Phi/\Phi_{\mathrm{edge}}$ the normalised toroidal magnetic flux, $\alpha=q\theta -\zeta$ the field line label with $\theta,\zeta$ the poloidal and toroidal Boozer angles, respectively, and $q_0 = q(s_0)$ the local safety factor, such that $\bm{B}=B_{\mathrm{ref}} \grad{x}\cross\grad{y}$, and $z$ is the field-aligned coordinate. \par
Subsequently, the use of the ballooning transform permits Fourier treatment of fluctuations in the $x,y$ directions with perpendicular wavevector of $\bm{k_\perp}=k_x \grad{x} + k_y \grad{y}$, while all equilibrium and geometric quantities are assumed to only vary along $z$, as valid when $\rho_{T\mathrm{i}}/a \ll 1$, where $a$ denotes the device minor radius. In particular, since \Cref{eq:FTcoo} defines a non-orthogonal coordinate system, this implies that the perpendicular wavenumber is
    \begin{align}
        k_\perp = \sqrt{k_x^2 g^{xx} + k_y^2 g^{yy} + 2 k_x k_y g^{xy}},
        \label{eq:FT_FLR_mag}
    \end{align}
where $g^{mn} = \grad{\xi^m}\vdot\grad{\xi^n}$ with $\{\xi^{1},\xi^{2},\xi^{3}\}=\{x,y,z\}$ are the components of the contravariant metric tensor of the flux-tube coordinate system, depending on $z$. The geometric terms related to the magnetic drift frequency and its derivatives appearing in \Cref{eq:Inmq_anasolve} are likewise given in terms of the components of this metric tensor as $\bm{k_\perp}\vdot\bm{\mathcal{L}} = k_x \mathcal{L}_x + k_y\mathcal{L}_y$, where \cite{Xanthopoulos2009,Gorler2011}
    \begin{align}   
        \begin{split}
            \mathcal{L}_x =& - \left(\left.\pdv{B_N}{y}\right|_{x,z} + \frac{\gamma^{2}}{\gamma^{1}} \left.\pdv{B_N}{z}\right|_{x,y}\right) \\
            \mathcal{L}_y =& \left.\pdv{B_N}{x}\right|_{y,z} + \frac{\gamma^{3}}{\gamma^{1}} \left.\pdv{B_N}{z}\right|_{x,y}
        \end{split}
    \label{eq:gradB-operators-GENE}
    \end{align}
with $B_{N} = B/B_\mathrm{ref}$ the normalised magnetic field strength, and $\gamma^{1} = g^{xx}g^{yy}-(g^{xy})^2$, $\gamma^{2} = g^{yz}g^{xx}-g^{xz}g^{xy}$ and $\gamma^{3} = g^{xz}g^{yy}-g^{xy}g^{yz}$. Analogously for the curvature component of the magnetic drift we have $\bm{k_\perp}\vdot\bm{\mathcal{K}}=k_x \mathcal{K}_x + k_y \mathcal{K}_y$, where due to the relation between the curvature vector and the magnetic pressure \cite{Helander2014}
\begin{align}
        \bmog{\kappa} = \grad_\perp{\ln B} + 2\beta \grad{\ln p},
\end{align}
where $\grad_\perp = \grad{} - \bm{e_b}(\bm{e_b}\vdot\grad)$ is the perpendicular gradient operator, it follows that $\mathcal{K}_x =\mathcal{L}_x$ and $\mathcal{K}_y = \mathcal{L}_y + \dv*{\beta_{\mathrm{ref}}}{x}/(2 B_N)$, where $\beta_{\mathrm{ref}} = 2\mu_0 p/B_{\mathrm{ref}}^2$. Lastly, upon adopting the coordinates from \Cref{eq:FTcoo}, the parallel derivative operator becomes $\gradpar{} = (1/(\sqrt{g}B_N)) \pdv*{}{z}$, where $\sqrt{g} = \left(\left(\grad{x}\cross\grad{y}\right)\vdot\grad{z}\right)^{-1}$ is the determinant of the metric tensor. \par
These geometric quantities are subsequently normalised by scaling equilibrium length scales to $L_{\mathrm{ref}}=\{a,R\}$, for stellarator and tokamak geometries, respectively, where $R$ denotes the major radius. Meanwhile, fluctuation length scales are normalised to the ion gyroradius at sound speed $\rho_{s} = c_s / \Omega_{i,\mathrm{ref}}$ where $c_s = \sqrt{T_{\mathrm{e}}/m_i}$  and $\Omega_{i,\mathrm{ref}} = e B_{\mathrm{ref}}/m_i$. Lastly, all frequencies are normalised to $c_{s}/L_{\mathrm{ref}}$. These normalisations are  directly compatible with the \textsc{Gene}/GIST package \cite{Xanthopoulos2009}, allowing for the treatment of realistic geometries based on MHD equilibria. \par 
As test cases for the various reduced ITG dispersion models, we consider the $\alpha=0$ flux tube geometries located at $s_0 = 0.5$ for the inward-shifted configuration of the Large Helical Device (LHD) stellarator\cite{Iiyoshi1990DesignDevice} and the high-mirror configuration of the Wendelstein 7-X (W7-X) stellarator \cite{Beidler1990}. In case of W7-X this particular flux tube is commonly known as the so-called \textit{bean flux tube}, which intersects the bean-shaped cross section of the flux surface at the outboard mid-plane. These flux tubes are characterised by relevant macroscopic equilibrium parameters of $q_0 = (1.43,  1.10)$ , $\hat{s}= (-1.36,-0.12)$, $ N_{fp} = (10,5)$,  $R = (3.673 \mathrm{m},\ 5.483 \mathrm{m})$ and $a = (0.636 \mathrm{m}, \ 0.521\mathrm{m})$, respectively, where $\hat{s}=\dv*{\ln q}{\ln x}$ denotes the global magnetic shear at the flux surface. \par
Meanwhile, for an axisymmetric geometry we consider the $\hat{s}$-$\alpha$ model for a large-aspect-ratio device with circular flux surfaces \cite{Connor1978}. This model is described by metric coefficients \cite{MerzThesis}
    \begin{align}
        g^{xx}=&1  &g^{xy} &= \hat{s}z -\alpha_{\mathrm{MHD}} \sin{z}  &g^{yy} &=  1+ (\hat{s}z -\alpha_{\mathrm{MHD}} \sin{z})^2\nonumber \\
        g^{xz} =&0  &g^{yz} &= 1/r_0 &g^{zz} &= 1/r_0^2
        \label{eq:s-a_metric}
    \end{align}
where $r_0$ is the minor radial coordinate of the flux tube position, and $\alpha_{\mathrm{MHD}}=-q(r_0)^2 R \left.\dv*{\beta_\mathrm{ref}}{r}\right|_{r_0}$ denotes the Shafranov shift. Note, however, that in \Cref{eq:s-a_metric} the definition of magnetic-shear parameter is taken as $\hat{s}= (r_0/q(r_0)) \left.\dv*{q}{r}\right|_{r0}$, which is slightly different from the definitions employed in the stellarator configurations, as $r_0$ is directly used as a radial coordinate in lieu of $x_0$, though the two definitions agree in case of concentric flux-surfaces. \Cref{eq:s-a_metric} is supplemented with $B_{N}=(1+\epsilon \cos z)^{-1}$, where $\epsilon=r_0/R$ is the local inverse aspect ratio, $\sqrt{g}=q(r_0) R/B_N$, while the magnetic drift components are taken to be
    \begin{align}
        \mathcal{L}_x = - \frac{\sin{z}}{R}, \quad \mathcal{L}_y = - \frac{\cos z + \sin{z} \left(\hat{s}z-\alpha_{\mathrm{MHD}}\sin z\right)}{R},
        \label{eq:s-a_drifts}
    \end{align}
which do not follow self-consistently from the metric tensor, since the model has $\order{\epsilon}$ inconsistencies with respect to an exact analytical description of circular flux surfaces, see Ref.~\cite{Lapillonne2009}. Despite these limitations, the $\hat{s}$-$\alpha$ model is commonly used for gyrokinetic benchmarks \cite{Dimits2000,Nevins2006}. Here we consider the specific instance of an $\hat{s}$-$\alpha$ geometry defined by the so-called Cyclone Base Case (CBC), for which $\hat{s} = 0.796, q_0 = 1.4, \epsilon=0.18$ \cite{Dimits2000}.\par
\begin{figure}
    \centering
        \includegraphics[width=0.33\textwidth]{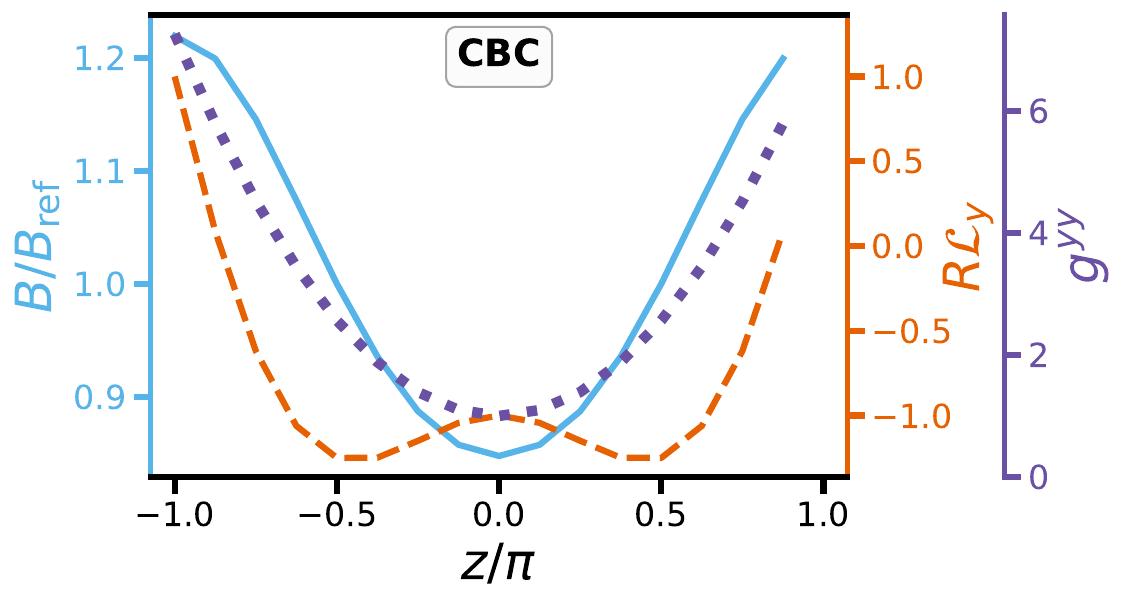}%
        \includegraphics[width=0.33\textwidth]{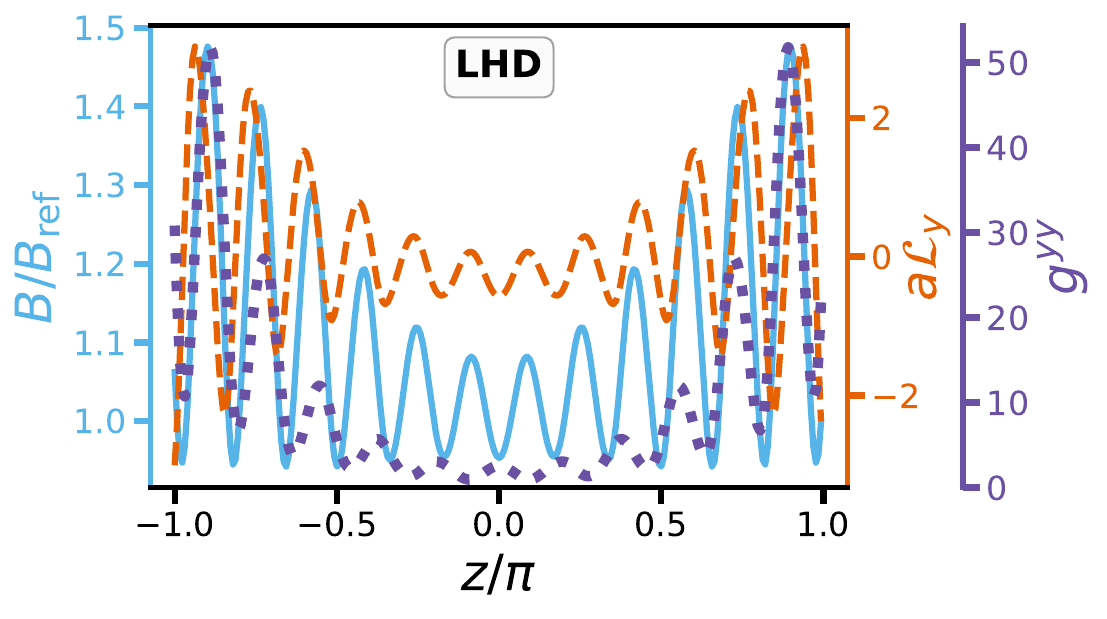}%
        \includegraphics[width=0.33\textwidth]{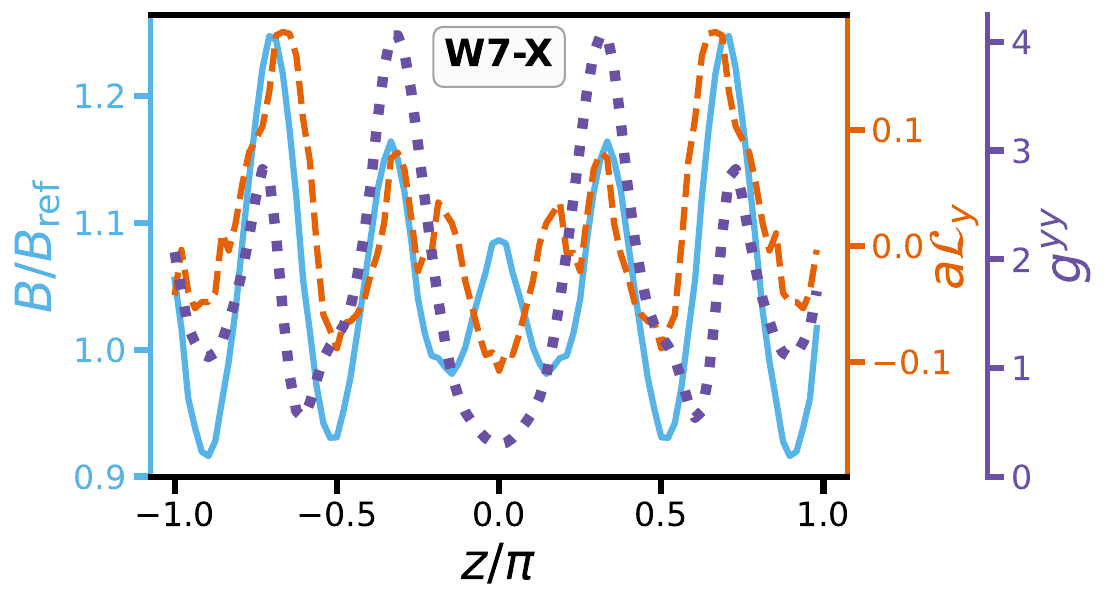}
    \caption{Variation of the geometric quantities $B$ (solid blue line), $g^{yy}$ (dotted purple line), and $\mathcal{L}_y$ (dashed orange line) along the magnetic field for the flux-tube geometries of an $\hat{s}$-$\alpha$ tokamak for CBC parameters (left), the inward-shifted configuration of the LHD stellarator (middle) and the high-mirror configuration of the W7-X stellatator (right) used for both the high-fidelity \textsc{Gene} simulations and to solve the dispersion relations from \Cref{sec:disp_deriv,sec:altmod}.}
    \label{fig:geometries}
\end{figure}

For consistency with the electrostatic approximation, we consider vacuum fields for both stellarator devices and take the $\alpha_{\textrm{MHD}}\rightarrow 0$ limit for the $\hat{s}$-$\alpha$ tokamak, reducing the geometry to concentric circular flux surfaces. The geometric quantities of interest that define the flux-tube geometries for the three main geometries considered in this work are shown in \Cref{fig:geometries}.

\subsection{Gyrokinetic simulation setup}
For each of the aforementioned configurations, we perform linear initial-value gyrokinetic simulations with the \textsc{Gene} code\cite{Jenko2000}, which serve a two-fold purpose; Firstly to establish a baseline of eigenfrequencies and electrostatic potential eigenmodes to verify the solutions of \Cref{eq:asymp_disp_rel}, and secondly to provide optimal guesses for the mode structure used in both the alternative reduced dispersion relation models from \Cref{sec:altmod}. In these \textsc{Gene} simulations, the ballooning transform permits the use of periodic boundary conditions in the $x,y$ directions, while a twist-and-shift boundary condition \cite{Beer1995Field-alignedTurbulence,Martin2018}
    \begin{align}
        \delta\tilde{A}_{k_x,k_y}(z,t) = \delta\tilde{A}_{k_x+p\Delta k_x,k_y}(z+2\pi p N_{\mathrm{pol}},t) \tilde{C}_{ky}^p
        \label{eq:twist-and-shift}
    \end{align}
is applied in the $z$ direction for every fluctuating quantity $\delta A$ decomposed into perpendicular Fourier modes $\delta \tilde{A}_{k_x,k_y}$, where $p\in\mathbb{Z}$, $\Delta k_x= 2\pi N_{\mathrm{pol}} k_y \hat{s}$ and $\tilde{C}_{ky}$ is a phase factor. \par
For each configuration, we vary the binormal wavenumber $k_y \rho_s=[0.1,0.2,\cdots,2]$ at fixed gradient drive of $a/L_n=2, a/L_T = 4$ for the stellarator configurations, while for the circular tokamak we modify the CBC gradients to $R/L_n = 7.86, R/L_T = 15.71$ such that these match with the equivalent gradients considered for the stellarator cases\footnote{Having used the fact that in the $\alpha_{\mathrm{MHD}}\rightarrow0$ limit the $\hat{s}$-$\alpha$ flux surfaces are perfectly concentric, hence the flux-surface label follows as $s_0 = (r_0/a)^2$, resulting in a scaling of the gradients $R/L_{n,T}=a/L_{n,T} \times \sqrt{s_0}/\epsilon$ in terms of the local inverse aspect ratio $\epsilon=r_0/R$. By taking the flux-surface label $s_0=0.5$, identical to the radial position used in stellarator flux tubes, this results in a gradient scaling factor of $\sqrt{s_0}/\epsilon\approx3.93$ for CBC geometry parameters.}, which we shall refer to as the CBC-like scenario for the remainder of the manuscript. We consider each flux tube to span one poloidal turn along the field line and centre the ballooning space of eigenmodes at $k_x =0$, which is where typically the most unstable modes are found in axisymmetric devices\cite{Duff2022EffectTurbulence} and the bean flux tube of stellarators \cite{Faber2015,Sanchez2021,Proll2013}. For each geometry, we performed convergence checks in the numerical resolutions until relative differences in the eigenfrequency were below $5\%$ across the full $k_y$-spectrum upon doubling the resolutions, resulting in the discretisations shown in \Cref{tab:GENE_res}.

\begin{table}[b]
\caption{Converged numerical resolutions of high-fidelity \textsc{Gene} simulations for flux-tube simulations of $\hat{s}$-$\alpha$ tokamak, LHD and W7-X stellarators. $N_z, N_{v\parallel}, N_{\mu}$ refer to the number of grid points used for the field-aligned coordinate, parallel velocity grid and magnetic moment, respectively, while $N_{kx}$ refers to the number of Fourier modes considered in the radial envelope of the fluctuations for the twist-and-shift boundary condition, where $N_\mathrm{pol}=1$ for each of the considered configurations.}
\centering
\begin{tabular}{l | c c c}
\hline
 & $\hat{s}$-$\alpha$ & LHD & W7-X \\
\hline
$N_z$ & 16 & 256 & 96 \\
$N_{kx}$ & 5 & 5 & 43 \\
$N_{v\parallel}$  & 32 & 24 & 32 \\
$N_{\mu}$ & 12 & 8 & 12 \\
\hline
\end{tabular}
\label{tab:GENE_res}
\end{table}

\subsection{Solving the dispersion relation} \label{sec:numerics-main}
The asymptotic dispersion relation \Cref{eq:adv-diff} is solved numerically via finite differences, calculating the geometric derivatives using fourth-order finite differences on the GIST equilibrium data and discretising the derivative operators acting on $\hat{\phi}$ using a second-order scheme with error correction, see \Cref{app:numerics_shenanigans} for details. \Cref{eq:adv-diff} is then reduced to 
    \begin{align}
        \underline{M}(\omega) \hat{\phi} = \bm{0}.
        \label{eq:NL_EVP}
    \end{align}
Self-consistent solutions $(\omega^\star,\hat{\phi}^\star)$ to \Cref{eq:NL_EVP} are determined by the nonlinear eigenvalue problem $\det \underline{M}(\omega^\star)=0$ and $\hat{\phi}^\star \in \ker \underline{M}(\omega^\star)$, with admissible eigenfunctions subjected to the boundary condition of $\lim_{L\rightarrow \infty} \hat{\phi}(L)=0$, which is solved separately for ballooning-parity (even $\hat{\phi}$) and tearing-parity (odd $\hat{\phi}$) eigenmodes. \par
Due to the equivalence between finite-$k_x$ modes within the flux tube and the extended ballooning representation of a $k_x=0$ mode beyond the original parallel extent of the flux tube represented by the twist-and-shift boundary condition [\Cref{eq:twist-and-shift}], we restrict ourselves to $k_x=0$ modes and solve \Cref{eq:adv-diff} on the extended ballooning domain. To construct $\underline{M}(\omega)$ an extension of the geometric elements into the extended ballooning domain is necessary, as given by \cite{Morren2026FastApproach}
\begin{align}
        \sqrt{g}(z+2\pi p N_{\mathrm{pol}}) =& \sqrt{g}(z) \nonumber \\
        g^{xx}(z+2\pi p N_{\mathrm{pol}})=& g^{xx}(z) \nonumber \\
        g^{xy}(z + 2\pi p N_{\mathrm{pol}}) =& g^{xy}(z)+ 2\pi p N_{\mathrm{pol}} \hat{s}g^{xx}(z) \nonumber \\
        g^{yy}(z + 2\pi p N_{\mathrm{pol}}) =& g^{yy}(z)+ 4\pi p N_{\mathrm{pol}} \hat{s} g^{xy}(z) +(2\pi p N_{\mathrm{pol}}\hat{s})^2 g^{xx}(z) \nonumber \\ 
        B(z+2\pi pN_{\mathrm{pol}}) =& B(z) \nonumber \\
        \mathcal{L}_x(z+2\pi p N_{\mathrm{pol}})=& \mathcal{L}_x(z) \nonumber \\
        \mathcal{K}_x(z+2\pi p N_{\mathrm{pol}}) = & \mathcal{K}_x(z) \nonumber \\
        \mathcal{L}_y(z+2\pi p N_{\mathrm{pol}}) = & \mathcal{L}_y(z) + 2\pi p N_{\mathrm{pol}} \hat{s} \mathcal{L}_x(z) \nonumber \\
        \mathcal{K}_y(z+2\pi p N_{\mathrm{pol}}) = & \mathcal{K}_y(z) + 2\pi p N_{\mathrm{pol}} \hat{s} \mathcal{L}_x(z).
        \label{eq:geometry-extend-formulae}
    \end{align}
Formally, both \Cref{eq:twist-and-shift,eq:geometry-extend-formulae} are only correct in axisymmetric configurations, as they neglect the variation of the geometry with toroidal angle. It has, however, been shown that despite this discrepancy in the geometry, reasonably accurate results can be obtained with this method if the flux tube is chosen with a sufficiently large $N_{\mathrm{pol}}$, which may necessitate $N_{\mathrm{pol}}>1$ at low global magnetic shear \cite{Martin2018,Faber2018}. In previous studies of Refs.~\cite{Sanchez2021,Martin2018}, $N_{\mathrm{pol}}=1$ has been observed to be sufficient for LHD and W7-X, and the use of \Cref{eq:geometry-extend-formulae} to represent the geometry in extended ballooning space has not been found restrictive in the previous study of Ref.~\cite{Morren2026FastApproach}. \par
Some care, however, needs to be taken in the evaluation of the resonant integrals $\mathbb{I}_{n,m}^{p,q}$ [\Cref{eq:Inmpq_integral}] when assembling the matrix $\underline{M}(\omega)$. These integrals are derived under the assumption that $\omega_{\kappa,s} \neq 0$, yet in limit of vanishing magnetic drift, the toroidal resonance disappears, and these integrals have the well-defined analytical limit (see \Cref{app:nonres_binom_lim}))
\begin{align}
    \lim_{\omega_{ds}\rightarrow 0} \mathbb{I}_{n,m}^{p,q} = \frac{\left(n-1\right)!!}{\omega^p} \frac{1+(-1)^n}{2} F_{m}^{q}(b_s)
    \label{eq:zero_drift_lim_integrals}
\end{align}
where $F_{m}^{q}(b_s)$ is a Weber-type integral incorporating the effects of FLR damping. The form of \Cref{eq:zero_drift_lim_integrals} has a relatively strong scaling of $\omega^{-p}$ with the eigenfrequency, which is particularly problematic at transport-relevant long-wavelength instabilities for which a drift-wave ordering of the frequency $\omega \propto k_y$ predicts a diverging tendency for $\mathbb{I}_{n,m}^{p,q}$ as $p$ increases. In order to guarantee confinement of passing particles, the magnetic drift must change direction along the field line, such that points with $\bm{v_{ds}}\approx \bm{0}$ are unavoidable, as exemplified by e.g. the ubiquitous zero-crossings of $\mathcal{L}_y$ in \Cref{fig:geometries}. The nature of this divergence is a direct consequence of the asymptotic expansion used to derive \Cref{eq:asymp_disp_rel}. While we have assumed $\omega_{\mathrm{ti}}/\omega \ll 1$ -- which is the correct expansion parameter when the magnetic drift is near-zero -- this condition requires $\omega$ to be large, in which case \Cref{eq:zero_drift_lim_integrals} is well-behaved. The drift-wave ordering, however, implies that below some critical wavenumber $\omega_{\mathrm{ti}}/\omega \sim \order{1}$ will be reached since $\omega_{\mathrm{ti}}$ is a fixed geometric property, in which case it is no longer valid to consider parallel streaming effects to have a negligible effect on the instability dynamics, especially locally near the zero crossings of the magnetic drift where parallel streaming is the only non-vanishing source of instability drive. Hence, the asymptotic expansion cannot be made to hold globally along the full domain at all wavenumbers, corresponding to the scenario when \Cref{eq:zero_drift_lim_integrals} will have diverging tendencies. \par
We have observed that when this discrepancy in the underlying physics is left unmended the system matrix $\underline{M}(\omega)$ becomes strongly (diagonally) dominated near the zero crossings of the magnetic drift as a result from the contribution by the correction terms in $h_\parallel^{(1)}, h_\parallel^{(0)}$ that are proportional to $\gradpar{\omega_{ds}},\gradpar[2]{\omega_{ds}}$ and $(\gradpar{\omega_{ds}})^2$ when the effect of the parallel inhomogeneity in the magnetic drift is included. These terms correspond to $\mathbb{I}_{n,m}^{p,q}$ integrals with $p=4$ and $p=5$, respectively, and are also characterised by the largest values of $n$ (see \Cref{eq:semi-ana-integrals_expand}), further exacerbating the divergence of \Cref{eq:zero_drift_lim_integrals}. The resulting eigenmodes were found to be numerical artefacts localised to the zero-crossings of the magnetic drift, while not solving the local dispersion relation $\htorloc(\omega^\star,l)\neq 0$. \par
To circumvent these issues with the breakdown of the asymptotic expansion, we locally apply a compliance filter which weights the integrals $\mathbb{I}_{n,m}^{p,q}(z;\omega) \rightarrow \mathbb{I}_{n,m}^{p,q}(z;\omega) W(z;w)$ in case of $p=4,5$, where the weight function $W(z;w)$ is chosen to be a product of so-called ``smootherstep'' functions, providing a sigmoid-like function which smoothly interpolates $[0,1]$, while maintaining continuity of the second derivative, see \Cref{app:weight_func_design}. These smooth S-curve filters are applied over a width of $w$ grid points both to the left and right of each zero crossing of the magnetic drift, where the widths were tuned to be $w=4,10,6$ for high-shear tokamak, LHD and W7-X cases, respectively. For widths much smaller than these, the eigenfrequency predictions of the model diverge, particularly at low wavenumber, while much larger widths effectively do not permit the inclusion of the physics associated with parallel inhomogeneities in the magnetic drift, as eigenfrequency predictions match those obtained from the model with these integrals suppressed altogether (i.e. $W(z;w)=0$), especially at high wavenumber, see \Cref{app:sensitivity_study}. \par
Another problem inherent to the asymptotic dispersion relation model is that only the physics associated with the toroidal drive and FLR damping are incorporated, which determine the growth rate and limit the extent of the eigenmode in ballooning space. From analysis of the asymptotic dispersion relation in the non-resonant limit for eigenmodes strongly localised to the outboard mid-plane, it is known that \Cref{eq:adv-diff} admits eigenmode solutions \cite{Plunk2014,Xie2016GlobalGradient,Horton1981ToroidalGradients}
\begin{align}
    \hat{\phi}(z)\sim H_l(\sqrt{\sigma(\omega)}z) \exp(-\sigma(\omega)z^2/2),
    \label{eq:hermite_solutions}
\end{align}
where $H_{l}$ denote the $l^{\textrm{th}}$ Hermite polynomial, with Hermite polynomial order $l$ and the Gaussian width $\sigma(\omega)$ determining the mode's fine-scale substructure and total extent along the field line, respectively. The lowest-order solutions $l=0,1$ are referred to as the ballooning-parity and tearing-parity ground-state, respectively, while larger even/odd $l$ correspond to higher excitation states of the ballooning/tearing-parity ITG instability, respectively. The growth rates of these excited states increases moderately with excitation number $l$ \cite{Plunk2014,Horton1981ToroidalGradients}, while these higher-excitation eigenmodes are also more strongly affected by parallel dynamics and thus greater Landau damping as a result of their finer scale substructure  \cite{Rodriguez2025TheLocalisation,Xie2016GlobalGradient}, resulting in a limit on finite-$k_\parallel$ structures that can be sustained by physical eigenmodes. Nonetheless, we remark that such higher-excitation state instabilities are proper solutions to the full gyrokinetic Vlasov-Poisson system and can dominate the ground-state ITG mode if the drive is sufficiently large to overcome parallel Landau damping, as is the case for steep-gradient regions \cite{Han2017MultipleBarriers,Pueschel2019OnDevices,Xie2016GlobalGradient}. Although this eigenmode classification is based on a non-resonant expansion in tokamak geometry, we find that it generally also applies to the eigenmodes obtained from \Cref{eq:adv-diff} when resonant effects are included and non-axisymmetric geometries are considered. Consequently, there is no inherent limit on the $k_\parallel$-spectrum of the resulting eigenmodes obtained from \Cref{eq:adv-diff}, as the effects of parallel Landau damping are lost in the asymptotic expansion of parallel streaming effects, the validity of which is therefore limited to low-$l$ solutions. \par
These issues do not persist in the two alternative dispersion relations considered in \Cref{sec:altmod}, which are characterised by a simple scalar root-finding problem $D(\omega|\hat{\phi})=0$ for a given $\hat{\phi}$, which should be chosen to correspond to a physically viable eigenmode. In case of the global model, an initial guess for the mode frequency is first obtained using a non-resonant expansion of $\htorloc$, reducing \Cref{eq:adhoc_aeitg_global_disp} to a simple quadratic dispersion relation, see Ref.~\cite{Morren2026FastApproach}. In case of the semi-local dispersion relation, evaluation of \Cref{eq:semiloc_disp} is significantly cheaper as the resonant integral needs only to be evaluated for fixed and precomputed values of $\expval{k_\parallel^2}, \expval{\omega_{ds}}, \expval{b_s}$. Consequently, \Cref{eq:semiloc_disp} is solved directly using a grid-search method for the eigenfrequency. 

\section{Comparison of dispersion model predictions} \label{sec:results}
Here we present the unstable ITG eigenfrequency predictions for each of the adiabatic-electron ITG dispersion relations discussed in \Cref{sec:disp_deriv,sec:altmod}. In case of the self-consistent asymptotic dispersion relation, we solve for three variants of \Cref{eq:adv-diff} to investigate the importance of the geometric derivatives which arise from ordering $\gradpar{\ln \mathrm{geo}}/\gradpar{\ln \hat{\phi}} \sim O(1)$. One version includes all the physics contained in the resonant integrals \Cref{eq:semi-ana-integrals_expand}, another neglects the influence of the parallel inhomogeneity in the magnetic drift (i.e. ${I}_{n,m}^{p,q}=0$ for $p=4,5$) allowing us to retain the remaining parallel geometric inhomogeneities without the need of a compliance filter, and lastly, one neglects the influence of geometric derivatives altogether, viz. considering $h_{\parallel}^{(1)}=h_{\parallel}^{(0)}=0$. These three variants will be referred to as the ``full'', ``hybrid'' and ``minimal'' dispersion models, respectively. In order to limit the growth rates and $k_\parallel$-spectrum of the eigenmodes obtained from \Cref{eq:adv-diff}, we consider an \textit{a posteriori} correction to the eigenfrequency predictions 
\begin{align}
    \omega' \rightarrow \omega -i \frac{\sqrt{2} v_{T\mathrm{i}} k_\parallel}{\pi},
    \label{eq:freq_Landau_correction}
\end{align}
which, as suggested by Ref.~\cite{Plunk2014}, is capable of providing an accurate estimate of the effects of transit-damping, where we have taken the mode width to be given by $\pi/k_\parallel$. In evaluating \Cref{eq:freq_Landau_correction}, we determine $k_\parallel$ from the corresponding eigenmodes using \Cref{eq:kpar_SL_norm}, analogous to the semi-local model, such that higher excitation states are subjected to a larger transit damping correction. \par 
In case of the heuristic alternative models, for both the ad-hoc global [\Cref{eq:adhoc_aeitg_global_disp}] and semi-local [\Cref{eq:semiloc_disp}] dispersion relations, the electrostatic potential eigenmodes obtained from the \textsc{Gene} simulations were used as a guess for the eigenfunction. Consequently, only the eigenfrequencies from \Cref{eq:adv-diff} are obtained self-consistently. For simplicity, we limit our test cases to an isothermal plasma $T_{\mathrm{i}}=T_{\mathrm{e}}$ with one singly charged ion species $Z_i=1$, though the theory is capable of including the effects of impurities or plasmas with weak thermal coupling between ions and electrons. \par
Note that there is a disparity in the sign of the propagation frequency between the high-fidelity \textsc{Gene} simulations and the analytical theory; in the latter ion drift waves are characterised by $\Re[\omega] \sim \omega_{*i}<0$ for conventionally peaked profiles, while ITGs in \textsc{Gene} are characterised by $\Re[\omega]>0$. In the following we adopt the \textsc{Gene} conventions for the sign of the propagation frequency. Additionally, for the asymptotic dispersion relation we only present results for the most unstable eigenmode, using $\Im[\omega']$ as the selection criterion in order to limit the solutions to lower excitation states for which the asymptotic expansion is valid, with the shown results below corresponding to the original uncorrected eigenfrequency predictions satisfying $\det \underline{M}(\omega)=0$.

\subsection{High-shear tokamak}
\begin{figure}[t]
        \centering
            \includegraphics[width=0.95\textwidth]{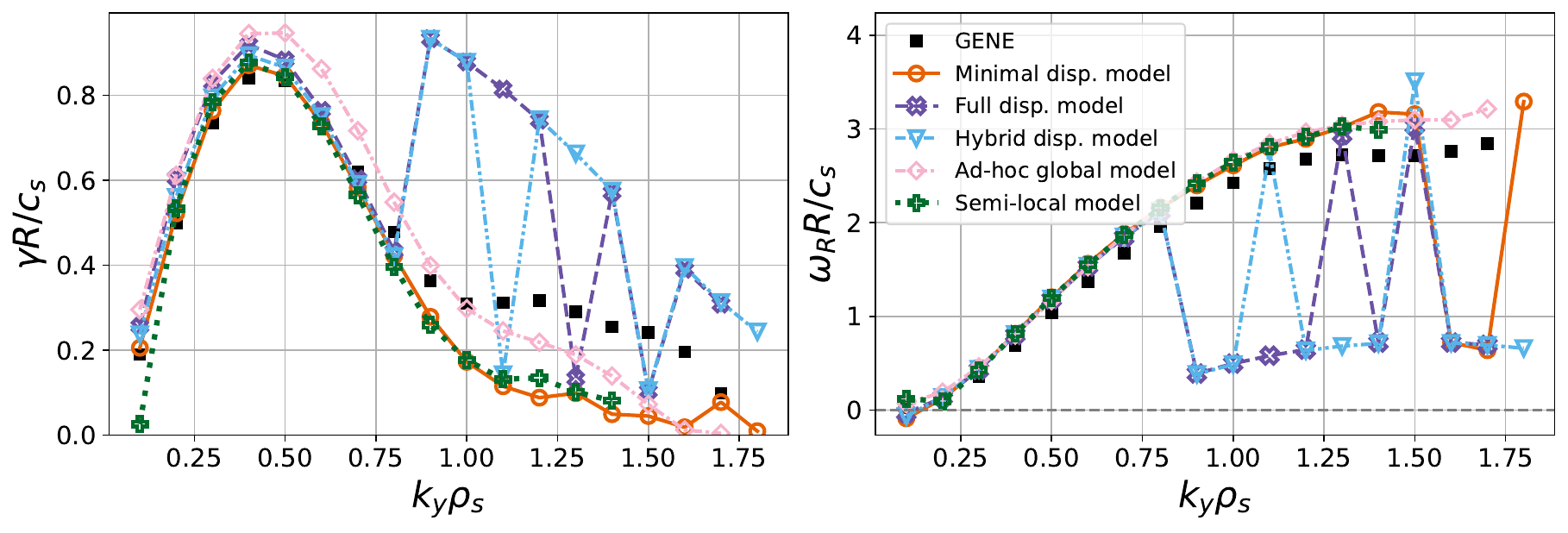}
        \caption{Growth rate and propagation frequency as functions of binormal wavenumber a CBC-like $\hat{s}$-$\alpha$ tokamak as obtained from the minimal (solid line, orange circles), hybrid (dash-dot-dotted line, blue triangles) and full (dashed line, purple crosses) variants of the asymptotic dispersion relation, the ad-hoc global model (dash-dotted line, pink diamonds) and the semi-local model (dotted line, green plusses), as well as reference solutions obtained by high-fidelity \textsc{Gene} simulations (black squares).}
        \label{fig:CBC_result}
\end{figure}

Results for the eigenfrequency predictions in the  CBC-like scenario are shown in \Cref{fig:CBC_result}. Note that for $k_y\rho_s > 1.4$, the semi-local dispersion relation -- even when using the exact \textsc{Gene} eigenfunctions -- is not able to find any unstable solutions with a sufficiently small residual for $D_{\mathrm{semi-loc}}(\omega)$ and its results were subsequently omitted there. With the exception of those few short-wavelength eigenmodes, both inconsistent models provide a roughly equal prediction for the propagation frequency $\omega_R = \Re[\omega]$. By contrast, the ad-hoc global model predicts consistently higher growth rates $\gamma = \Im[\omega]$ than the semi-local model, which may be attributed to the lack of parallel ion Landau damping in the former, while the salient features thereof are included in the semi-local model through an effective parallel wavenumber based on \Cref{eq:kpar_SL_norm}. Nonetheless, both models tend to capture the trends observed in \textsc{Gene} simulations, though they predict a faster rate of stabilisation for $k_y \rho_s > 1$. The consistent asymptotic dispersion relation also provides generally accurate predictions for the eigenfrequency, with growth rates in all three of its variants roughly between the predictions of the two heuristic models up to $k_y \rho_s = 0.8$. The minimal asymptotic model, in particular, provides a slightly better match to the high-fidelity \textsc{Gene} simulations than the semi-local model for $k_y \rho_s<0.5$. For $k_y \rho_{s}>0.8$ the predictions of the three asymptotic model variants diverge, with the minimal model closely following the results for the semi-local model, while the hybrid and full models display transitions to a more unstable mode, with a significantly smaller propagation frequency. Most notably, the inclusion of the various correction terms due to geometric derivatives consistently predicts higher growth rates than the minimal model, especially beyond $k_y \rho_s = 0.8$. The greater influence of the geometric derivatives for the higher-wavenumber instabilities is to be expected from \Cref{eq:semi-ana-integrals_expand}, as the prefactors for nearly all terms in $h_\parallel^{(1)}, h_\parallel^{(0)}$ are proportional to either $k_y$ or $k_y^2$, with the exception of the $\gradpar{\ln B} \{\mathbb{I}_{0,3}^{3.0}\}$ term in $h_{\parallel}^{(1)}$ being scale-independent.

\begin{figure}[t]
    \centering
        \includegraphics[width=0.95\textwidth]{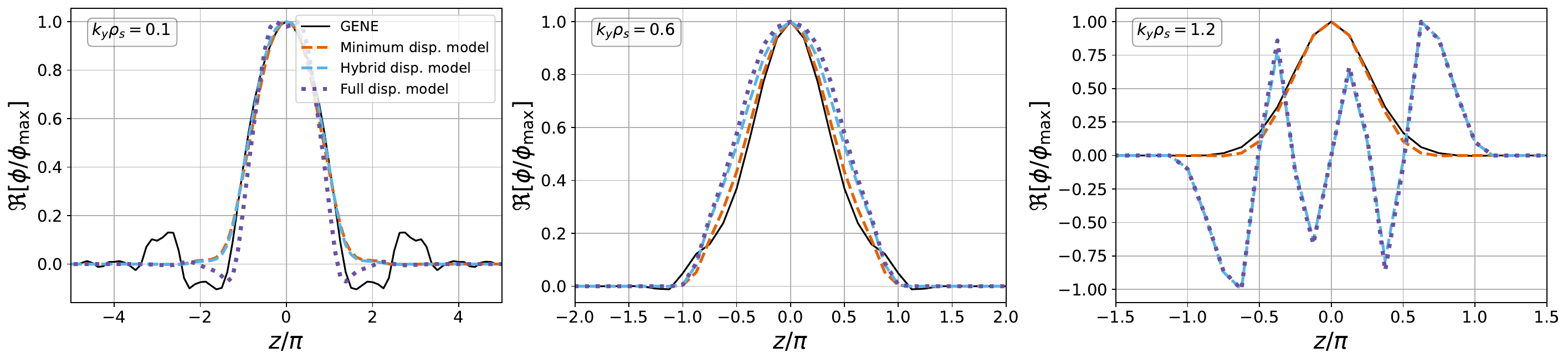}
    \caption{Eigenfunctions of the electrostatic potential $\Re[\phi]$ along the field line in the extended ballooning representation for the CBC-like tokamak geometry. From left to right, three different wavenumbers are shown with increasing $k_y\rho_s$. The high-fidelity \textsc{Gene} simulation results (solid black lines) are compared with predictions from the minimal (dashed orange lines), hybrid (dash-dotted blue lines) and full (dotted purple lines) variants of the asymptotic dispersion relation. Note that all eigenfunctions have been normalised with respect to their maximum.}
    \label{fig:CBC_phi_comparison}
\end{figure}

When examining the corresponding electrostatic potential eigenmodes, as shown in \Cref{fig:CBC_phi_comparison}, a clear distinction between the $k_y \rho_s \leq 0.8$ and $k_y \rho_s > 0.8$ modes can be made. The longer-wavelength modes closely capture the high-fidelity \textsc{Gene} eigenfunctions, with the inclusion of the geometric derivatives yielding a slight improvement in the tails of the lowest-wavenumber eigenfunction ($|z/\pi|>1$) at the cost of the development of a weak double-hump structure near the outboard mid-plane ($|z/\pi|\ll 1$) instead of a monotonic Gaussian feature as observed in both the minimal model and the high-fidelity simulation. At intermediate wavelengths, all three variants of the model accurately predict the increased localisation of the eigenfunction to the outboard mid-plane under influence of FLR damping, though the inclusion of parallel magnetic inhomogeneities tends to slightly widen the eigenfunction. Meanwhile, at shorter wavelengths beyond the observed mode transition, the minimum model continues to yield an $l=0$ ground state eigenmode matching the behaviour of the high-fidelity eigenmodes despite increased discrepancy in the growth rate, whereas the inclusion of geometric derivatives induces an $l=5$ excited state as the most unstable eigenmode. We remark that without the inclusion of the parallel Landau damping correction [\Cref{eq:freq_Landau_correction}], such a transition to excited states of the ITG mode as the most dominant instability would already occur at $k_y \rho_s = 0.6$ for all three model variants, including the minimum model where the growth rate of an $l=3$ mode overtakes that of the ground state ITG. In all three model variants, however, the $l=0$ ITG remains a valid subdominant eigenmode solution to the asymptotic dispersion relation at shorter wavelengths, as explicitly exemplified for the minimum model by \Cref{fig:CBC_result,fig:CBC_phi_comparison}.

\subsection{Large Helical Device} \label{sec:LHD}
\begin{figure}[t]
    \centering
        \includegraphics[width=0.95\textwidth]{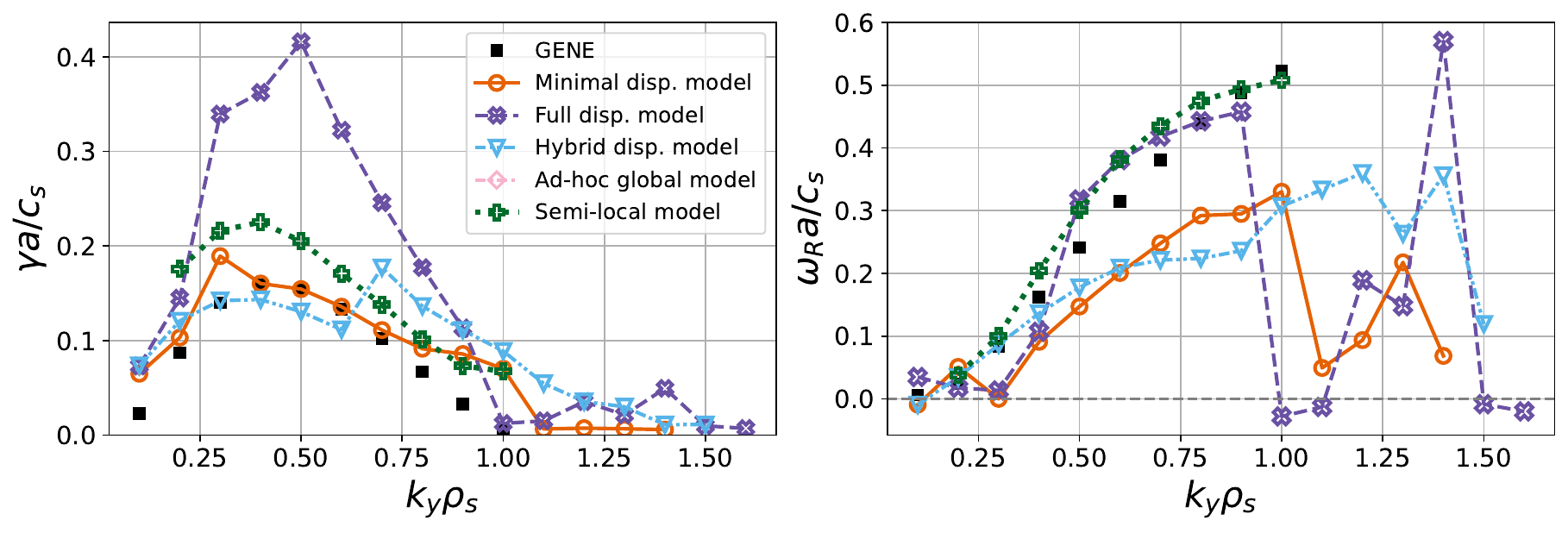}
    \caption{Growth rate and propagation frequency as functions of binormal wavenumber in the inward-shifted configuration of LHD as obtained from the minimal (solid line, orange circles), hybrid (dash-dot-dotted line, blue triangles) and full (dashed line, purple crosses) variants of the asymptotic dispersion relation, the ad-hoc global model (dash-dotted line, pink diamonds) and the semi-local model (dotted line, green plusses), as well as reference solutions obtained by high-fidelity \textsc{Gene} simulations (black squares). Note the lack of unstable solutions from the ad-hoc global dispersion model.}
    \label{fig:LHD_result}
\end{figure}

Results for the ITG eigenfrequencies obtained for LHD are shown in \Cref{fig:LHD_result}. Note that no unstable solutions were found with the ad-hoc global model, similarly to the semi-local model at $k_y \rho_s = 0.1$, and have hence been omitted from the plot. Likewise, it was observed that when conventional Hermitian norms are used for $\langle k_\parallel^2 \rangle, \expval{\omega_{ds}}$ and $\expval{b_s}$, the semi-local model was not able to find any unstable solutions, while remarkably accurate results are obtained when non-Hermitian norms are employed. The inability of the ad-hoc model and semi-local model with Hermitian operator norm to obtain any solutions emphasises not only the importance of retaining some aspects of the particle streaming, but also the need for an accurate representation of the underlying operator by the heuristic parallel wavenumber. \par
Concerning the asymptotic dispersion relation, both the minimal and hybrid models yield more accurate predictions for the growth rate than the semi-local model up to $k_y \rho_s = 0.9$, and $k_y \rho_s = 0.7$, respectively, although their prediction for the propagation frequency are roughly smaller by a factor of two, with the hybrid model yielding smaller discrepancies at lower wavenumber. The full asymptotic relation dispersion model, generally yields the most accurate predictions for the propagation frequency, though it significantly overestimates the growth rates for $k_y \rho_s \in [0.3,0.8]$. All three asymptotic model variants, however, predict instability to persist up to shorter wavelengths than predicted by the high-fidelity simulations, where ITGs are stabilised beyond $k_y \rho_s = 1$. This discrepancy may be attributed to the stabilising influence of parallel Landau damping in the high-fidelity simulation (as will be demonstrated explicitly in \Cref{sec:toroidal_vs_slab_competition}), which is inherently absent from the asymptotic model\footnote{Recall that the correction for finite-transit damping \Cref{eq:freq_Landau_correction} is only applied to the model predictions \textit{a posteriori} to limit the $k_\parallel$ of the most dominant eigenmode. The growth rate determined by the condition $\det{\underline{M}(\omega)}=0$ follows from an asymptotic treatment of parallel streaming, and is thus determined by a balance between toroidal resonant drive and FLR damping.}, which instead relies exclusively on FLR damping to quench the instability. Unlike the high-shear tokamak case, the inclusion of geometric derivatives does not have a uniformly destabilising effect, although including the parallel inhomogeneity in the magnetic drift results in additional destabilisation up to $k_y \rho_s = 0.8$, while providing additional stabilisation at shorter wavelengths, with exception of $k_y \rho_s = 1.4$. The greater differences between the minimal model and both the hybrid and full models compared with the axisymmetric case is in accordance with the scaling argument that $\gradpar{\ln \mathrm{geo}} \sim N_{fp}/(qR)$ is considerably larger in many-field-period stellarators such as LHD. We additionally remark that without the inclusion of the correction factor for finite-transit damping in \Cref{eq:freq_Landau_correction}, the growth rate of the most unstable eigenmode would be larger than the prediction from the semi-local model across the full spectrum for all three variants of the model.

\begin{figure}
    \centering
        \includegraphics[width=0.95\textwidth]{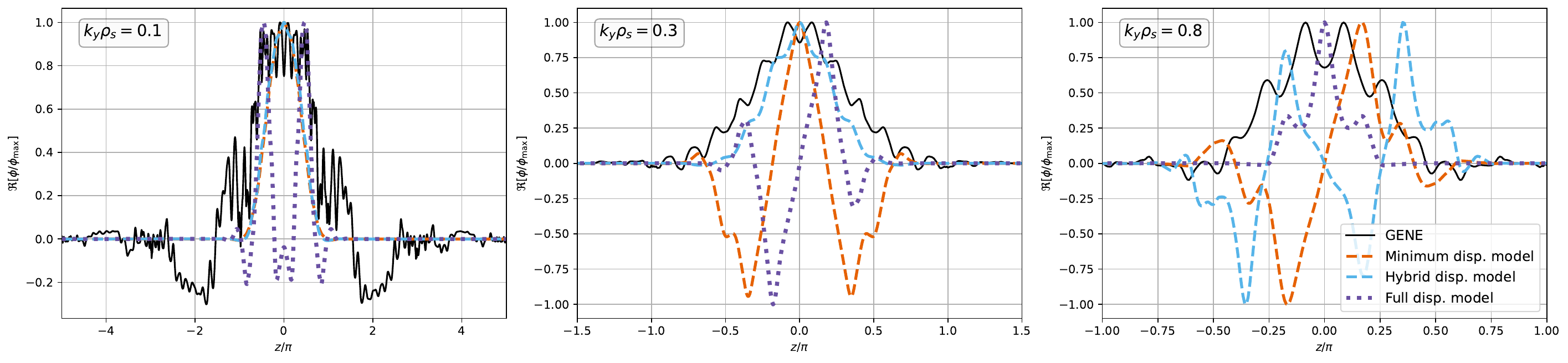}
    \caption{Eigenfunctions of the electrostatic potential $\Re[\phi]$ along the field line in the extended ballooning representation for the LHD geometry. From left to right, three different wavenumbers are shown with increasing $k_y\rho_s$. The high-fidelity \textsc{Gene} simulation results (solid black lines) are compared with predictions from the minimal (dashed orange lines), hybrid (dash-dotted blue lines) and full (dotted purple lines) variants of the asymptotic dispersion relation. Note that all eigenfunctions have been normalised with respect to their maximum.}
    \label{fig:LHD_phi_comparison}
\end{figure}

When examining the eigenfunctions in \Cref{fig:LHD_phi_comparison} we note that -- with the exception of $k_y \rho_s = 0.1$ -- the envelope of the eigenfunctions predicted by the minimal and hybrid models have a similar extent along the field line as the envelope of the high-fidelity eigenmodes, albeit with a mismatch in their substructure. The high-fidelity eigenfunctions predominantly show the existence of relatively small and narrow ripples along the envelope -- which corresponds to an $l=0$ ground state mode -- with the ripples presumably bound to the geometric features of the flux tube (compare \Cref{fig:LHD_phi_comparison} with \Cref{fig:geometries}). Meanwhile, the predictions from the asymptotic dispersion relation yield eigenmodes consisting of undulations significantly wider than the ripples observed in the high-fidelity \textsc{Gene} simulations, corresponding to higher excitation states of the instability. At low wavenumber both the minimal model and hybrid model yield an $l=0$ ground state mode, and subsequently undergo a transition to $l=2$ and $l=3$ excited states at $k_y \rho_s = 0.3$ and $k_y \rho_s = 0.7$, respectively. The full model displays the opposite behaviour, starting from an excited state at low wavenumber, before returning to a ground-state-like mode for $k_y \rho_s \geq 0.5$, albeit with an envelope width that is generally narrower than the high-fidelity eigenmodes. Hence, it must be concluded that the inclusion of geometric derivative terms in the asymptotic dispersion relation -- especially the terms related to the parallel inhomogeneity in the magnetic drift -- does not necessarily lead to an improvement in the predictions for either eigenfrequency or eigenmode. \par 
Returning to the more peculiar $k_y \rho_s = 0.1$ case, we observe that the high-fidelity eigenmode is significantly more extended along the field and more prone to the aforementioned ripples. Additionally, the \textsc{Gene} eigenfunction does not have a monotonically decaying envelope, but rather develops a plateau extending for roughly one poloidal turn around the outboard mid-plane. Neither of these effects are reproduced by any of the variants of the asymptotic dispersion model, which instead predict eigenmodes more akin to what is observed at other wavenumbers. This eigenmode is particular to the $k_y \rho_s = 0.1$ wavenumber, with all other \textsc{Gene} eigenmodes more closely resembling those shown for $k_y \rho_s = 0.3,0.8$. This odd structure of the eigenmode may be attributed to a greater importance of parallel streaming effects, which are more likely to dominate over the instability drive from magnetic drift at such long wavelengths since $\omega_{ds}\propto k_y$. \par 

\subsection{Wendelstein 7-X}
\begin{figure}[b!]
    \centering
        \includegraphics[width=0.95\textwidth]{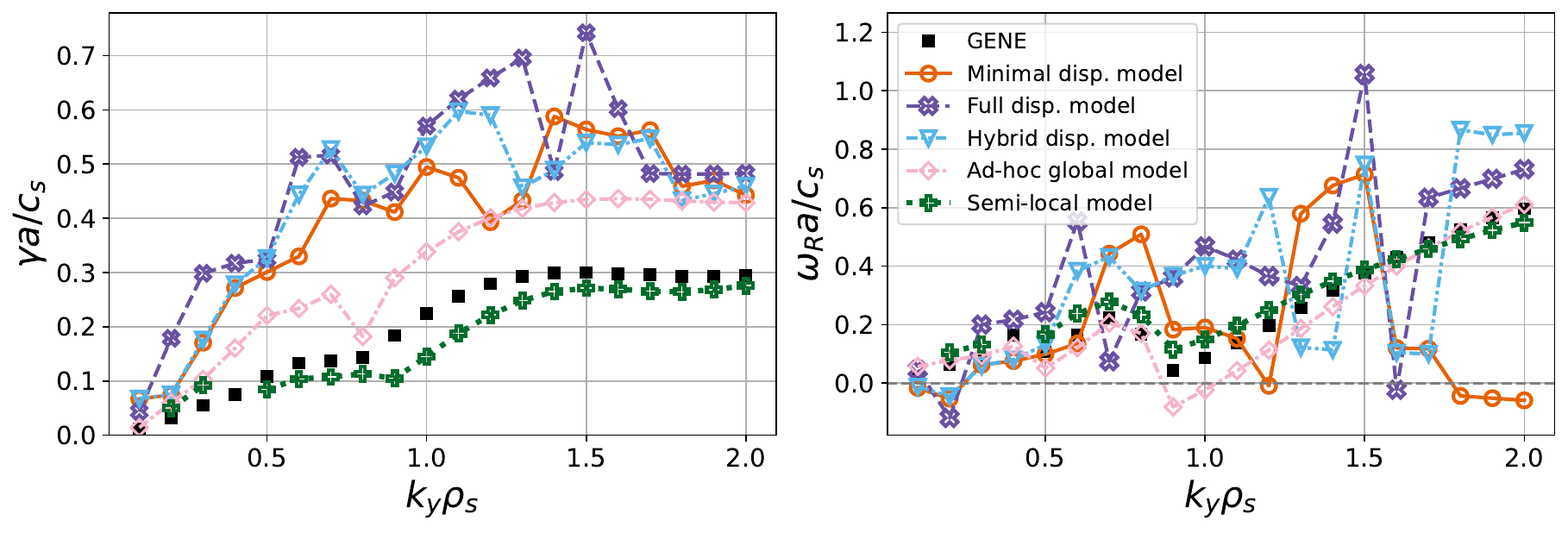}
    \caption{Growth rate and propagation frequency as functions of binormal wavenumber in the high-mirror configuration of W7-X as obtained from the minimal (solid line, orange circles), hybrid (dash-dot-dotted line, blue triangles) and full (dashed line, purple crosses) variants of the asymptotic dispersion relation, the ad-hoc global model (dash-dotted line, pink diamonds) and the semi-local model (dotted line, green plusses), as well as reference solutions obtained by high-fidelity \textsc{Gene} simulations (black squares).}
    \label{fig:W7X_result}
\end{figure}

The eigenfrequency predictions obtained from the reduced dispersion relations in the high-mirror configuration of W7-X are shown in \Cref{fig:W7X_result}. Note that the semi-local model did not yield any unstable solutions for either $k_y \rho_s = 0.1, 0.4$ and is hence there omitted from the plot; the lack of unstable solutions persists regardless of whether one uses Hermitian or non-Hermitian definitions for the operator norms. At the remaining wavenumbers, however, the predictions for the eigenfrequency obtained from the semi-local model are significantly more accurate than those obtained from the ad-hoc global model. Whereas the former only marginally underpredicts the growth rates, the latter substantially overpredicts instability, which hints at the relative importance of parallel streaming effects in W7-X.  \par 

The trends observed in the high-fidelity eigenfrequency are somewhat qualitatively reproduced by the asymptotic dispersion model up to roughly $k_y \rho_s \approx 0.5$, with a relatively small error in the propagation frequency, but a significant overestimation of the growth rate. Meanwhile, at shorter wavelengths the predictions by all three variants of the asymptotic dispersion model become significantly more erratic, with strong non-monotonic behaviour in the growth rate, accompanied by frequent mode transitions, even though the latter are distinctly absent from the high-fidelity above $k_y \rho_s  = 0.9$. Up to $k_y \rho_s = 1.4$, however, the inclusion of the geometric derivatives results in an additional destabilisation with respect to the minimal asymptotic dispersion model, which at low wavenumber is mostly associated with the parallel inhomogeneity of the magnetic drift, whereas at higher wavenumbers this is attributable to the variation of the $k_\perp$ and $B$ along the field line. The additional destabilisation by the derivative of the magnetic drift is, however, not as severe as observed in LHD. This may be explained from \Cref{fig:geometries} as the geometric features of W7-X are more spread out (poloidally) along the field line compared to LHD, and is thus conform with the qualitative scaling argument $\gradpar{\ln \mathrm{geo}} \sim N_{fp}/(qR)$ noting that the value of $qR$ is similar between the two configurations, though the number of field periods differs by a factor of two. \par

\begin{figure}[t]
    \centering
        \includegraphics[width=0.95\textwidth]{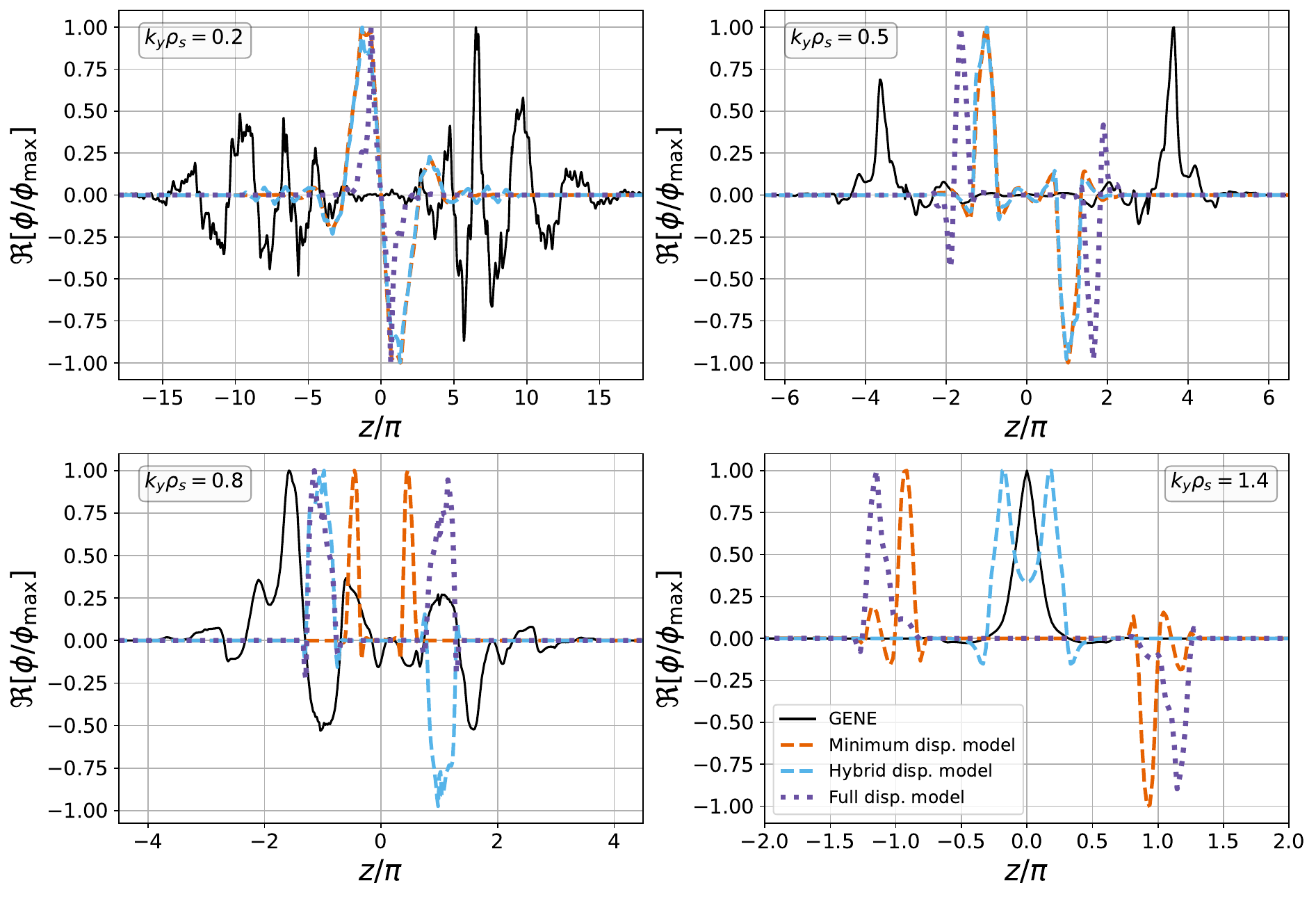}
    \caption{Eigenfunctions of the electrostatic potential $\Re[\phi]$ along the field line in the extended ballooning representation for the W7-X geometry. From left to right, top to bottom four different wavenumbers are shown with increasing $k_y\rho_s$. The high-fidelity \textsc{Gene} simulation results (solid black lines) are compared with predictions from the minimal (dashed orange lines), hybrid (dash-dotted blue lines) and full (dotted purple lines) variants of the asymptotic dispersion relation. Note that all eigenfunctions have been normalised with respect to their maximum.}
    \label{fig:W7X_eigenmodes}
\end{figure}

The predictions for the eigenmodes by all three variants of the asymptotic dispersion relation, as shown in \Cref{fig:W7X_eigenmodes}, deviate significantly from the high-fidelity \textsc{Gene} eigenfunctions, especially at long wavelengths where the latter yields eigenmodes significantly extended along the field line, whereas the former yields strongly localised eigenmodes in comparison, with the area of localisation not aligning with any particular features of the high-fidelity eigenmodes. It should be noted, however, that the high-fidelity eigenfunctions have a tendency to be somewhat asymmetric up to $k_y \rho_s = 0.9$, whereas the asymptotic dispersion model is configured to solve for modes with a well defined parity, such that an exact match here is not expected. At shorter wavelengths, however, the high-fidelity simulations reveal the instabilities to be tokamak-like, corresponding to an $l=0$ ground state ballooning-parity mode with a strong localisation to the outboard mid-plane. Although this feature is not reproduced by the dominant eigenmode obtained by any of the variants of the asymptotic dispersion model, an eigenmode with similar features exists as a subdominant eigenmode. Unlike the CBC-like and LHD scenarios, the eigenmodes obtained from the asymptotic dispersion relation model cannot readily be classified according to the paradigm of \Cref{eq:hermite_solutions}, with eigenmode solutions in W7-X instead typically corresponding to a pair of localised peaks occurring (anti)symmetrically around $z=0$, with a near vanishing amplitude around the outboard mid-plane. \par

\subsection{Geometric sensitivity}
The results in the preceding sections have focussed on different classes of toroidal confinement devices, and consequently the underlying differences in the salient geometric features between the configurations can be substantial. In order to investigate whether the asymptotic dispersion relation model contains sufficient sensitivity to more subtle and continuous changes to the magnetic geometry, we revisit the $\hat{s}$-$\alpha$ tokamak and consider varying the global magnetic shear $\hat{s}$. For this exercise, we consider a range of shear values $\hat{s}=[-1.5,-1.3,\cdots,1.5]$ at a fixed binormal wavenumber of $k_y \rho_s = 0.2$, while maintaining the same drive of $R/L_n= 7.86,\ R/L_T =15.71$ as considered for the CBC-like scenario above. This range of shear values was chosen to bridge the gap between conventional tokamaks, typically operating at large positive shear, and the stellarator configurations of LHD and W7-X which are characterised by a strong and weak negative global shear, respectively. However, we note that moderately negative-shear scenarios may also be obtained in tokamaks by means of current control to create a hollow current profile, and may be of interest for e.g. continuous power plant operation\cite{Taylor1997PhysicsTokamaks,Kessel1994ImprovedShear,Goldston1994AdvancedProspects}. \par
\begin{figure}
    \centering
    \includegraphics[width=0.95\textwidth]{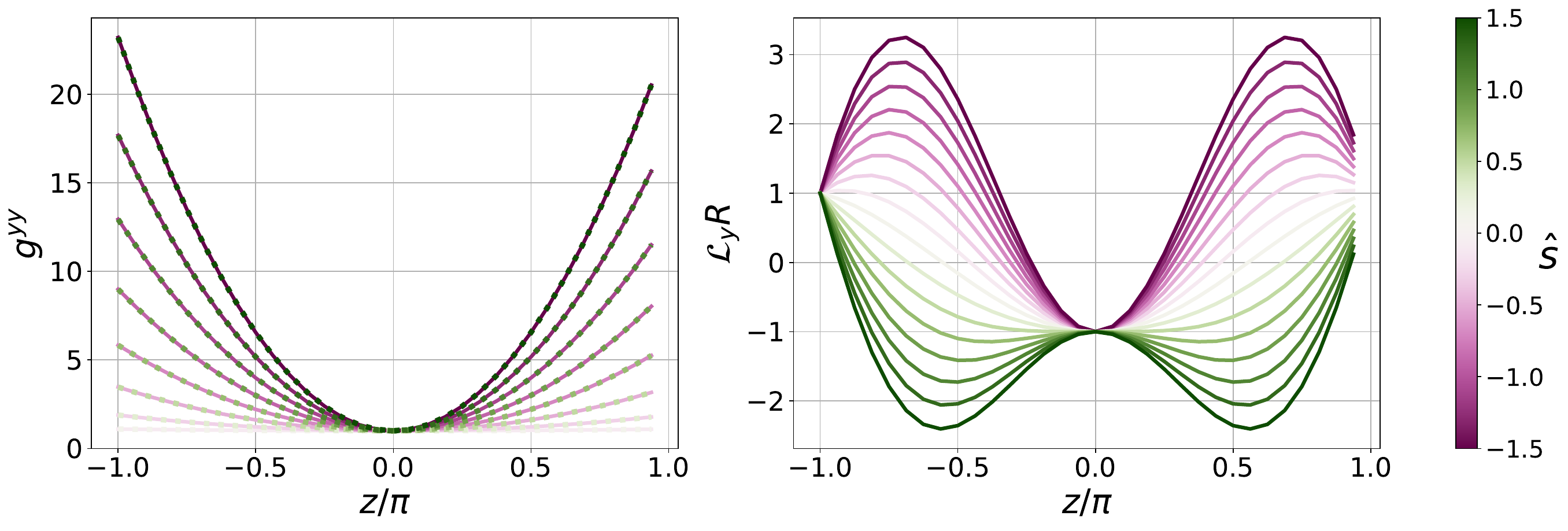}
    \caption{Variation of the metric coefficient $g^{yy}$ and bi-normal component of the magnetic drift $\mathcal{L}_y$ for an $\hat{s}$-$\alpha$ tokamak with global magnetic shear along one poloidal turn. Note that we have taken $\alpha_{\textrm{MHD}}=0$ in accordance with the electrostatic limit, rendering the influence of shear symmetric in the FLR term but not in the magnetic drift.}
    \label{fig:shat_geometry_effect}
\end{figure}

Modifying the global shear $\hat{s}$ directly affects the metric tensor components and the binormal component of the curvature drift, see \Cref{eq:s-a_metric,eq:s-a_drifts}, respectively. Modifications to the magnetic shear will subsequently affect the balance between FLR damping and instability drive by toroidal resonances. In the limit of $k_x=0$ instabilities, this will be determined by a competition of $g^{yy}$ versus $\mathcal{L}_y$, the variation of which with $\hat{s}$ is shown in \Cref{fig:shat_geometry_effect}. Note that the sign of the global magnetic shear does not affect the FLR damping, but does impact the shape of the drift well. In particular, the effect of magnetic shear on the shape of the drift well is considerable, with both the extent and depth of the so-called bad-curvature region $\mathcal{L}_y<0$, where toroidal resonances with drift waves in the ion diamagnetic direction $\omega_{*i}$ possible, being significantly reduced.

As the salient features of the geometry, and thus the underlying driving and damping mechanisms, change substantially throughout the scan, the numerical resolutions used for previous CBC-like \textsc{Gene} simulations (see \cref{tab:GENE_res}) were found to be inadequate, and instead we employed resolutions of $N_z=32, N_{kx}=15, N_{v\parallel}=32, N_\mu=12$, along with a flux tube extent of $N_{\mathrm{pol}}=1$ permitted by axisymmetry. Likewise, for the evaluation of the full model asymptotic dispersion relation with the effects of the parallel inhomogeneity of the magnetic drift included, the width of the compliance filter was increased to $w=8$ to maintain the same physical distance along the field line along which the interpolation of the weight function occurs compared with the CBC-like scenario. Observe from \Cref{fig:shat_geometry_effect} that the distance between zero crossings of the magnetic drift around the outboard mid-plane gradually decreases with magnetic shear, although subsequent zero crossings at larger ballooning angles as predicted by \Cref{eq:geometry-extend-formulae} tend to be more consistently separated by roughly $14$ grid points. \par
\begin{figure}[t]
    \centering
        \includegraphics[width=0.95\textwidth]{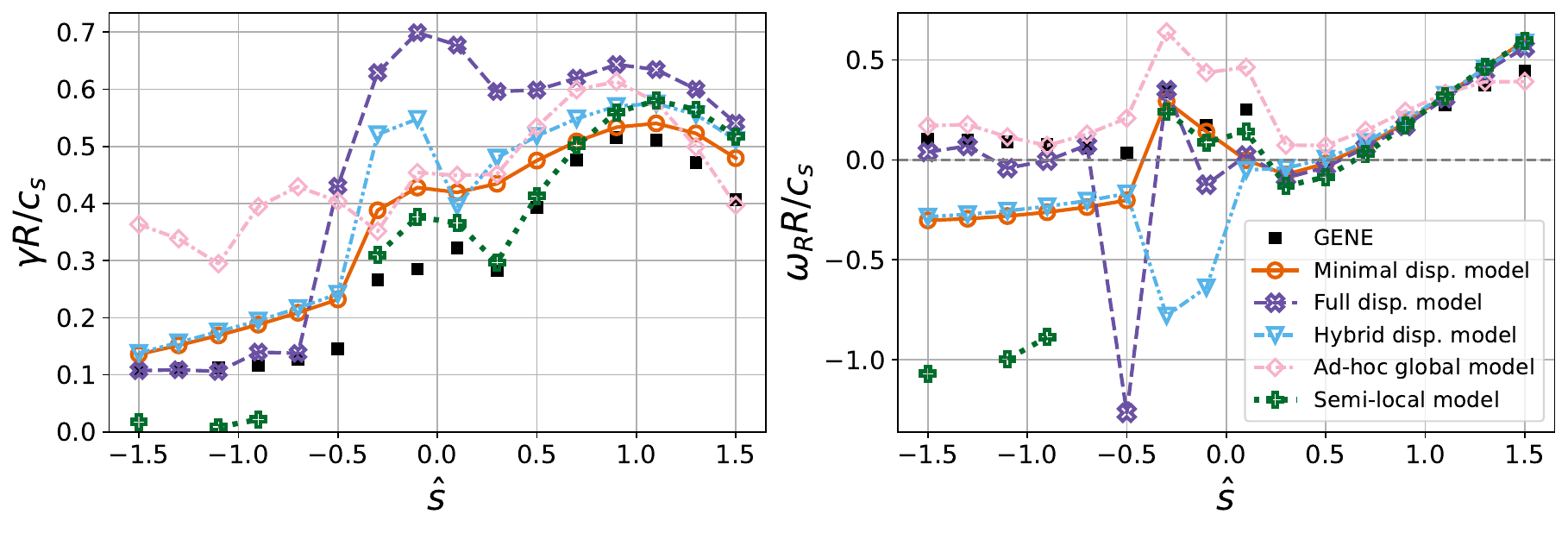}
    \caption{Growth rate and propagation frequency as functions of global magnetic shear at binormal wavenumber of $k_y \rho_s = 0.2$ for an an $\hat{s}$-$\alpha$ tokamak, as obtained from the minimal (solid line, orange circles), hybrid (dash-dot-dotted line, blue triangles) and full (dashed line, purple crosses) variants of the asymptotic dispersion relation, the ad-hoc global model (dash-dotted line, pink diamonds) and the semi-local model (dotted line, green plusses), as well as reference solutions obtained by high-fidelity \textsc{Gene} simulations (black squares).}
    \label{fig:shear_scan_result}
\end{figure}
\begin{figure}[t]
    \centering
        \includegraphics[width=0.95\textwidth]{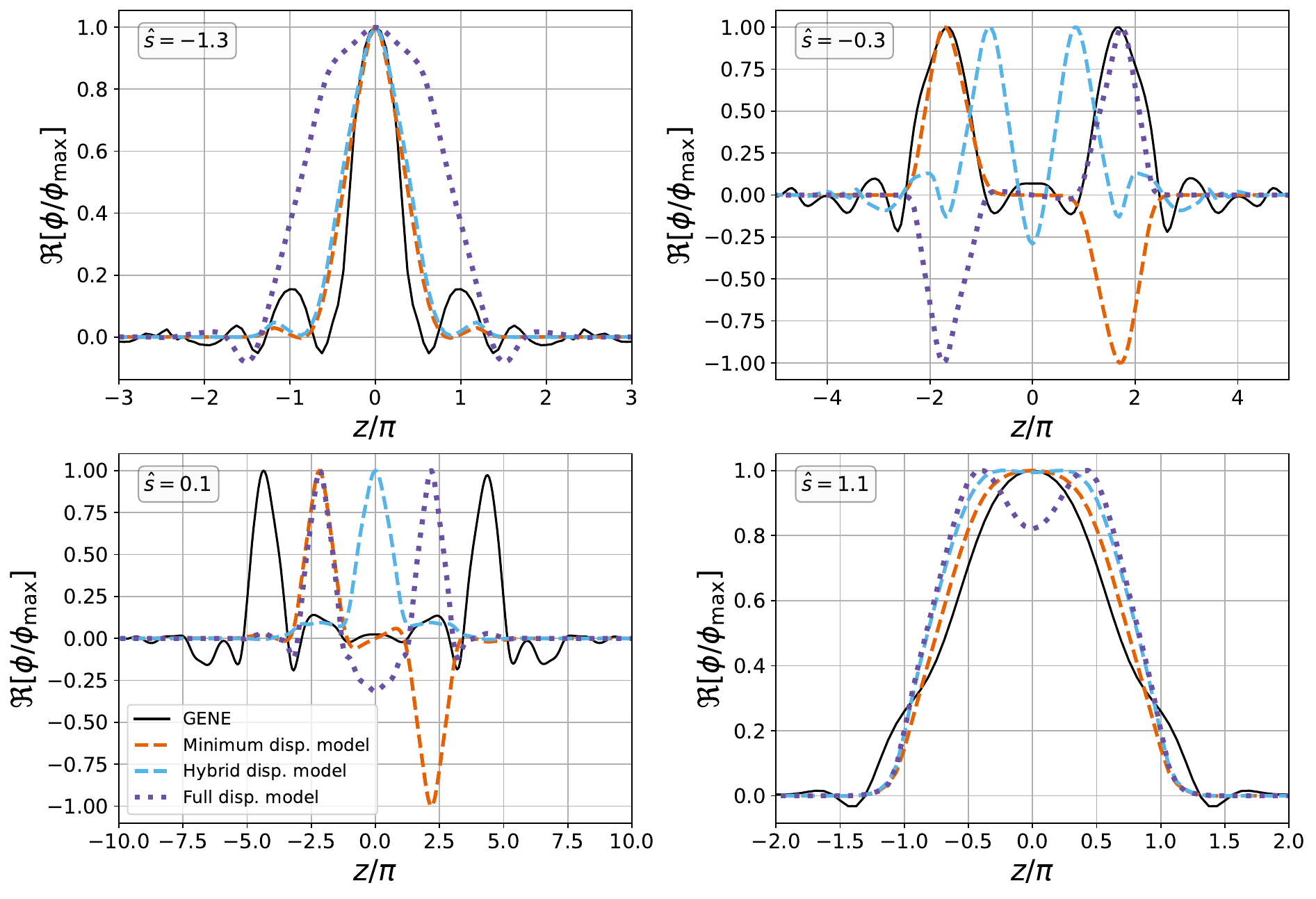}
    \caption{Eigenfunctions of the electrostatic potential $\Re[\phi]$ along the field line in the extended ballooning representation for the circular concentric tokamak geometry at $k_y \rho_s = 0.2$. From left to right, top to bottom, four different global shears are shown with increasing $\hat{s}$. The high-fidelity \textsc{Gene} simulation results (solid black lines) are compared with predictions from the minimal (dashed orange lines), hybrid (dash-dotted blue lines) and full (dotted purple lines) variants of the asymptotic dispersion relation. Note that all eigenfunctions have been normalised with respect to their maximum.}
    \label{fig:shear_scan_eigenmodes}
\end{figure}

Results for the eigenfrequencies and eigenfunctions of the shear scan are shown in \Cref{fig:shear_scan_result,fig:shear_scan_eigenmodes}, respectively. With regards to the eigenfrequency, we note that the semi-local model is only able to provide eigenfrequency predictions down to $\hat{s}=-0.3$; above that value, however, it is consistently more accurate than the ad-hoc global model, with the exception of $\hat{s}=1.3,1.5$. At lower negative shear, the semi-local model only finds a handful of marginally unstable modes, in contrast to the ad-hoc global model, which predicts the continued existence of instability down to $\hat{s}=-1.5$ and beyond, albeit with a growth rate significantly larger than the high-fidelity \textsc{Gene} simulations, despite a close match in the propagation frequency. \par

The self-consistent asymptotic dispersion solver provides quantitatively accurate eigenfrequencies down to a magnetic shear of $\hat{s}=0.5$, whereafter the inclusion of the geometric derivatives results in a strong disparity between the minimal and full variants of the model, particularly closer to zero shear. With exception of the $\hat{s}=-0.3$ and $\hat{s}=-0.1$ cases, the hybrid model and the minimal are in close agreement, with the inclusion of geometric derivatives yielding a small additional destabilisation and shift in the propagation at positive and negative shear, respectively, indicating a weak effect of parallel inhomogeneities, consistent with the scaling argument that $\gradpar{\ln \mathrm{geo}} \sim 1/(qR)$ is generally small in axisymmetric configurations. Further including the effect of the parallel inhomogeneity in the magnetic drift results in a further inflation of growth rates down to $\hat{s}=-0.5$, as also seen in the CBC-like, LHD and W7-X scenarios. By contrast, at even lower negative shear the net effect of the derivative of the magnetic drift is actually stabilising, along with a significant shift in the propagation frequency, both effects resulting in a closer agreement of the full model eigenfrequency with the high-fidelity predictions compared with the minimal and hybrid models. All three variants of the asymptotic dispersion model qualitatively follow the trends of a (local) maximum destabilisation around $\hat{s}\approx 1$, and a sharp decrease in growth rates towards large negative shear for $\hat{s}<-0.5$, as observed in the high-fidelity simulations. Both the hybrid and minimal models, however, more accurately predict the existence of a region of intermediate growth rates near zero magnetic shear for $\abs{\hat{s}}\leq 0.3$, though the full model severely overpredicts the growth rates in this region, instead resulting in a global destabilisation maximum at $\hat{s}=-0.1$. \par 

With regards the eigenfunction, both the minimal and hybrid asymptotic models accurately predict the tendency of the eigenfunction to be a strongly localised $l=0$ ground state at the outboard mid-plane for both large positive and large negative shear, down to $\hat{s}=0.3$ and up to $\hat{s}=-0.5$, respectively. In the intermediate region, as the magnetic shear approaches zero, the high-fidelity simulations predict that the modes become localised at finite ballooning angle, away from the outboard mid-plane, due to a weakening influence of FLR damping. The eigenmode predictions from the minimal asymptotic model display a similar quality in this region, albeit disagreeing in the exact localisation at $\hat{s}=0.1$ -- corresponding to the larger discrepancy in propagation frequency observed in \Cref{fig:shear_scan_result} -- and being of tearing-parity rather than ballooning-parity at $\hat{s}=-0.3$. In the latter case, an equivalent ballooning-parity mode with nearly identical eigenfrequency is also found by the model. Likewise, for the $\hat{s}=0.1$ case, the minimal model also yields an eigenmode localised near $z/\pi \approx \pm 4.5$ similar to the \textsc{Gene} results with a comparable growth rate as the mode localised at $z/\pi \approx \pm 2.3$ displayed in \Cref{fig:shear_scan_eigenmodes}, though this mode is characterised by a moderately larger $k_\parallel$, such that after applying the transit damping correction of \Cref{eq:freq_Landau_correction} it becomes subdominant. The eigenmode for $\hat{s}=-0.1$ (not shown) agrees in both localisation and parity with the high-fidelity predictions, in accordance with the close match of the propagation frequency for the minimal asymptotic model. In this intermediate region of low magnetic shear, the hybrid model results in the largest discrepancy in the eigenfunction, which corresponds to an $l=0$ and $l=2$ ballooning-parity mode for $\hat{s}=0.1$ and $\hat{s}=-0.1,\ -0.3$, respectively, coinciding with the larger disparity in propagation frequency predictions shown in \Cref{fig:shear_scan_result}. Including the effects from the parallel inhomogeneity of the magnetic drift results in a modest modification of the eigenfunctions at positive shear, where the shape of the eigenfunction becomes more susceptible to the shape of the steepening drift wells, as particularly noticeable for the high-shear eigenfunction. At the other extreme of large negative shear, the full dispersion model -- despite greater accuracy in the prediction of the eigenfrequency -- severely overestimates the eigenmode width compared with the minimal and hybrid models. Near zero magnetic shear, however, the full model predictions for the eigenfunction agree with the minimal model for $\hat{s}=-0.3$, yields a ballooning-parity equivalent of the eigenmode found by the minimal model at $\hat{s}=0.1$, and wrongly predicts an $l=4$ excited state mode at $\hat{s}=-0.1$. Likewise at $\hat{s}=-0.5$, corresponding to the largest propagation frequency discrepancy in \Cref{fig:shear_scan_result}, the full model wrongly predicts the eigenmode to be an $l=1$ tearing-parity ground state mode, in contrast to the $l=0$ ballooning-parity ground state prediction by both the minimal and hybrid models.  \par

\subsection{Influence of toroidal versus slab drive} \label{sec:toroidal_vs_slab_competition}
The asymptotic expansion used to obtain \Cref{eq:asymp_disp_rel} is based on the assumption that the influence of parallel streaming on the dynamics of the instability drive is negligible with respect to the influence of toroidal effects. This assumption fails to be satisfied in regions along the field line close to where the magnetic drift vanishes, or, as shown in \Cref{app:asymp_breakdown}, in certain regions of velocity-space depending on the (local) eigenmode properties. However, such local deviations may be relatively insignificant for the overall instability dynamics as determined by the global properties along the full flux tube, if, e.g., locally the mode amplitude is small in such regions. \par
\begin{figure}[t]
    \centering
        \includegraphics[width=0.95\textwidth]{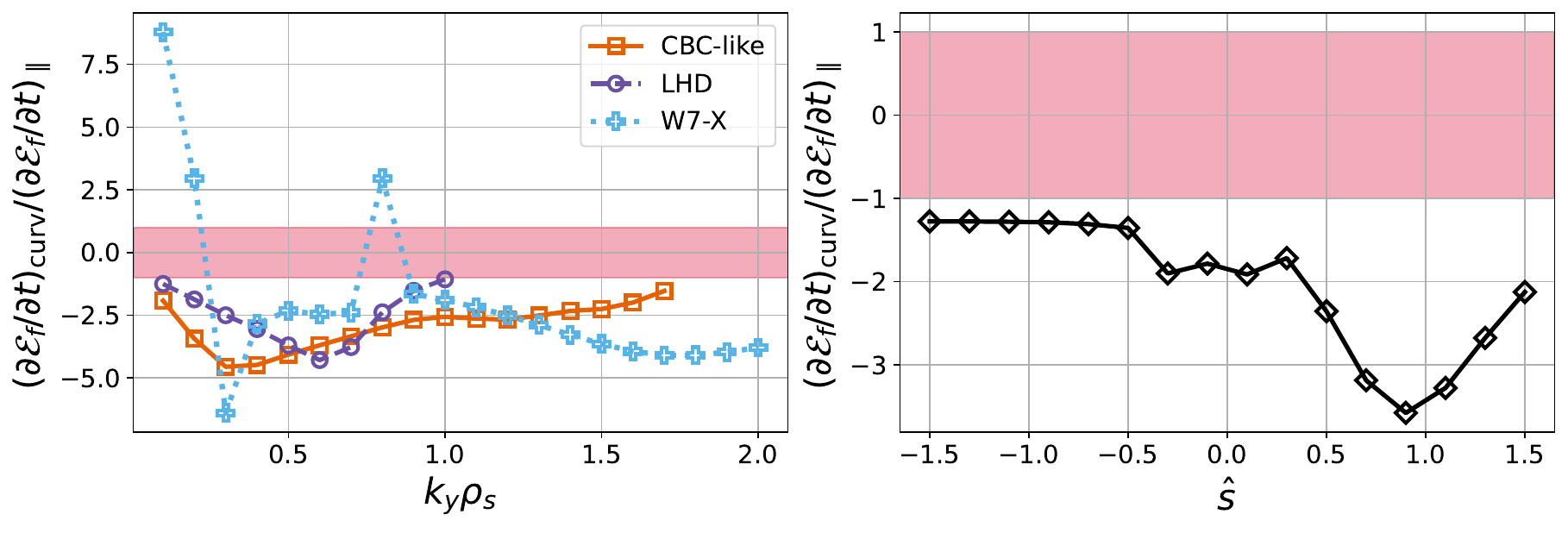}
    \caption{Ratio of free-energy transfer induced by the magnetic drift and parallel streaming terms of the gyrokinetic Vlasov equation (first and third terms of \Cref{eq:lin_GKVE}, respectively) in high-fidelity \textsc{Gene} simulations. Shown are the variation of this ratio as a function of binormal wavenumber (left) for CBC-like tokamak (solid line, orange squares), inward-shifted configuration of LHD (dashed line, purple circles) and high-mirror configuration of W7-X geometries (dotted line, blue plusses), and as a function of global magnetic shear (right) for the geometry scan of $\hat{s}$-$\alpha$ tokamaks (solid line, black diamonds). The red coloured band corresponds to the region where $|\left(\pdv*{\mathcal{E}_f}{t}\right)_{\mathrm{curv}}/\left(\pdv*{\mathcal{E}_f}{t}\right)_{\mathrm{\parallel}}|\leq 1$ and subsequently the effect of parallel streaming is dominant over the effects from toroidal curvature. Note that a negative ratio implies that one term exerts a stabilising influence on the mode, while the other destabilises it.}
    \label{fig:Free-energy}
\end{figure}

To investigate the relative importance of the neglected role of parallel streaming term with respect to the role of toroidal effects over the full domain, we perform an additional diagnostic in the high-fidelity \textsc{Gene} simulations where we measure the rate of change of the electrostatic free-energy $\mathcal{E}_{\phi}$ as a result of each term in the gyrokinetic Vlasov equation [\Cref{eq:lin_GKVE}] \cite{BanonNavarro2011}. For a linear unstable eigenmode this quantity satisfies $\pdv*{\ln \mathcal{E}_{\phi}}{t} = 2 \gamma > 0$ and the individual contributions of the curvature and parallel streaming terms to the pumping of $\mathcal{E}_{\phi}$ thus give a measure for the magnitude of their respective driving/damping influence on the mode\footnote{Additionally $\mathcal{E}_{\phi}$ is also subject to weak damping from numerical dissipation which is typically about an order of magnitude smaller than the pumping of $\mathcal{E}_{\phi}$ by either streaming or curvature and is subsequently neglected in the analysis here.}. Hence, we measure the ratio of $\pdv*{\mathcal{E}_{\phi}}{t}$ attributed to the parallel streaming and magnetic drift terms in the simulation and average this quantity over the last $10\%$ of timesteps before simulations converge onto the most unstable eigenmode to obtain a proxy measure for the importance of streaming effects to the ITG instabilities. As shown in \Cref{fig:Free-energy} -- where the red-shaded band corresponds to the region where transit effects would dominate the pumping of electrostatic free energy -- most of the instabilities tend to adhere to the default paradigm of being driven unstable by magnetic drifts, while the parallel streaming counteracts this drive via Landau damping. A few noticeable exceptions occur in W7-X, particularly at $k_y\rho_s=0.1,0.2,0.8$ where the parallel streaming term is actually observed to contribute to the instability drive. Note that in all geometries the ratio of free-energy transfer is marginally below $-1$ close to marginal stability, indicating a close balance between toroidal drive from magnetic drift and parallel Landau damping. And indeed, as anticipated in \Cref{sec:LHD}, the mode at $k_y \rho_s = 0.1$ in LHD is characterised by a free-energy-transfer ratio of $-1.25$, indicating that parallel streaming plays an almost equally important role for the instability drive of the peculiar eigenmode characterised by a larger extent along the field and a wide plateau of near-constant $\hat{\phi}$ near the outboard mid-plane (see \Cref{fig:LHD_phi_comparison}). \par 
For the geometry scan in circular tokamaks, the free-energy pumping ratio between curvature and parallel streaming is seen to plateau at roughly $-2$ and $-1.3$ close to zero shear and at large negative shear, respectively. The latter indicates that transit effects are nearly comparable to drift effects for $\hat{s}\leq-0.5$, thus rendering an asymptotic expansion in the smallness of $\omega_{\mathrm{ti}}/\omega$ certainly inaccurate, which may explain the greater discrepancies with respect to large positive shear observed in the predictions for the eigenfrequency and eigenmode width for both the minimal and hybrid models, and the full dispersion model, respectively. Additionally, we note that this close balance between toroidal and transit effects at large negative shear coincides with the region where the eigenfrequency predictions from the ad-hoc global model suffer from a large overestimation of the growth rate, notwithstanding the use of exact high-fidelity eigenfunctions in \Cref{eq:adhoc_aeitg_global_disp}, which may thus be attributed to it lack of any parallel dynamics. \par 
For all considered cases, however, even when the eigenmodes are appreciably destabilised, the effects of toroidal drift tend to exceed those of parallel streaming by only a factor of $\approx 2\mathrm{-}4$. If this ratio were taken as a proxy for our ordering of $\omega_{\mathrm{ti}}/\omega_{di}$, this renders $\omega_{\mathrm{ti}}/\omega_{di} \sim 0.25\mathrm{-}0.5$, which may be insufficient to warrant an asymptotic expansion. \par

These observations may partially explain why the solutions of the asymptotic dispersion relation model have a stronger agreement in the axisymmetric case compared to the stellarator cases, in particular compared with W7-X where growth rates are substantially overpredicted and the eigenmodes differ significantly from their high-fidelity counterpart. However, given that this measure does not justify the use of an asymptotic expansion of parallel streaming, it cannot explain why the asymptotic dispersion model is capable of achieving close agreement in the eigenfunctions and/or eigenfrequencies in the high-shear tokamak and LHD, respectively, nor why the model is capable of achieving close quantitative accuracy at negative shear in the geometry scan of $\hat{s}$-$\alpha$ tokamaks. However, since $\pdv*{\ln\mathcal{E}_\phi}{t}=2\gamma$, this proxy measure disproportionally skews the relative importance of curvature/transit effects in favour of modes characterised by large/small growth rates, respectively, as may be inferred by comparing \Cref{fig:Free-energy} with the shapes of the high-fidelity growth rate curves in \Cref{fig:CBC_result,fig:LHD_result,fig:W7X_result,fig:shear_scan_result}. To more directly assess the impact of toroidal effects on the unstable eigenmodes we consider an additional numerical experiment where we perform supplementary \textsc{Gene} simulations with the magnetic drifts artificially suppressed, thus removing the toroidal drive altogether, as discussed in \Cref{app:toroidal_vs_slab}. These additional simulations show that the instabilities observed in the high-fidelity \textsc{Gene} simulation of LHD and the circular tokamaks, including those at low and negative global shear, are intrinsically of toroidal nature. Consequently, an asymptotic dispersion relation based solely on drift-resonances may be expected to retain some of the pertinent aspects of those instabilities, even if the effects of parallel streaming are not necessarily small in the sense of an $\omega_{\mathrm{ti}}/\omega_{d\mathrm{i}}\ll 1$ expansion. Meanwhile, the instabilities observed in the W7-X case are found to bare close resemblance to their drift-less counterparts, particularly at low wavenumber, suggesting they are of hybrid slab-toroidal nature, or even almost completely slab-like, and can therefore not accurately be reproduced by a model based on asymptotic expansion of weak transit effects. This may also explain why the semi-local model is capable of providing such accurate eigenfrequency predictions in W7-X, since the effects of transit resonances are maintained by that model, albeit via a heuristic parallel wavenumber. \par

\section{Conclusions and outlook} \label{sec:discussion}
We have examined a simplified ITG dispersion relation based on an asymptotic expansion for small transit frequency with respect to the mode frequency of the gyrokinetic Vlasov-Poisson system -- a simplification often considered in tokamak literature under the additional assumptions of highly localised eigenmodes and a non-resonant limit to obtain analytical predictions -- to self-consistently describe the toroidal branch of ITG modes in the adiabatic-electron limit, as a potential candidate for a fast and accurate reduced model for predicting and understanding the influence of magnetic geometry on microinstability. Three variants of this model are considered; one where it is assumed that the length scale of parallel variation in the eigenmode is significantly shorter than the length scale of parallel variation in the magnetic geometry, and two where the latter are deemed comparable and additional corrections to the dispersion relation are taken into account, differing only in their inclusion of the parallel inhomogeneity of the magnetic drift or lack thereof. \par
It is found that, counter-intuitively, the asymptotic dispersion relation model generally provides results in closer agreement with \textsc{Gene} predictions when the variation of the magnetic geometry is neglected. When the effects of derivatives of the magnetic geometry are taken into account, this generally results in further destabilisation, with the effect being strongest for stellarator geometries where the connection length is considerably shorter than one poloidal turn and the geometric variation is thus larger. Noticeable exceptions, however, occur in LHD and for large negative-shear tokamaks, where their effects are stabilising. The asymptotic model provides quantitatively accurate predictions for the eigenfrequency and eigenmode structure in conventional tokamaks for long-wavelength instabilities, though at shorter wavelengths the accuracy deteriorates, associated with the emergence of higher excitation states of the ITG stability which infringe on the validity limits of the asymptotic expansion. For stellarator geometries, the agreement in eigenfrequency is decreased with respect to the tokamak case, although qualitative trends observed in the \textsc{Gene} simulations are reproduced. In the case of LHD, the predictions for the eigenfunctions obtained with the model accurately capture the general width of the mode envelope, though they fail to capture the fine-scale ripple sub-structure observed in high fidelity simulations. Meanwhile, in the case of W7-X, eigenmode predictions fail to represent the eigenfunctions obtained from high-fidelity simulations, coinciding with substantial overestimation of the growth rates compared with LHD. Additionally, a scan in magnetic shear in the case of a circular tokamak shows that the model is capable of quantitatively predicting changes in the stability of ITG modes as the magnetic geometry is varied, though the quality of these predictions deteriorates once the magnetic shear becomes negative. Compared with recent non-asymptotic ITG dispersion model from Ref~.\cite{Rodriguez2025TheLocalisation} for simplified geometry, we find that our model tends to mostly overestimate rather then underestimate the ITG growth rates, which may be attributed to a combination of the lack of parallel Landau damping along with the inclusion of resonances due to the $\grad{B}$ component of the magnetic drift in our model, and potentially the treatment of the geometry.  \par
Additionally, we compare the predictions for the eigenfrequency with two alternative non-selfconsistent reduced models for adiabatic-electron ITG, which rely on a guess for the potential eigenfunction. In the evaluation of these alternative models, the eigenfunction guesses are taken to be the high-fidelity \textsc{Gene} eigenmodes in order to provide an optimal comparison point. Between these alternative dispersion relations, the semi-local model [\Cref{eq:semiloc_disp}] provides more accurate eigenfrequencies than the ad-hoc global model [\Cref{eq:adhoc_aeitg_global_disp}], especially in cases of stellarator geometries, though the former does not always find an instability. As expected, compared to the asymptotic dispersion relation model, these alternative models more closely match the eigenfrequency predictions from \textsc{Gene}, especially in W7-X. Two notable exceptions, however, occur at $\hat{s}<-0.5$ in circular tokamaks, where the asymptotic dispersion relation model is found to more accurately predict eigenfrequencies than both heuristic models, and the LHD stellarator, where the ad-hoc global model does not predict the existence of instability at all, though the semi-local model closely reproduces the \textsc{Gene} eigenfrequency. These findings, particularly the strong agreement of the semi-local model with the \textsc{Gene} eigenfrequencies, indicates that parallel streaming not just determines the  eigenfunctions, but may also plays a substantial role in the instability drive. \par
The latter effect was more closely examined in the high-fidelity simulations, where toroidal effects typically exceed parallel streaming in strength by only a factor of $\approx 2 \mathrm{-} 4$, thus placing stringent limitations on the applicability of the asymptotic expansion of the dispersion relation. 
The fact that despite these limitations, the asymptotic dispersion relation is capable of providing reasonably accurate predictions in circular tokamaks and LHD may be attributed instabilities observed in the high-fidelity simulations intrinsically being of toroidal nature, such that a dispersion relation based solely on drift-resonances may be expected to retain some of the pertinent aspects of the instability. Meanwhile, the larger discrepancies for W7-X are directly attributed to the more hybrid slab-toroidal nature of the instabilities in the high-fidelity simulations. \par 
Consequently, the effect of parallel streaming on adiabatic-electron ITG is shown to be more significant than formally admissible for an expansion in asymptotically small transit frequency, especially in low-(negative-)shear devices; though qualitative trends in eigenfrequency -- particularly the growth rate -- may be reproduced, considerable errors in the prediction of the corresponding eigenfunctions are common. The greater sensitivity of the eigenfunction compared to the eigenfrequency to the use of this asymptotic expansion may be explained from the fact that the eigenmodes remain determined to $\order{1}$ by the effects of parallel streaming, just like in the exact non-asymptotic case, even though the regularisation term is only $\order{(\omega_{\mathrm{ti}}/\omega)^2}$ and thus formally an ostensibly small correction to the dispersion relation if the underlying assumption on the smallness of transit effects holds. By contrast, even in the limit of zero parallel streaming, a local estimates for the ITG eigenfrequency may already be obtained from the local dispersion relation $\htorloc(\omega,l)=0$, with the other terms in \Cref{eq:adv-diff} being $\order{(\omega_{\mathrm{ti}}/\omega)^2}$ smaller and thus presumably providing a small correction to the eigenfrequency. Therefore, weighing these various limitations and constraints against the relatively low computational costs of linear high-fidelity adiabatic-electron simulations, the use of a reduced dispersion relation model based an asymptotic expansion for weak transit effects for the purpose of fast and accurate predictions on how changes to the geometry affect microinstability for the purpose of stellarator optimisation is, at least in its current form, discouraged. Rather, fast and accurate predictions for the eigenfrequencies may be more reliably achieved from the semi-local model if quasi-empirical proxies for the mode structure $\hat{\phi}$ depending on the geometry could be devised, though that is outside of the scope of this work. Nevertheless, the asymptotic dispersion relation has proven to be a useful tool to enhance our understanding of the importance of parallel streaming effects for toroidal ITG instabilities, which may be used to make better-informed assumptions on the parameter regime in future reduced ITG models. \par
For future work, we consider improving the validity range of the asymptotic dispersion model by separately treating the regions near vanishing magnetic drift -- where the underlying assumption of weak transit effects break down -- from the bulk of domain where the asymptotic expansion may be reasonable to maintain. In the former regions, by assuming the magnetic drift to be vanishingly small, the linear gyrokinetic Vlasov equation may be reduced to lowest order to \cite{Rodriguez2025TheLocalisation}
    \begin{align}
        \frac{v_\parallel}{\omega} \gradpar{v_\parallel \gradpar{\hat{g}_s^{+}}}+\omega\hat{g}^+_s \approx \frac{q_s}{T_s} F_{Ms} \left(\omega-\omega_{*s}^T\right)J_{0s}\hat{\phi},
        \label{eq:zero-drift-asymptotic}
    \end{align}
where $\hat{g}_s^{+}$ is the part of the non-adiabatic gyrocenter distribution function which is even in $v_\parallel$, and thus yields a non-vanishing density fluctuation in the quasi-neutrality condition [\Cref{eq:QN_cond}]. Obtaining a viable solution to \Cref{eq:zero-drift-asymptotic} for $\hat{g}_s^+$ may be challenging and results locally in an integral equation for the dispersion relation, which should subsequently be matched to the asymptotic solution obtained from \Cref{eq:asymp_disp_rel} sufficiently far away from the zero crossings. \par
Furthermore, we also consider reintroducing the effects of trapped electrons as an alternative approach to incorporate the role of parallel streaming. Due to the mass disparity, electron-transit effects represent the dominant timescale in the linear gyrokinetic Vlasov equation; through the bounce-average, this provides another source for non-local effects in the dispersion relation, which is $\order{f_t}$, with $f_t$ the trapped-particle fraction, and thus under conventional assumptions dominates over the $\order{(\omega_{\mathrm{ti}}/\omega)^2}$ non-local contribution from the ions in \Cref{eq:asymp_disp_rel}. The lifting of the adiabatic-electron approximation -- which stringently limits the application to realistic fusion parameter regimes -- would thus allow us to retain the effects of parallel streaming on the ITG instability to dominant order and simultaneously describe trapped-electron modes in the same dispersion relation, which are known to contribute significantly to anomalous heat losses on the ion gyroradius scale in conjunction to ITGs \cite{Horton1999DriftTransport,Merz2010}. This latter improvement of the model, if successful, could open the path to attractive fast and accurate reduced modelling of geometric effects on microstabilities, as the computational costs of high-fidelity kinetic-electron simulations is currently a prohibitive factor for the direct inclusion of gyrokinetic simulations in stellarator optimisation.

\ack{
The authors would like to thank P. Helander, A. Zocco, P. Costello, P.T. Mulholland, G.T. Roberg-Clark, T.J. Bogaarts and J.H.M. Prins for insightful physics and numerics discussions, as well as the \textsc{Gene} development team for their support.

This work has been carried out within the framework of the EUROfusion Consortium, funded by the European Union via the Euratom Research and Training Programme (Grant Agreement No 101052200 — EUROfusion). Views and opinions expressed are however those of the author(s) only and do not necessarily reflect those of the European Union or the European Commission. Neither the European Union nor the European Commission can be held responsible for them. 

Simulations were performed on the HPC Pitagora at the CINECA consortium.
}

\roles{
\textbf{M.C.L. Morren:} Conceptualization (lead); Formal analysis (lead); Investigation (lead); Methodology (lead); Resources (equal); Software (lead); Supervision (supporting); Validation (lead); Visualization (lead); Writing - original draft (lead).
\textbf{M.J. Pueschel:} Conceptualization (equal), Formal analysis (supporting); Investigation (equal); Methodology (equal); Resources (equal), Writing - review \& editing (equal).
\textbf{J.H.E. Proll:} Conceptualization (equal); Funding acquisition (lead); Investigation (equal); Methodology (equal); Resources (equal); Supervision (lead); Writing - review \& editing (equal).
\textbf{G.G. Plunk:} Formal analysis (supporting); Methodology (supporting).
\textbf{Ö.D. Gürcan:} Formal analysis (supporting); Methodology (supporting); Validation (supporting). 
\textbf{A. Manole:} Formal analysis (supporting); Investigation (supporting); Validation (supporting).
}

\data{The data that support the findings of this study are available upon reasonable request from the authors.}

\section*{AI disclosure statement}
OpenAI's chatGPT-5.X family (GPT-5.1 up to and including GPT-5.6) was used to aid in the development and improvement of the code used to solve the nonlinear eigenvalue problem of the asymptotic dispersion relation [\Cref{eq:NL_EVP}]. The analytical derivations, high-fidelity simulations, data analysis and preparation of the manuscript were  carried first hand by the authors.

\bibliographystyle{unsrt}
\bibliography{references.bib}

\appendix

\renewcommand\thefigure{\thesection\arabic{figure}} 
\renewcommand\theequation{\thesection\arabic{equation}} 
\renewcommand\thetable{\thesection\arabic{table}}

\let\oldsection\section 
\def\section{\setcounter{figure}{0}\setcounter{equation}{0}\setcounter{table}{0}\oldsection} 

\crefalias{section}{appendix}
\crefalias{subsection}{appendix}

\section{Derivation of the dispersion relation} \label{app:disp_rel_deriv_details}
\subsection{Asymptotic expansion of the gyrokinetic Vlasov-Poisson system} \label{app:asymptoti-GKVE-expansion}
We start from \Cref{eq:lin_GKVE} -- for notational convenience dropping the species' index $s$ -- and perform a subsidiary expansion in the unknowns
\begin{align}
    \hat{g} = \sum_{j=0}^{\infty} \hat{g}^j, \quad
    \hat{\phi} = \sum_{j=0}^{\infty} \hat{\phi}^{j}
    \label{eq:phi_g_expansions}
\end{align}
where each subsequent term in the expansion is taken to be $\hat{g}^{j+1}/\hat{g}^{j}\sim \omega_{\mathrm{ti}}/\omega\ll1$, and likewise for the electrostatic potential. Substituting \Cref{eq:phi_g_expansions} into the GKVE (\Cref{eq:lin_GKVE}) and ordering $v_\parallel \gradpar{}/\omega \sim \omega_{\mathrm{ti}}/\omega \ll 1$ then yields a hierarchy of equations, with the $k^{\mathrm{th}}$ term of the expansion being
\begin{align}
    v_\parallel \gradpar{\hat{g}^{k-1}} - i\left(\omega-\omega_d\right) \hat{g}^k = - i\frac{q}{T} F_{M} \left(\omega-\omega_{*}^T\right) J_0 \hat{\phi}^{k}
    \label{eq:GKVE_expand_ordj}
\end{align}
with the leading order solution 
\begin{align}
    \hat{g}^0 = \frac{q}{T} F_{M} \frac{\omega-\omega_{*}^T}{\omega-\omega_d} J_0 \hat{\phi}^0.
    \label{eq:g0_sol}
\end{align}
\Cref{eq:GKVE_expand_ordj} may subsequently be solved iteratively for the non-adiabatic gyrocenter distribution function at each order $j$, yielding by induction the solution to $N^{\mathrm{th}}$ order as
\begin{align}
    \hat{g}^N = \sum_{p=0}^{N} \left(\frac{-iv_\parallel}{\omega-\omega_d}\gradpar{}\right)^p \frac{q}{T}F_{M} \frac{\omega-\omega_{*}^T}{\omega-\omega_d} J_0 \hat{\phi}^{N-p}.
    \label{eq:gN_sequence_sol}
\end{align}
Subsequently inserting \Cref{eq:phi_g_expansions,eq:gN_sequence_sol} into the right-hand side of the quasi-neutrality condition \Cref{eq:QN_cond} yields
\begin{align}
    \int J_0 \hat{g_s} \dd^3{\bm{v}} = & \sum_{j=0}^{\infty} \sum_{p=0}^{j} \int J_0 \left(\frac{-iv_\parallel}{\omega-\omega_d}\gradpar{}\right)^p \frac{q}{T}F_{M} \frac{\omega-\omega_{*}^T}{\omega-\omega_d} J_0 \hat{\phi}^{j-p} \dd^3{\bm{v}} \nonumber \\
    = & \frac{q}{T} \Biggl[\sum_{j=0}^{\infty} \int \frac{\omega-\omega_{*}^T}{\omega-\omega_d} F_M J_0^2 \hat{\phi}^{j} \dd^3{\bm{v}} -i \sum_{j=1}^{\infty} \int \frac{J_0 v_\parallel}{\omega-\omega_{d}} \gradpar{} \frac{\omega-\omega_{*}^T}{\omega-\omega_d} J_0 \hat{\phi}^{j-1} F_M \dd^3{\bm{v}}\nonumber \\
    -& \sum_{j=2}^{\infty} \int J_0 \left(\frac{v_\parallel \gradpar{}}{\omega-\omega_d}\right)^2 \frac{\omega-\omega_{*}^T}{\omega-\omega_d} F_M J_0 \hat{\phi}^{j-2} \dd^3{\bm{v}} + i \sum_{j=3}^{\infty} \int J_0 \left(\frac{v_\parallel \gradpar{}}{\omega-\omega_d}\right)^3 \frac{\omega-\omega_{*}^T}{\omega-\omega_d} F_M J_0 \hat{\phi}^{j-3} \dd^3{\bm{v}} \nonumber \\
    +& \sum_{j=4}^{\infty} \int J_0 \left(\frac{v_\parallel \gradpar{}}{\omega-\omega_d}\right)^4 \frac{\omega-\omega_{*}^T}{\omega-\omega_d} F_M J_0 \hat{\phi}^{j-4} \dd^3{\bm{v}} + \cdots  \Biggr] \nonumber \\
    = & \frac{q}{T} \sum_{p=0}^{\infty} \int J_0 \left(\frac{-iv_\parallel}{\omega-\omega_d}\gradpar{}\right)^{2p} \frac{q}{T}F_{M} \frac{\omega-\omega_{*}^T}{\omega-\omega_d} J_0 \left(\sum_{j=0}^{\infty-2p} \hat{\phi}^{j}\right) \dd^3{\bm{v}}
    \label{eq:asymptotic_velint}
\end{align}
where in the second step we explicitly expanded the first few terms in the sum over $p$ and grouped terms with similar order of the parallel derivative. Upon realising that all of $\omega_d, \omega_{*}^T, F_M,J_0$ are even functions of $v_\parallel$, it follows that upon integration over all of velocity space every term with an odd parallel derivative vanishes as a result of symmetry, which is subsequently used in the last step where only terms with even $p$ are retained, additionally the summation index for the various sums over $j$ has been relabelled to start from zero, and linearity of the derivative operator was used to exchange the order of derivative and $j$-summation. Meanwhile the right-hand side of the quasi-neutrality condition is not directly affected by the expansion and simply becomes $\specprefac \left(\sum_{j=0}^{\infty}\hat{\phi}^j\right)$. \par 
Substituting \Cref{eq:asymptotic_velint} into \Cref{eq:QN_cond} and truncating the asymptotic expansion to second order in $(\omega_{\mathrm{ti}}/\omega)^2$ to keep only the lowest order correction then subsequently yields \Cref{eq:asymp_disp_rel} from the main text, where \Cref{eq:phi_g_expansions} has been used to reintroduce the full electrostatic potential $\hat{\phi}$ in favour of the infinite sum on both sides of the equation, and the retained $p=0,1$ terms have been explicitly expanded. Note that each subsequent neglected term of the sum over $p$ in \Cref{eq:asymptotic_velint} is $\order{(\omega_{\mathrm{ti}}/\omega)^2}$ smaller than the previous, and thus presumably ostensibly small. 

\subsection{Validity limits of the asymptotic expansion} \label{app:asymp_breakdown}
\Cref{eq:GKVE_expand_ordj} is formally the correct way to treat expand the linear gyrokinetic Vlasov equation if the effect of parallel streaming on the left-hand side of \Cref{eq:lin_GKVE} is negligibly small with respect to $(\omega-\omega_{d})\hat{g}$. This condition is more restrictive than merely demanding $\omega_{\mathrm{ti}}/\omega \ll 1$ if we order $\omega_{di}/\omega \sim \order{1}$ for kinetic effects to be relevant, for (near-)resonances can render this presumably large term on the left-hand side of \Cref{eq:lin_GKVE} comparable to the role of parallel streaming. Writing  $\gradpar{\hat{g}}=ik_\parallel \hat{g}$, where $k_\parallel(\bm{x})$ is formally a derivative operator, though for a given $\hat{g}(\bm{x},\bm{v})$ will be associated with a complex number, a solution of \Cref{eq:lin_GKVE} follows from
\begin{align}
    \hat{g} = \left[\left(\omega-\omega_{d}\right)\left(1-\epsilon\right)\right]^{-1}\left(\frac{q}{T} F_{Ms} \left(\omega-\omega_{*}^T\right) J_0 \hat{\phi}\right)
    \label{eq:app_g_GKVE_eps_sol}
\end{align}
where we have continued to drop the species' index $s$, $\left[\cdots\right]^{-1}$ denotes formal inversion of the linear gyrokinetic propagator, and introduced 
\begin{align}
    \epsilon(\bm{x},\bm{v})= \frac{k_\parallel(\bm{x}) v_\parallel}{\omega-\omega_d(\bm{x},\bm{v})}.
    \label{eq:app_eps_param}
\end{align}
The full solution for $\hat{g}$ in the asymptotically small streaming limit, as obtained by combining \Cref{eq:gN_sequence_sol,eq:phi_g_expansions}, corresponds to the geometric series expansion of \Cref{eq:app_g_GKVE_eps_sol} 
\begin{align*}
    \hat{g} = \sum_{j=0}^{\infty} \epsilon^j \frac{q}{T}F_{M} \frac{\omega-\omega_{*}^{T}}{\omega-\omega_d} J_0 \hat{\phi}
\end{align*}
which converges only to the exact solution if $\abs{\epsilon}<1$, and subsequently a second order truncation of this expansion as considered for the dispersion relation in the main text furthermore requires $\abs{\epsilon}\ll 1$ to be accurate. \par
\begin{figure}
    \centering
    \includegraphics[width=0.85\linewidth]{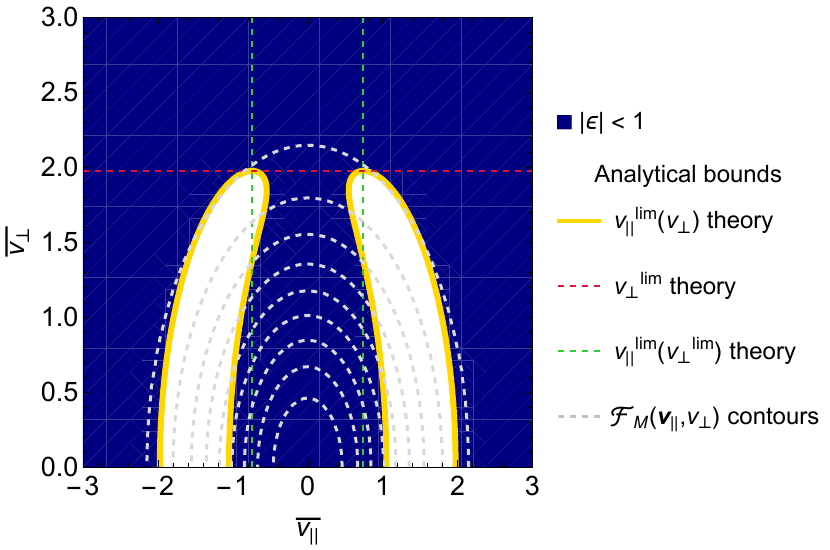}
    \caption{Contour of the validity condition of the asymptotic expansion $\abs{\epsilon}<1$ in velocity-space (blue) in case of parameters $\abs{k_\parallel(\bm{x}) v_T/\tilde{\omega}_d}^2 = 1, \ \Re\left[\omega/\tilde{\omega}_d\right]=2, \ \Im\left[\omega/\tilde{\omega}_d\right]^2=3/10$, chosen such that neither \Cref{eq:app_freqres_cond,eq:app_gamma_cond} is satisfied and a finite phase-space region exists where the convergence condition of the asymptotic expansion is violated. The limiting boundary layer where $\abs{\epsilon}=1$ as predicted by \Cref{eq:app_phase-space_lim} is indicated in yellow, while the red and green dashed lines indicate $\bar{v}_\perp^{\mathrm{lim}}$ and $\bar{v}_\parallel^{\mathrm{lim}}(\bar{v}_\perp^{\mathrm{lim}})$, respectively, as predicted by \Cref{eq:app_vperp_limit,eq:app_vparlim_at_vperplim}. Silver dashed lines indicate contours for the Maxwellian distribution $\mathcal{F}_{M}$.}
    \label{fig:validity_contours}
\end{figure}

The validity of the asymptotic expansion for weak-transit effects is therefore bound by $\abs{\epsilon} < 1$. The expansion parameter $\epsilon(\bm{x},\bm{v})$ is, however, formally phase-space dependent and the condition $\abs{\epsilon} < 1$ may thus be violated by certain populations of particles. To investigate the potential severity of these violations, this validity condition may be written as,
\begin{align}
    \left(\Re\left[\frac{\omega}{\tilde{\omega}_d(\bm{x})}\right]-\left(\bar{v}_\parallel^2+\frac{\bar{v}_\perp^2}{2}\right)\right)^2+\Im\left[\frac{\omega}{\tilde{\omega}_d(\bm{x})}\right]^2-\abs{\frac{k_\parallel(\bm{x}) v_T}{\tilde{\omega}_d(\bm{x})}}^2 \bar{v}_\parallel^2 > 0,
    \label{eq:eps_val_cond}
\end{align}
where we made use of the equivalent but easier to manipulate condition $\abs{\epsilon}^2<1$, we assumed $\omega_{\kappa}(\bm{x})\approx \omega_{\grad{B}}(\bm{x})\equiv\tilde{\omega}_d(\bm{x})\neq 0$ as appropriate for low-$\beta$ plasmas considered in the electrostatic limit, and introduced normalised velocities $\bar{v}_{\parallel,\perp}=v_{\parallel,\perp}/v_T$. In the extreme limit of vanishing streaming effects $k_\parallel(\bm{x})\rightarrow 0$ it follows that \Cref{eq:eps_val_cond} is automatically valid throughout all of phase-space (provided that $\Im[\omega]\neq0$), as in the corresponding limit $\epsilon \rightarrow 0$. When streaming effects remain finite, the limiting phase-space boundary where \Cref{eq:eps_val_cond} ceases to be valid is characterised by
\begin{align}
    \hspace*{-1.5cm}
    \left(\bar{v}_{\parallel}^{\mathrm{lim}}\right)^2 = \frac{1}{2} \left(\acond + 2\bcond - \bar{v}_\perp^2 \pm \sqrt{\left(\acond+2\bcond\right)^2-4 \frac{\abs{\omega}^2}{\tilde{\omega}_d^2} - 2 \acond \bar{v}_\perp^2}\right).
    \label{eq:app_phase-space_lim}
\end{align}
The requirement that $\bar{v}_{\parallel}^{\mathrm{lim}}\in\mathbb{R}$ places several restrictions on the possible phase-space regions where it would be possible to violate $\abs{\epsilon}<1$. First, in order for the radicand to be positive, the condition can only be violated for perpendicular velocities satisfying
\begin{align}
    \bar{v}_\perp^2 \leq \left(\bar{v}_\perp^{\lim}\right)^2  =\frac{\left(\acond+2\bcond\right)^2-4 \frac{\abs{\omega}^2}{\tilde{\omega}_d(\bm{x})^2}}{2\acond}.
    \label{eq:app_vperp_limit}
\end{align}
When $\bar{v}_\perp$ takes its limiting value of \Cref{eq:app_vperp_limit}, the condition where $\abs{\epsilon}<1$ is violated reduces to a set of two isolated points in phase-space characterised by 
\begin{align}
    v_\parallel^{\mathrm{lim}}(v_\perp^{\mathrm{lim}})=\pm \sqrt{\left(\frac{\abs{k_{\parallel}(\bm{x}) v_T}}{2\tilde{\omega}_d(\bm{x})}\right)^2+\left(\frac{\Im[\omega]}{\abs{k_\parallel(\bm{x})v_T}}\right)^2},
    \label{eq:app_vparlim_at_vperplim}
\end{align}
while for velocities exceeding $v_\perp>v_\perp^{\mathrm{lim}}$ the influence of $\grad{B}$-drift is sufficiently large to overcome the influence of particle streaming, such that $\abs{\epsilon}<1$ is satisfied regardless of the parallel velocity. The limiting perpendicular velocity [\Cref{eq:app_vperp_limit}] itself also imposes an additional condition on the instability properties for which $\abs{\epsilon}<1$ can be guaranteed. When the mode's growth rate satisfies
\begin{align}
    \Im[\omega]^2 > (\gamma^{\mathrm{lim}})^2 = \abs{k_\parallel(\bm{x}) v_T}^2 \left(\frac{1}{4}\acond+\bcond\right)
    \label{eq:app_gamma_cond}
\end{align}
there is no real-valued $\bar{v}_\parallel^{\mathrm{lim}}$. Likewise \Cref{eq:app_gamma_cond} by itself proposes another even stricter condition on the mode's propagation frequency
\begin{align}
    \Re\left[\omega \tilde{\omega}_d(\bm{x})\right] < - \frac{\abs{k_\parallel(\bm{x})v_T}^2}{4},
    \label{eq:app_freqres_cond}
\end{align}
as straightforwardly obtained from requiring $\Im[\omega]^2\geq 0$.
For all eigenfrequencies satisfying either \Cref{eq:app_gamma_cond} or \Cref{eq:app_freqres_cond}, it follows that $\abs{\epsilon}<1$ will be satisfied for all particles. The physical interpretations of \Cref{eq:app_gamma_cond,eq:app_freqres_cond} are for the mode's growth rate to be sufficiently large to compensate for any potential toroidal resonance ($\Re[\omega]\approx \omega_d$) such that the mode's exponential growth dominates over the dynamics of parallel streaming, and for the mode's frequency to be sufficiently far off-resonance with the magnetic drift such that $k_\parallel v_T$ can never become comparable to $\omega-\omega_{d}$, respectively. These limits on the validity region of the asymptotic expansion in phase-space are illustrated in \Cref{fig:validity_contours}, where the blue-shaded area corresponds to the region where $\abs{\epsilon}<1$ is satisfied, with the yellow curve denoting its boundary according to \Cref{eq:app_phase-space_lim}. The green and red dotted lines denote the limiting values of \Cref{eq:app_vperp_limit,eq:app_vparlim_at_vperplim}, respectively, corresponding to the apex of the $v_\parallel^{\mathrm{lim}}$ boundary. Note that for this particular $k_\parallel,\tilde{\omega}_d,\omega$ were chosen to guarantee that neither conditions of \Cref{eq:app_freqres_cond,eq:app_gamma_cond} are satisfied, and area enclosed by the yellow curve is large for visualisation purposes. 
However, note that both these aforementioned conditions are still phase-space dependent through the spatial dependence of the magnetic geometry and the profile of the eigenmode along the field line, such that $\abs{\epsilon}<1$ may still be violated by some populations of particles elsewhere along the field line. \par 

To probe the potential impact of these violations, we use $\int_{\abs{\epsilon(\bm{v})}\geq 1} \mathcal{F}_{M} \dd^3{\bm{v}}$ as a proxy measure for the fraction of particles for which the asymptotic expansion of the linear gyrokinetic Vlasov equation under weak transit-effects would fail, where $\mathcal{F}_{M}=F_{Mi}/n_i$ as introduced in the main text. Note that this measure corresponds to the area enclosed by the yellow curve as weighted by the Maxwellian, whose contours are indicated by the silver semi-circular dashed lines in \Cref{fig:validity_contours}. For the specific set of parameters corresponding to \Cref{fig:validity_contours}, this measure yields a particle population of 26\% which violates $\abs{\epsilon}<1$. To investigate what typical values of $\int_{\abs{\epsilon(\bm{v})}\geq 1} \mathcal{F}_{M} \dd^3{\bm{v}}$  are, we randomly generate $10^{4}$ samples for $\abs{k_\parallel v_T/\tilde{\omega}_d}\in\mathcal{U}[0,6], \Re\left[\omega/\tilde{\omega}_d\right]\in\mathcal{U}[-1,1], \Im[\omega/\tilde{\omega}_d]^2\in\mathcal{U}[0,1]$, where $\mathcal{U}[a,b]$ denotes the uniform distribution for values between $a$ and $b$. The maximum fraction of particles for which $\abs{\epsilon}\geq 1$ particles capped at $71.5\%$, while only in $11.6\%$ of cases $\int_{\abs{\epsilon(\bm{v})}\geq 1} \mathcal{F}_{M} \dd^3{\bm{v}}$ exceeded $0.5$. Most commonly, however, in nearly $30\%$ of cases $\abs{\epsilon}<1$ was found to be satisfied by all particles, despite samples being chosen to favour higher values of $\abs{k_\parallel v_T/\tilde{\omega}_d}$ compared with $\Re[\omega/\tilde{\omega}_d],\Im[\omega/\tilde{\omega}_{d}]$ which make the conditions of \Cref{eq:app_gamma_cond,eq:app_freqres_cond} harder to satisfy. Increasing the ranges of $\Re[\omega/\tilde{\omega}_d] \in \mathcal{U}[-4,4]$ or $\Im[\omega/\tilde{\omega}_d]^2 \in \mathcal{U}[0,3]$ were both observed to result in an the number of cases for which $\abs{\epsilon}<1$ is valid for all particles to increase to $44\%$ and $46\%$, respectively, while the number of cases for which $\int_{\abs{\epsilon(\bm{v})}\geq 1} \mathcal{F}_{M} \dd^3{\bm{v}}>0.5$ dropped to $9.5\%$ and $3.8\%$, respectively, without any significant changes to maximum value of $\int_{\abs{\epsilon(\bm{v})}\geq 1} \mathcal{F}_{M} \dd^3{\bm{v}}$. \par
While the present analysis has focused on $\tilde{\omega}_d \neq 0$ for interest of toroidally driven ITG, it is straightforward to show that in the limit of vanishing drifts instead of \Cref{eq:eps_val_cond}, the condition $\abs{\epsilon}<1$ simply reduces to $\omega_{\mathrm{ti}} \bar{v}_\parallel/\omega < 1$, which is the velocity-space dependent equivalent of the $\omega_{\mathrm{ti}}/\omega$ expansion parameter considered in the main text, with the connection length $L_\parallel$ replaced by $1/k_\parallel$, which may be taken as an estimate for the characteristic scale of parallel variation in the eigenmode. Consequently, if the assumption $\omega_{\mathrm{ti}}/\omega \ll 1$ holds -- ergo if the instability is characterised by a sufficiently slowly varying parallel mode structure -- then $\abs{\epsilon}\geq1$ can occur only for suprathermal particles in the tail of the distribution, of which there are few.     

\subsection{Expanding geometric derivatives} \label{app:geom_derivs}
In order to solve the asymptotic dispersion relation \Cref{eq:asymp_disp_rel} it is imperative to consider how the $\gradpar{}$ operator affects the various geometry-related terms. While the magnetic drift $\omega_{ds}$ and the Bessel function $J_{0s}$ are explicitly geometry dependent, a more subtle issue lies in the treatment of the velocity coordinates. Usually in local theory of ITG, cylindrical velocity coordinates $v_\parallel,v_\perp,\theta$ are adopted to most conveniently express the resonant integrals. However, due to the conservation of kinetic energy $E=m_s v^2/2$ and adiabatic invariance of the magnetic moment $\mu = m_s v_\perp^2/2B$ the unperturbed orbits of particles are modulated by the variation in the magnetic field strength according to
\begin{align}
    v_\parallel(l)=\sigma_\parallel \sqrt{\frac{2E}{m_s}\left(1-\frac{\mu B(l)}{E}\right)} \quad
    v_\perp(l) = \sqrt{\frac{2\mu B(l)}{m_s}}
    \label{eq:orbmod}
\end{align}
with $\sigma_\parallel=v_\parallel/\abs{v_\parallel}$ accounting for the direction of travel along the field line. As the ordering in our model does not permit to describe trapped-ion modes, we neglect trapping effects by considering all ions as passing particles and thus consider $\sigma_\parallel$ as constant. Consequently, the orbit modulation results in an additional implicit geometry dependence in the velocity coordinates that has to be accounted for, with the relevant derivatives being
\begin{align}
    \begin{split}
        \gradpar{v_\parallel} &= -\frac{v_\perp(l)^2}{2v_\parallel(l)} \gradpar{\ln B} \\
        \gradpar{v_\perp} &= \frac{v_\perp(l)}{2} \gradpar{\ln B}.
    \end{split}
    \label{eq:vel_grads}
\end{align}
Note that orbit modulation of the velocity coordinates may be neglected in the special case of $v_\parallel^2+v_\perp^2 = 2E/m_s$ as appears in the Maxwellian distribution and the temperature-dependent diamagnetic drift frequency. Additionally, due to the existence of flux-surfaces it follows that $\gradpar{\mathcal{F}_{Ms}}=0$ as the density and temperature are flux functions. Upon further adopting Clebsch-based coordinates like \Cref{eq:FTcoo} such that the magnetic field and perpendicular wavenumber are expressed  as $\bm{B}=B_{\mathrm{ref}}\grad{x}\cross\grad{y}$ and $\bm{k_\perp}=k_x \grad{x}+k_y \grad{y}$, respectively, it follows that $\gradpar{\omega_{*s}^T}=0$ as well, since the diamagnetic frequency -- despite the explicit appearance of $\bm{k_\perp}(l), \bm{e_b}(l), B(l)$ terms in its coordinate-independent form -- is given by $\omega_{\grad{n_s}}= (T_s/q_s B_\mathrm{ref}) k_y \dv*{\ln n_s}{x}$. \par 
Consequently, taking into account orbit modulation it follows that the variation of the Bessel function along the field line is given by
\begin{align}
    \gradpar{J_{0}\left(\frac{k_\perp v_\perp}{\Omega_s}\right)} =& -J_{1}\left(\frac{k_\perp v_\perp}{\Omega_s}\right) \frac{k_\perp v_\perp}{\Omega_s} \left(\gradpar{\ln k_\perp} + \gradpar{\ln v_\parallel}-\gradpar{\ln \Omega_s}\right) \nonumber \\
    =& -J_{1}\left(\frac{k_\perp v_\perp}{\Omega_s}\right) \frac{k_\perp v_\perp}{2\Omega_s} \left(\gradpar{\ln k_\perp^2}-\gradpar{\ln B}\right)
    \label{eq:gradJ0}
\end{align}
having made use of the recursive relations between the Bessel functions \cite{Abramowitz1968HandbookTables}
\begin{align*}
    \dv{J_{\alpha}(\xi)}{\xi} = \frac{1}{2}\left(J_{\alpha-1}(\xi)-J_{\alpha+1}(\xi)\right) \\
    2\alpha J_\alpha(\xi)=\xi \left(J_{\alpha-1}(\xi)+J_{\alpha+1}(\xi)\right)
\end{align*}
and having used $\gradpar{\ln \Omega_s}=-\gradpar{\ln B}$, \Cref{eq:vel_grads} and upon favouring the derivative of $k_\perp^2$ over the derivative of $k_\perp$ to simplify the dependence on the metric tensor elements according to \Cref{eq:FT_FLR_mag}, whilst simultaneously yielding a common numerical prefactor. \par
Likewise, the variation of the magnetic drift along the field line is given by
\begin{align}
    \gradpar{\omega_{ds}} =& \frac{\omega_{\grad{B},s}}{2v_{Ts}^2} \gradpar{v_\perp^2} + \frac{v_\perp^2}{2v_{Ts}^2} \left(-\omega_{\grad{B},s}\gradpar{\ln \Omega_s}+\frac{v_{Ts}^2}{\Omega_{s}}\gradpar{[\bm{k_\perp}\vdot(\bm{e_b}\cross\grad{B})]}\right) \nonumber \\
    &+ \frac{\omega_{\kappa,s}}{v_{Ts]^2}} \gradpar{v_\parallel^2} + \frac{v_\parallel^2}{v_{Ts}^2} \left(-\omega_{\kappa,s}\gradpar{\ln \Omega_s}+\frac{v_{Ts}^2}{\Omega_s} \gradpar{[\bm{k_\perp}\vdot\left(\bm{e_b}\cross\bmog{\kappa}\right)]}\right) \nonumber \\
    =& \frac{\omega_{\grad{B},s}v_\perp^2}{2v_{Ts}^2} \gradpar{\ln{[\bm{k_\perp}\vdot(\bm{e_b}\cross\grad{B})]}} + \frac{\omega_{\kappa,s}v_\parallel^2}{v_{Ts}^2} \left(\gradpar{\ln{[\bm{k_\perp}\vdot(\bm{e_b}\cross\grad{B})]}}-\gradpar{\ln B}\right) \nonumber \\
    &- \frac{\omega_{\kappa,s}v_\perp^2}{v_{Ts}^2} \gradpar{\ln B}
    \label{eq:grad_omd}
\end{align}
where the variation of the thermal drift components $\omega_{\grad{B},s},\omega_{\kappa,s}$ arises both due to the variation of the magnetic field strength as a result of gyrofrequency scaling $\bm{v}_{ds}\propto 1/\Omega_s$ but also has a complicated geometric term due to arrangement of the curvature wells along the field line and its coupling to the variation of the perpendicular wavenumber. Note that unlike the other geometry-related derivatives, this latter term is not strictly positive, see e.g. \Cref{fig:geometries}, and thus formally may have an ill-defined logarithm in some regions along the field line. In the context of \Cref{eq:grad_omd}, however, this quantity is, however, not to be understood literally but is only introduced to allow for a more compact notation with common prefactor by making use of the relation $\gradpar{\ln f(l)}=\gradpar{f(l)}/f(l)$, and logarithmic derivatives of all quantities are only ever evaluated in this way, rather than by evaluating the derivative directly on the log function. \par
Note that as a peculiar side-effect of the orbit modulation, the asymmetry that exists between the curvature and $\grad{B}$ components of the magnetic drift is further enhanced in its derivative, as the variation of $v_\perp(l)^2$ with the magnetic field strength is exactly opposite to the variation of $1/\Omega_{s}$, thus leaving the only geometric variation in the $\grad{B}$ drift component due to the arrange of curvature wells. Meanwhile a similar cancellation does not occur for the curvature drift component depending on $v_\parallel^2$, leaving the orbit modulation of $v_\parallel$ not fully accounted for an resulting in the counter-intuitive $\omega_{s,\kappa}v_\perp^2/v_{Ts}^2$ term. \par
As they shall appear momentarily when expanding the nested parallel derivative operator in \Cref{eq:asymp_disp_rel}, we analogously evaluate the second derivatives of the Bessel function and magnetic drift frequency as 
\begin{align}
    \gradpar[2]{J_{0}\left(\frac{k_\perp v_\perp}{\Omega_s}\right)} =& - J_{1}\left(\frac{k_\perp v_\perp}{\Omega_s}\right) \frac{k_\perp v_\perp}{2\Omega_s} \left(\left[\gradpar[2]{\ln k_\perp^2}-\gradpar[2]\ln B\right]+\frac{1}{2} \left[\gradpar{\ln k_\perp^2}-\gradpar{\ln B}\right]^2\right) \nonumber \\
    &- \frac{J_{0}\left(\frac{k_\perp v_\perp}{\Omega_s}\right)-J_{2}\left(\frac{k_\perp v_\perp}{\Omega_s}\right)}{8} \left(\frac{k_\perp v_\perp}{\Omega_s}\right)^2 \left(\gradpar{\ln k_\perp^2}-\gradpar{\ln B}\right)^2
    \label{eq:grad2_J0}
\end{align}
and
\begin{align}
    \gradpar[2]{\omega_{ds}} &= \frac{\omega_{\grad{B},s}v_\perp^2}{2 v_{Ts}^2}\left\{\gradpar[2]{\ln[\bm{k_\perp}\vdot(\bm{e_b}\cross\grad{B})]} +\left(\gradpar{\ln[\bm{k_\perp}\vdot(\bm{e_b}\cross\grad{B})]}\right)^2\right\} \nonumber \\
    &+\frac{\omega_{\kappa,s}v_\parallel^2}{v_{Ts}^2} \left\{\gradpar[2]{\ln[\bm{k_\perp}\vdot(\bm{e_b}\cross\bmog{\kappa})]}-\gradB[2]+\left(\gradpar{\ln[\bm{k_\perp}\vdot(\bm{e_b}\cross\bmog{\kappa})]}-\gradB\right)^2\right\} \nonumber \\
    &- \frac{\omega_{\kappa,s}v_\perp^2}{v_{Ts}^2} \left\{\gradB[2]-\left[\gradB\right]^2 + 2 \gradB \gradpar{\ln[\bm{k_\perp}\vdot(\bm{e_b}\cross\bmog{\kappa})]}\right\}
    \label{eq;grad2_omd}
\end{align}
as follow straightforwardly from a repeated parallel derivative of \Cref{eq:gradJ0,eq:grad_omd}, respectively, conveniently reusing some of the previously calculated derivatives for quantities that reappear as a consequence from the introduction of the logarithmic derivatives, and in the case of \Cref{eq:grad2_J0} once more making use of the recurrent relations between the Bessel functions. \par

\subsection{Resonant integrals} \label{app:res_int_calculations}
Starting from \Cref{eq:ODE_ints} we will find it useful to explicitly expand the parallel derivatives of the geometry-related terms which yield for $h_{\parallel}^{(1)}$ and $h_{\parallel}^{(0)}$
\begin{align}
    h_{\parallel}^{(1)} =& \specprefac \int \mathcal{F}_{Ms} \frac{\omega-\omega_{*s}^{T}}{\omega-\omega_{ds}} J_{0s} v_\parallel \left(\frac{2v_\parallel}{\resfacspec[2]} \gradpar{J_{0s}} + \frac{3 v_\parallel J_{0s}}{\resfacspec[3]} \gradpar{\omega_{ds}}+\frac{J_{0s}}{\resfacspec[2]} \gradpar{v_\parallel}\right) \nonumber \\
    \equiv &\specprefac \left[h_{s}^{\gradpar{J_{0s}}}+h_s^{\gradpar{\omega_{ds}}}+h_s^{\gradpar{v_\parallel}}\right], \nonumber \\
    h_{\parallel}^{(0)} = & \specprefac \int \mathcal{F}_{Ms} \frac{\omega-\omega_{*s}^{T}}{\omega-\omega_{ds}} J_{0s} v_\parallel \left(\frac{\gradpar{v_\parallel}\gradpar{J_{0s}}}{\resfacspec[2]} + \frac{J_{0s}}{\resfacspec[3]}\gradpar{v_\parallel}\gradpar{\omega_{ds}} + \frac{3v_\parallel}{\resfacspec[3]} \gradpar{J_{0s}}\gradpar{\omega_{ds}}\right. \nonumber \\
    & \left. + \frac{v_\parallel}{\resfacspec[2]} \gradpar[2]{J_{0s}} + \frac{v_\parallel J_{0s}}{\resfacspec[3]} \gradpar[2]{\omega_{ds}} + \frac{v_\parallel J_{0s}}{\resfacspec[4]} \left(\gradpar{\omega_{ds}}\right)^2\right) \nonumber \\
    \equiv & \specprefac \left[h_{s}^{\gradpar{v_\parallel}\gradpar{J_{0s}}} + h_{s}^{\gradpar{v_\parallel}\gradpar{\omega_{ds}}} + h_{s}^{\gradpar{J_{0s}}\gradpar{\omega_{ds}}} + h_{s}^{\gradpar[2]{J_{0s}}} + h_{s}^{\gradpar[2]{\omega_{ds}}} + h_{s}^{\gradpar{\omega_{ds}}^2}\right],
    \label{eq:geom_deriv_int_defs}
\end{align}
respectively, where for convenience we have introduced the nine separate integrals, each related to one of the magnetic inhomogeneity terms. \par
Upon substituting the various expressions for the  derivatives of the geometry-related expressions determined in \Cref{app:geom_derivs}, all of these integrals share the same basic structure involving some higher-order resonance of the mode with the magnetic drift which may be parameterised as
\begin{align}
    \mathbb{I}_{n,m}^{p,q} = \int \frac{\left(\frac{\bar{v}_\parallel}{v_{Ts}}\right)^n \left(\frac{v_\perp}{v_{Ts}}\right)^{m-1}}{\resfacspec[p]} \mathcal{F}_{Ms} J_{0}\left(\frac{k_\perp v_\perp}{\Omega_s}\right) J_{q}\left(\frac{k_\perp v_\perp}{\Omega_{s}}\right) \dd^3{\bm{v_s}},
    \label{eq:Inmpq_main_def}
\end{align}
where the $m-1$ factor accounts for an additional $v_\perp$ arising from expanding the velocity differential $\dd^3\bm{v_s}$, which is subsequently multiplied by a prefactor relating to the derivative of a purely geometric quantity. For example, using \Cref{eq:gradJ0} it follows that
\begin{align*}
    h_{s}^{\gradpar{J_{0s}}}=&-\frac{k_\perp v_{Ts}^3}{\Omega_s} \left(\gradpar{\ln k_\perp^2}-\gradpar{\ln B}\right) \int \mathcal{F}_{Ms} \frac{\omega-\omega_{*s}^{T}}{\resfacspec[3]} \frac{v_\parallel^2 v_\perp}{v_{Ts}^3} J_{0s}J_{1s} \dd^3{\bm{v_s}} \nonumber \\
    =& - \frac{k_\perp v_{Ts}^3}{\Omega_{s}} \left(\gradpar{\ln k_\perp^2}-\gradpar{\ln B}\right) \left[\left(\omega-\omega_{\grad{n_s}}+\frac{3 \omega_{\grad{T_s}}}{2}\right) \mathbb{I}_{2,2}^{3,1}-\frac{\omega_{\grad{T_s}}}{2} \left(\mathbb{I}_{4,2}^{3,1}+\mathbb{I}_{2,4}^{3,1}\right)\right]
\end{align*}
which agrees with the first term for $h_{\parallel}^{(1)}$ given in \Cref{eq:semi-ana-integrals_expand} in the main text, where we have compactified notation by introducing $b_s = (k_\perp v_{Ts}/\Omega_s)^2$ and \Cref{eq:Inmpq_triplet_shorthand} as a shorthand for the term in square brackets, and used the rule for the difference of logarithm in the prefactor. In similar vein, by further using \Cref{eq:grad_omd,eq:vel_grads,eq:grad2_J0,eq;grad2_omd} for the other integrals introduced in \Cref{eq:geom_deriv_int_defs} yields \Cref{eq:semi-ana-integrals_expand} after a bit of straightforward algebra, where the same procedure of introducing the paramterised integral \Cref{eq:Inmpq_main_def} was also applied to $h_{\parallel}^{(2)}$ and $\htorloc$ terms. \par
Note that the velocity integrals of \Cref{eq:Inmpq_main_def} -- which in the dispersion relation \Cref{eq:adv-diff} provide a contribution to the perturbed local ion density $\delta n_s(l)$ -- are to be performed at fixed $l$, and by virtue of expanding all the parallel derivatives only depend on the local velocities $v_\parallel(l), v_\perp(l)$. Consequently, for the purpose of evaluating those integrals, we may treat $v_\parallel, v_\perp$ again as cylindrical velocity variables, as their orbit modulation by the magnetic field strength \Cref{eq:orbmod} is locally irrelevant. In order to then solve \Cref{eq:Inmpq_main_def} analytically, we proceed by writing the higher-order resonance as
\begin{align*}
    \frac{1}{\resfacspec[p]} = \frac{(-1)^{p-1}}{(p-1)!} \left(\pdv{}{\omega}\right)^{p-1} \frac{1}{\resfacspec}
\end{align*}
such that subsequently all integrals may be reduced to the usual first-order resonance form by
\begin{align}
    \mathbb{I}_{n,m}^{p,q} = \frac{(-1)^{p-1}}{(p-1)!} \left(\pdv{}{\omega}\right)^{p-1} \mathbb{I}_{n,m}^{1,q}.
    \label{eq:Inmpq_deriv_trick}
\end{align}
Finally solving for this latter type of resonant integral, we introduce normalised velocity coordinates $u_{\parallel,\perp} = v_{\parallel,\perp}/(\sqrt{2}v_{Ts})$ yielding after along the gyrophase-angle
\begin{align}
    \mathbb{I}_{n,m}^{1,q}= \frac{\sqrt{2}^{n+m+2}}{\sqrt{2\pi}} \int_{-\infty}^{+\infty} \dd{u_\parallel} \int_{0}^{\infty} \dd{u_\perp} \frac{u_\parallel^n u_\perp^m \exp(-u_\parallel^2)\exp(-u_\perp^2)}{\left(\omega - (2\omega_{\kappa,s} u_\parallel^2 + \omega_{\grad{B},s} u_\perp^2)\right)}  J_{0}\left(\sqrt{2b_s}u_\perp\right)J_{q}\left(\sqrt{2b_s}u_\perp\right).
    \label{eq:Inmq_base_def}
\end{align}
Assuming $\omega_{\kappa,s}\neq0$, and exchanging the order of integration between $u_\parallel,u_\perp$ subsequently yields
\begin{align}
    \hspace*{-2cm}
    \mathbb{I}_{n,m}^{1,q} =& -\frac{\sqrt{2}^{n+m-1}}{\omega_{\kappa,s} \sqrt{\pi}} \int_{0}^{\infty} \dd{u_\perp} u_\perp^m \exp(-u_\perp^2)  J_{0}\left(\sqrt{2b_s}u_\perp\right)J_{q}\left(\sqrt{2b_s}u_\perp\right) \int_{-\infty}^{\infty} \dd{u_\parallel} \frac{u_\parallel^n \exp(-u_\parallel^2)}{u_\parallel^2+\frac{\omega_{\grad{B},s}}{2\omega_{\kappa,s}}u_\perp^2-\frac{\omega}{2\omega_{\kappa,s}}} \nonumber \\
    =& - \frac{\sqrt{2}^{n+m-1}}{\omega_{\kappa,s}} \int_{0}^{\infty} \dd{u_\perp} u_\perp^m \exp(-u_\perp^2)J_{0}\left(\sqrt{2b_s}u_\perp\right)J_{q}\left(\sqrt{2b_s}u_\perp\right) G_n\left(z_s(u_\perp,\omega),-z_s(u_\perp,\omega)\right)
    \label{eq:Inmq_anasolve}
\end{align}
where we have introduced the short-hand notation $z_s(u_\perp,\omega) = \sqrt{\frac{\omega}{2\omega_{\kappa,s}}-\frac{\omega_{\grad{B},s}}{2\omega_{\kappa,s}}u_\perp^2}$ to characterise the pair of poles of the denominator in the $u_{\parallel}$ integral, with the latter exactly taking on the form of the two-pole generalisation of the plasma dispersion function \Cref{eq:Gn_int}. Combining \Cref{eq:Inmpq_deriv_trick,eq:Inmq_anasolve} then results in \Cref{eq:Inmpq_integral} from the main text, where additionally a redundancy in the expression of the two-pole generalisation of the plasma dispersion function was used in the special case of $z_2=-z_1$ (see \Cref{eq:smollgn_z}), to wrap the effect of the resonance into a simpler function $\mathsf{g}_n(z_s)=G_n(z_s,-z_s)$.

\section{Properties of the generalised plasma dispersion function} \label{app:Gn_props}
Starting from the general definition of $G_n(z_1,z_2)$ \Cref{eq:Gn_int} we may rewrite $G_n$ into several equivalent recurrence relations by performing partial fraction decomposition on the denominator as
\begin{align}
    G_n(z_1,z_2) =& \frac{1}{\sqrt{\pi}} \int_{-\infty}^{\infty} x^{n-1} \exp(-x^2) \left(\frac{1}{x-z_1}+\frac{z_2}{(x-z_1)(x-z_2)}\right) = Z_{n-1}(z_1) + z_2 G_{n-1}(z_1,z_2) \nonumber \\
    =& \frac{1}{\sqrt{\pi}} \int_{-\infty}^{\infty} x^{n-1} \exp(-x^2) \left(\frac{1}{x-z_2}+\frac{z_1}{(x-z_1)(x-z_2)}\right) =  Z_{n-1}(z_2) + z_1 G_{n-1}(z_1,z_2)  \nonumber \\
    =& \frac{1}{\sqrt{\pi}} \int_{-\infty}^{\infty} x^{n-1} \exp(-x^2) \left(\frac{z_1/(z_1-z_2)}{x-z_1}-\frac{z_2/(z_1-z_2)}{(x-z_2)}\right) = \frac{z_1 Z_{n-1}(z_1)-z_2 Z_{n-1}(z_2)}{z_1-z_2} \nonumber \\
    =& \frac{1}{\sqrt{\pi}} \int_{-\infty}^{\infty} x^n \exp(-x^2) \left(\frac{1/(z_1-z_2)}{x-z_1} - \frac{1/(z_1-z_2)}{x-z_2}\right) = \frac{Z_{n}(z_1)-Z_n(z_2)}{z_1-z_2}
    \label{eq:Gn_recurr}
\end{align}
where we have introduced
\begin{align}
    Z_n(z) = \frac{1}{\sqrt{\pi}} \int_{-\infty}^{\infty} \frac{x^n\exp(-x^2)}{x-z} \dd{x}
    \label{eq:Zn_int}
\end{align}
as an extension to the usual plasma dispersion integral \Cref{eq:Z0int} equivalent to the definition of the two-pole generalisation $G_n(z_1,z_2)$, and used \Cref{eq:Gn_int}. Note that in particular the last two recurrence relations in \Cref{eq:Gn_recurr} are only valid for $z_1 \neq z_2$, which is the main case of interest for this work. To further simplify these two recurrent relations, we similarly use partial fraction decomposition on \Cref{eq:Zn_int} to yield
\begin{align}
    Z_n(z) = \frac{1}{\sqrt{\pi}} \int_{-\infty}^{\infty} x^{n-1} \exp(-x^2) \left(1+\frac{z}{x-z}\right) = \frac{1}{\sqrt{\pi}} \Gamma\left(\frac{n}{2}\right) \frac{1}{2} \left(1+(-1)^{n-1}\right) + z Z_{n-1}(z)
    \label{eq:Zn_recurr}
\end{align}
where we used the coordinate transform $x=\pm \sqrt{y}$ to write Gaussian integral in terms of the Euler Gamma function $\Gamma(p)=\int_{0}^{\infty} y^{p-1} e^{-y}\dd{y}$. Note that the $(1+(-1)^{n-1})/2$ factor, emerging as a result of using $x=-\sqrt{y}$ in the negative half-domain and $x=\sqrt{y}$ in the positive half-domain, accounts for the vanishing of this Gaussian integral in the case of even $n$ due to asymmetry in the integrand, as overlooked in Ref.~\cite{Gurcan2014NumericalCurvature}. Using \Cref{eq:Zn_recurr} it directly follows that the last two recurrence relations of \Cref{eq:Gn_recurr} are indeed equal.  Further iterating on \Cref{eq:Zn_recurr} then yields by induction the following generating sequence for $Z_n(z)$ in terms of \Cref{eq:Z0int}
\begin{align}
    Z_n(z) = \frac{1}{\sqrt{\pi}} \sum_{k=0}^{n-1}\Gamma\left(\frac{n-k}{2}\right)z^k \frac{1+(-1)^{n-k-1}}{2} + z^n Z_0(z).
    \label{eq:Zn_sequence}
\end{align}
Inserting \Cref{eq:Zn_recurr} into the last of the recurrence relations from \Cref{eq:Gn_recurr} then yields \Cref{eq:Gn_sequence} from the main text. \par
Two special cases deserve elaboration. First, as most useful for the evaluation of the resonant integrals that appear in the asymptotic dispersion relation, is the special case of $z_2=-z_1\equiv-z$, in which case \Cref{eq:Gn_sequence} may be simplified to
\begin{align}
    \mathsf{g}_n(z)=G_n(z,-z)=&\frac{1}{2\sqrt{\pi}} \sum_{k=0}^{n-1} z^{k-1} \left(1+(-1)^{k+1}\right) \Gamma\left(\frac{n-k}{2}\right) \frac{1+(-1)^{n-k-1}}{2} \nonumber \\
    &+ \frac{z^{n-1}}{2} Z_0(z)\left(1+\left(-1\right)^{n}\right)
    \label{eq:smollgn_z}
\end{align}
having used \Cref{eq:Z0-Faddeeva_rel} to simplify $Z_0(-z)=-Z_0(z)$. As a direct consequence from \Cref{eq:smollgn_z}, it follows that $\mathsf{g}_{2n+1}(z)=0$, thereby embedding the symmetry property of the original integral \Cref{eq:Gn_int} in the limit of $z_2=-z_1$. And indeed, our asymptotic dispersion relation only contains $\mathsf{g}_{2n}(z)$ terms, as the parts of the distribution function that are odd in $v_\parallel$ result in a vanishing contribution to the density fluctuation in the quasi-neutrality condition \Cref{eq:QN_cond} when integrating over all of velocity space. For these surviving resonant integrals over the even part of the distribution function we may simplify \Cref{eq:smollgn_z} even further as
\begin{align}
    \mathsf{g}_{2n}(z)=\frac{1}{\sqrt{\pi}} \sum_{k=0}^{2n-1} z^{k-1} \Gamma\left(\frac{2n-k}{2}\right) \frac{1+(-1)^{k+1}}{2}+z^{2n-1}Z_0(z)
    \label{eq:smollg2n}
\end{align}
having used $(-1)^{2n-k-1}=(-1)^{k+1}$ along with the observation that $\left[\left(1+(-1)^{k+1}\right)/2\right]^2 = \left(1+(-1)^{k+1}\right)/2$ as the term equals either unity or zero depending whether $k$ is odd or even. Noting that in \Cref{eq:smollg2n} only terms for odd $k$ survive in the sum, such that consequently $\Gamma(n-k/2)$ will be half-integer, and subsequently $\mathsf{g}_{2n}(z)$ will consist of a simple even polynomial with rational coefficients, combined with a monomial term weighting the plasma dispersion function, as exemplified by \Cref{tab:g2n_func_overview}. 

\begin{table}
\caption{Overview of $\mathsf{g}_{2n}(z)$ functions that are used in \Cref{eq:semi-ana-integrals_expand} to describe the effects of drift-resonances in the asymptotic dispersion relation, revealing the simple structure of consisting of the plasma dispersion integral multiplied by a monomial term and a simple polynomial with rational coefficients. Note that both the degree of the monomial and the polynomial increase with $n$}
\centering
\begin{tabular}{l | c }
\hline
$n$ & $\mathsf{g}_{n}(z)$ \\
\hline
$0$ & $\frac{Z_0(z)}{z}$ \\
$2$ & $1 + zZ_0(z)$ \\
$4$  & $\frac{1}{2}+z^2 + z^3 Z_0(z)$ \\
$6$ & $\frac{3}{4} + \frac{z^2}{2}+ z^4 + z^5Z_0(z)$  \\
$8$ & $\frac{15}{8} + \frac{3z^2}{4} + \frac{z^4}{2}+z^6 + z^7 Z_0(z)$ \\
\hline
\end{tabular}
\label{tab:g2n_func_overview}
\end{table}

Lastly, for completeness, another special case deserving additional attention is the case of $z_1=z_2\equiv z$. In this limit \Cref{eq:Gn_sequence} is invalid. However, we may return to \Cref{eq:Gn_recurr} and use either of the first two recurrence relations to obtain a generally valid sequence for $G_n(z_1,z_2)$, which follows by induction as
\begin{align}
    G_n(z_1,z_2)=\sum_{k=0}^{n-1} z_2^k Z_{n-1-k}(z_1) + z_1^n G_0(z_1,z_2) = \sum_{k=0}^{n-1} z_1^k Z_{n-1-k}(z_2) + z_2^n G_0(z_1,z_2),
    \label{eq:Gn_to_G0_sequence}
\end{align}
which in the limit of $z_1=z_2=z$ are indeed identical. In principle, given a closed form expression for $G_0(z,z)$, $G_n(z,z)$ may be obtained by combining \Cref{eq:Zn_sequence,eq:Gn_to_G0_sequence}, however this results in quite cumbersome expression due to the recursive sequences. Rather, we find it more straightforward to start from \Cref{eq:Gn_int} and define as special case
\begin{align}
    \mathcal{G}_n(z)=G_n(z,z)=\frac{1}{\sqrt{\pi}}\int_{-\infty}^{\infty}\frac{x^m \exp(-x^2)}{(x-z)^2}\dd{x}
    \label{eq:Gnzz_int}
\end{align}
from which we obtain via integration by parts
\begin{align}
    \mathcal{G}_n(z)=nZ_{n-1}(z)-2Z_{n+1}(z).
    \label{eq:Gnzz_recurr}
\end{align}
Inserting \Cref{eq:Zn_sequence} then yields
\begin{align}
    \hspace*{-3cm}
    \mathcal{G}_n(z)=&\frac{1}{\sqrt{\pi}} \sum_{k=0}^{n-2} \frac{1+(-1)^{n-k}}{2} \left(n \Gamma\left(\frac{n-k-1}{2}\right) - 2 \Gamma\left(\frac{n+1-k}2{}\right)\right)z^k  - \frac{2\Gamma(1/2)z^n}{\sqrt{\pi}} + Z_0(z) z^{n-1}\left(n-2z^2\right) \nonumber \\
    =& \frac{1}{\sqrt{\pi}} \sum_{k=0}^{n-2} \Gamma\left(\frac{n-k-1}{2}\right) \left(1+k\right)z^k \frac{1+(-1)^{n-k}}{2} + n z^{n-1}Z_0(z) - 2z^n + Z_0(z) z^{n-1}\left(n-2z^2\right)
    \label{eq:Gnzz_sequence}
\end{align}
where the first $n-1$ terms of the sum involving the Gamma functions were lumped together upon using $(-1)^{n-k-2}=(-1)^{n-k}$, with the two remaining terms in the sequence for $Z_{n+1}(z)$ explicitly split off, and in the second step we used the defining properties of the Gamma function $\Gamma(z+1)=z\Gamma(z)$ along with $\Gamma(1/2)=\sqrt{\pi}$. Alternatively, it also follows directly from \Cref{eq:Gnzz_int} that $\mathcal{G}_n(z)= \dv*{Z_n(z)}{z}$, and it readily verified that \Cref{eq:Gnzz_sequence} is indeed the derivative of \Cref{eq:Zn_sequence}, where the derivative of $Z_0$ is straightforwardly obtained by making use of a special limiting case of \Cref{eq:Gnzz_recurr} to find
\begin{align}
    \mathcal{G}_0(z)=\dv{Z_0}{z} = - 2Z_1(z) = -2 \left(1+zZ_0(z)\right)
\end{align}
having once more made use of \Cref{eq:Zn_sequence}, which is indeed the well known ODE which, when solved in the upper-half plane to stay away from the discontinuity along the real-axis, yields $Z_0(z)$ to be the plasma dispersion function $i\sqrt{\pi}\exp(-z^2)\erfc(-iz)$ \cite{FriedContePlasmaZ}.

\section{Non-resonant limit of dispersion relation} \label{app:nonres_binom_lim}
One possibly stringent limitation to the use of \Cref{eq:Inmpq_deriv_trick,eq:Inmq_anasolve} is the assumption of non-vanishing curvature drift $\omega_{\kappa,s}\neq0$. While this can be guaranteed in a guaranteed in simplified local model for ITG based on simplified Z-pinch geometry \cite{Ivanov2023AnField}, in a toroidal configuration the magnetic drift must at some point reverse. Though analytically \Cref{eq:Inmq_anasolve} remains valid even in the limit of infinitesimally small magnetic drift, subsequent numerical implementations of \Cref{eq:Inmpq_deriv_trick} may loose accuracy as the $\omega$-derivative of \Cref{eq:Inmq_anasolve} introduces an additional $1/\omega_{\kappa,s}^{p-1}$ term to the prefactor as a result of the chain rule. \par
Meanwhile the original definition of this integral \Cref{eq:Inmpq_main_def} is very well behaved, even when the magnetic drift vanishes altogether, where the integral is almost trivially solvable. In order to investigate the behaviour of these integrals in the close vicinity to zero crossings of the magnetic drift, rather than neglecting the drift altogether, we consider expanding the denominator for $\omega_{ds}/\omega\ll 1$ \\
\begin{align}
    \frac{1}{\resfacspec[p]}= &\frac{1}{\omega^p} \sum_{k=0}^{\infty} \binom{p+k-1}{k} \left(\frac{\omega_{ds}}{\omega}\right)^k \nonumber \\
    \approx & \frac{1}{\omega^p} \left(1+p\frac{\omega_{ds}}{\omega}+\frac{p(p+1)}{2} \left(\frac{\omega_{ds}}{\omega}\right)^2+\frac{p(p+1)(p+2)}{6}  \left(\frac{\omega_{ds}}{\omega}\right)^3 + \order{\left(\frac{\omega_{ds}}{\omega}\right)^4}\right)
    \label{eq:drift-denom-fluid-lim}
\end{align}
where $\binom{l}{m} = l!/(m!(l-m)!)$ is the binormial coefficient, which have been explicitly expanded up to fourth order. It should be noted, however, that the $\omega_{ds}/\omega \ll 1$ cannot be made strictly rigorous throughout the full extend of velocity space as particles in the tail in the distribution may have a considerable magnetic drift, even if drift of the particles in the thermal bulk is small enough to warrant expansion. The number of particles which could possibly resonate is, however, exponentially small, such that upon integration over the distribution we may reasonably expect the non-resonant effect of the thermal bulk to dominate. Inserting \Cref{eq:drift-denom-fluid-lim} into \Cref{eq:Inmpq_main_def} then subsequently yields the behaviour of the resonant integrals for near-zero magnetic drift as 
\begin{align}
    \mathbb{I}_{n,m}^{p,q} \approx & \frac{1}{\omega^p} \left( \mathbb{I}^{n,m,q}_{\textrm{fluid}}+ p \left[ \frac{\omega_{\kappa,s}}{\omega} \mathbb{I}^{n+2,m,q}_{\textrm{fluid}}+\frac{\omega_{\grad{B},s}}{2\omega} \mathbb{I}^{n,m+2,q}_{\textrm{fluid}}\right] + \frac{p(p+1)}{2} \biggl[\left(\frac{\omega_{\kappa,s}}{\omega}\right)^2 \mathbb{I}^{n+4,m,q}_{\textrm{fluid}}+ \right. \nonumber \\
    & + \left. \frac{\omega_{\kappa,s}\omega_{\grad{B},s}}{\omega^2} \mathbb{I}^{n+2,m+2,q}_{\textrm{fluid}}+\frac{1}{4} \left(\frac{\omega_{\grad{B},s}}{\omega}\right)^2 \mathbb{I}^{n,m+4,q}_{\mathrm{fluid}}\biggr] + \frac{p(p+1)(p+2)}{6} \biggl[\left(\frac{\omega_{\kappa,s}}{\omega}\right)^3 \mathbb{I}_{\mathrm{fluid}}^{n+6,m,q} + \right. \nonumber \\ 
    &\left. + \frac{3}{2} \frac{\omega_{\kappa,s}^2\omega_{\grad{B},s}}{\omega^3} \mathbb{I}^{n+4,m+2,q}_{\textrm{fluid}} + \frac{3}{4}\frac{\omega_{\kappa,s}\omega_{\grad{B},s}}{\omega^3} \mathbb{I}^{n+2,m+4,q}_{\textrm{fluid}} + \frac{1}{8} \left(\frac{\omega_{\grad{B},s}}{\omega}\right)^3 \mathbb{I}^{n,m+6,q}_{\mathrm{fluid}} \biggr] \right)
    \label{eq:Inmpq_cubic_expand}
\end{align}
having used the velocity-dependence of the magnetic drift $\omega_{ds} = \omega_{\grad{B},s} (v_\perp /v_{Ts})^2 /2 + \omega_{\kappa,s} (v_\parallel / v_{Ts})^2$ and having introduced the following emergent parameterised integral
\begin{align}
    \mathbb{I}^{n,m,q}_{\textrm{fluid}} = \frac{1}{\sqrt{2\pi}} \int_{-\infty}^{\infty} \dd{x_\parallel} \int_{0}^{\infty} \dd{x_\perp} x_{\parallel}^{n} x_\perp^m \exp(-\frac{x_\parallel^2}{2})\exp(-\frac{x_\perp^2}{2}) J_0(\sqrt{b_s}x_\perp)J_q(\sqrt{b_s}x_\perp)
    \label{eq:Inmq_fluid_def}
\end{align}
with normalised velocity coordinates $x_{\parallel,\perp} =  v_{\parallel,\perp}/v_{Ts}$. The velocity integral of \Cref{eq:Inmq_fluid_def} neatly factorises into a parallel and perpendicular part, which are given in terms of moment of the standard normal distribution 
\begin{align*}
    \frac{1}{\sqrt{2\pi}}\int_{-\infty}^{\infty} \dd{x_\parallel} x_\parallel^n \exp(-\frac{x_\parallel^2}{2}) = \begin{cases}
        0 & \mathrm{odd} \ n \\
        (n-1)!! & \mathrm{even} \ n
    \end{cases}
\end{align*}
and Weber-type integrals\cite{Plunk2022EnergeticGrowth}
\begin{align*}
    F_m^q(b_s) = \int_{0}^{\infty} x_\perp^m \exp(-\frac{x_\perp^2}{2}) J_0(\sqrt{b_s}x_\perp)J_q(\sqrt{b_s}x_\perp),
\end{align*}
respectively. Note that for the integrals involved in \Cref{eq:semi-ana-integrals_expand} we always have an even exponent $n$ as odd moments of the perturbed distribution function do not contribute to the density fluctuation in Poisson's law, and subsequently we may use the recurrent relation $(n+2)!!=(n+2)n!!$ to further simplify \Cref{eq:Inmpq_cubic_expand} to
\begin{align}
    \mathbb{I}_{2n,m}^{p,q} = & \frac{(2n-1)!!}{\omega^p} \left(F_m^{q}(b_s) \biggl[1 + \frac{p \omega_{\kappa,s}}{\omega}(2n+1) + \frac{p(p+1)}{2}\left(\frac{\omega_{\kappa,s}}{\omega}\right)^2(2n+1)(2n+3)+\frac{p(p+1)(p+2)}{6} \times \right. \nonumber \\
    &\left. \left(\frac{\omega_{\kappa,s}}{\omega}\right)^3(2n+1)(2n+3)(2n+5) \biggr] + F_{m+2}^{q}(b_s) \frac{p}{2}\frac{\omega_{\grad{B},s}}{\omega} \biggl[1+(p+1)\frac{\omega_{\kappa,s}}{\omega}(2n+1) \right. \nonumber \\
    & \left. +\frac{(p+1)(p+2)}{2} \left(\frac{\omega_{\kappa,s}}{\omega}\right)^2 (2n+1)(2n+3) \biggr] + F_{m+4}^{q}(b_s) \frac{p(p+1)}{8} \left(\frac{\omega_{\grad{B},s}}{\omega}\right)^{2} \times \right. \nonumber \\ 
    &\left.\biggl[1+ (p+2)(2n+1) \frac{\omega_{\kappa,s}}{\omega}\biggr] + F_{m+6}^{q}(b_s) \frac{p(p+1)(p+2)}{48} \left(\frac{\omega_{\grad{B},s}}{\omega}\right)^3\right)
    \label{eq:cubic_Inmpq_final}
\end{align}
whose leading-order term in the limit of vanishing magnetic drift coincides with \Cref{eq:zero_drift_lim_integrals} of the main text, where the prefactor has been slightly modified to more generally also account for the vanishing of the integral for odd $n$. For completeness, the various Weber functions encountered in \Cref{eq:cubic_Inmpq_final} upon using the values for ${n,m,p,q}$ the appear in \Cref{eq:semi-ana-integrals_expand} are given in \Cref{tab:Fmq_func_overview}, where $\Gamma_n(b)=\exp(-b)I_n(b)$ denotes the exponentially scaled modified Bessel function of the first kind. \par

\begin{table}[b!]
\caption{Overview of Weber-type integrals governing FLR damping as appear in the cubic expansion of drift-resonances in the vicinity of vanishing drift.}
\centering
\hspace*{-3cm}
\begin{tabular}{l l | c }
\hline
$q$ & $m$ & $F_{m}^{q}(b)$ \\
\hline
$0$ & $1$ & $\Gamma_0(b)$ \\
$0$ & $3$ & $2 \left(\Gamma_0(b) + b \left(\Gamma_1(b)+\Gamma_0(b)\right)\right)$\\
$0$ & $5$ & $4 \left( (2-b) \Gamma_0(b) + \left(3-2b\right) b \FLRdiff[] \right)$\\
$0$ & $7$ & $8 \left\{ (2b-3)(b-2) \Gamma_0(b) -\left(1-b\right) \left(4b-11\right)b \FLRdiff \right\}$\\
$0$ & $9$ & $-16\Big(
        \Gamma_0(b)\left[4b^3-27b^2+46b-24\right]
        +2\left(\Gamma_1(b)-\Gamma_0(b)\right)
        \left[4b^3-28b^2+52b-25\right]
        \Big)$ \\
$0$ & $11$ & $
        32\Big(
        \Gamma_0(b)\left[8b^4-88b^3+293b^2-326b+128\right]
        +b\left(\Gamma_1(b)-\Gamma_0(b)\right)
        \left[16b^4-108b^3+627b^2-728b+274\right]
        \Big)$\\
$0$ & $13$ & $
        -64\Big(
        \Gamma_0(b)\left[16b^5-260b^4+1419b^3-3114b^2+2256b-720\right]
        +b\left(\Gamma_1(b)-\Gamma_0(b)\right)
        \left[32b^5-528b^4+2962b^3-6045b^2+6264b-1764\right]
        \Big)$\\
$1$ & $2$ & $- \sqrt{b} \FLRdiff$ \\
$1$ & $4$ & $2 \sqrt{b}  \left(\Gamma_0(b) - (1-2b)\FLRdiff\right)$\\
$1$ & $6$ & $4 \sqrt{b} (2-b) \left(2\Gamma_0(b)-(1-4b) \FLRdiff\right)$  \\
$1$ & $8$ & $8\sqrt{b}\left(
        \Gamma_0(b)\left[4b^2-19b+10\right]
        +\left(\Gamma_1(b)-\Gamma_0(b)\right)
        \left[8b^3-40b^2+44b-6\right]
        \right)$ \\
$1$ & $10$ & $-16\sqrt{b}\left(
        \Gamma_0(b)\left[8b^3-68b^2+158b-36\right]
        +\left(\Gamma_1(b)-\Gamma_0(b)\right)
        \left[16b^4-148b^3+347b^2-258b+24\right]
        \right)$ \\
$1$ & $12$ & $32\sqrt{b}\left(
        \Gamma_0(b)\left[16b^4-212b^3+891b^2-1356b+600\right]
        +\left(\Gamma_1(b)-\Gamma_0(b)\right)
        \left[32b^5-432b^4+1882b^3-3087b^2+1644b-128\right]
        \right)$ \\
$2$ & $3$ & $-2 \left(\Gamma_0(b)+(1+b)\FLRdiff\right)$\\
$2$ & $5$ & $-4(1-b) \left(\Gamma_0(b) + (1+2b) \FLRdiff\right)$ \\
$2$ & $7$ & $
        -8\left(
        \Gamma_0(b)(2b-1)(b-2)
        +\left(\Gamma_1(b)-\Gamma_0(b)\right)
        \left(4b^3-11b^2+2b+2\right)
        \right)$ \\
$2$ & $9$ & $16\left(
        \Gamma_0(b)\left[4b^3-23b^2+27b-6\right]
        +2\left(\Gamma_1(b)-\Gamma_0(b)\right)
        \left[4b^4-24b^3+32b^2-3b-3\right]
        \right)$ \\
$2$ & $11$ & $-32\left(
        \Gamma_0(b)\left[8b^4-82b^3+225b^2-160b+24\right]
        +\left(\Gamma_1(b)-\Gamma_0(b)\right)
        \left[16b^5-164b^4+487b^3-423b^2+24b+24\right]
        \right)$ \\
\hline
\end{tabular}
\label{tab:Fmq_func_overview}
\end{table}

Certain ``standard'' results for the asymptotic dispersion relation may be obtained upon using \Cref{eq:cubic_Inmpq_final} in lieu of the resonant integrals in \Cref{eq:semi-ana-integrals_expand} by truncating the expansion to first order in $\omega_{\kappa,s}/\omega \sim \omega_{\grad{B},s}/\omega\ll 1$, for example using the integral naming convention of \Cref{eq:geom_deriv_int_defs}
\begin{align*}
    h_{s}^{\gradpar{J_{0s}}} \approx& \frac{v_{Ts}^2}{\omega^2} b_s \gradpar{\ln \frac{k_\perp^2}{B}} \Biggl\{\left(1-\frac{\omega_{\grad{n_s}}}{\omega}\right) \biggl[\FLRdiff[s] \left(1+\frac{9\omega_{\kappa}}{\omega}\right)-\frac{3\omega_{\grad{B},s}}{\omega} \times \nonumber \\
    &\left(\Gamma_0(b_s)-(1-2b_s)\FLRdiff[s]\right)\biggr]+\frac{\omega_{\grad{T_s}}}{\omega} \biggl[\left(\Gamma_0(b_s)-(1-2b_s) \right. \times \nonumber \\ 
    & \left.\FLRdiff[s]\right) \left(1+\frac{9\omega_{\kappa,s}}{\omega}\right)-\FLRdiff[s]\frac{9\omega_{\kappa,s}}{\omega} + (2-b_s) \times \nonumber \\
    & \left(2\Gamma_0(b_s)-(1-4b_s)\FLRdiff[s]\right) \frac{3\omega_{\grad{B},s}}{\omega} \biggr]\Biggr\}.
\end{align*}
However, some care must be taken regarding the smallness of the drift, as this expansion is being carried out in conjunction to the $\omega_{\mathrm{t}s}/\omega\ll 1$ expansion of the gyrokinetic Vlasov-Poisson system, with terms of the latter kind becoming more explicit in the non-resonant expansion, as apparent from the above prefactor. To simultaneously treat both expansions on equal footing, we formally consider $\sqrt{\omega_{\mathrm{t}s}/\omega} \sim \omega_{ds}/\omega \sim \delta \ll 1$ and retain only terms to $\order{\delta}$, which after accounting for the appearance of $\gradpar{\hat{\phi}}$ factors multiplying the expanded integrals of \Cref{eq:semi-ana-integrals_expand} in \Cref{eq:adv-diff} then yields after some straightforward algebra the coefficients of the dispersion relation in the non-resonant limit as
\begin{align}
    h_{\parallel,\mathrm{non-res}}^{(2)} =& \specprefac \frac{v_{Ts}^2}{\omega^2} \left(\left(1-\frac{\omega_{\grad{p_s}}}{\omega}\right) \Gamma_0(b_s) - \frac{\omega_{\grad{T_s}}}{\omega} b \FLRdiff[s]\right) \nonumber \\
    h_{\parallel,\mathrm{non-res}}^{(1)} =& \specprefac \frac{v_{Ts}^2}{\omega^2} \Biggl( \gradpar{\ln(\frac{k_\perp^2}{B})} b_s \biggl[\diamagterm \FLRdiff[s]+\frac{\omega_{\grad{T_s}}}{\omega}\times  \nonumber \\ 
    & \left(\Gamma_{0}(b_s)+ 2b_s \FLRdiff[s]\right) \biggr] - \gradpar{\ln B} \biggl[\diamagterm \left(\Gamma_0(b_s)+b_s \FLRdiff[s]\right) \nonumber \\
    &+ \frac{\omega_{\grad{T_s}}}{\omega} b_s \left(\Gamma_0(b_s)+(2b_s-1) \FLRdiff[s]\right)\biggr]\Biggr) \nonumber \\
    h_{\parallel,\mathrm{non-res}}^{(0)} = & \specprefac \frac{v_{Ts}^2}{2\omega} b_s \Biggl( \left(\gradFLR[2]+\frac{1}{2} \left(\gradFLR\right)^2\right) \times \biggl[\diamagterm \FLRdiff[s] \nonumber \\
    & + \frac{\omega_{\grad{T_s}}}{\omega} \left(\Gamma_0(b_s) + 2b_s \FLRdiff[s] \right)\biggr] - \frac{1}{2} \left(\gradFLR\right)^2 \biggl[\diamagterm \times \nonumber \\
    & \left(2 \Gamma_0(b_s) + (1+2b_s) \FLRdiff[s]\right) - \frac{\omega_{\grad{T_s}}}{\omega} (1-2b_s) (\Gamma_0(b_s) + 2b_s \FLRdiff[s] \biggr] \nonumber \\
    & -\gradB \gradFLR \times \biggl[\diamagterm \biggl((1-2b_s)\FLRdiff[s]-\Gamma_0(b_s)\biggr) - \frac{\omega_{\grad{T_s}}}{\omega} \times \nonumber \\ 
    & \left(\FLRdiff[s] b_s \left(4b_s - 5 \right) - 2 \Gamma_0(b_s) \left(1-b_s\right)\right)\biggr]\Biggr)  \nonumber \\
    h_{\mathrm{tor},\mathrm{non-res}}^{\mathrm{loc}} = & 1 + \specprefac \Biggl(1-\biggl[\Gamma_0(b_s) \left(1-\frac{\omega_{\grad{n_s}}}{\omega}+\frac{\omega_{\kappa,s}+\omega_{\grad{B},s}}{\omega}-\frac{\left(\omega_{\grad{n_s}}+\omega_{\grad{T_s}}\right)\omega_{\grad{p_s}})}{\omega^2} \right. \nonumber \\ 
    & \left. + b_s \frac{\omega_{\grad{T_s}}\omega_{\grad{B},s}}{\omega^2}\right) + b_s \FLRdiff[s] \left(\frac{\omega_{\grad{B},s}}{\omega} - \frac{\omega_{\grad{T_s}}}{\omega} - \frac{\omega_{\grad{p_s}}\left(\omega_{\grad{B},s}+\omega_{\kappa,s}\right)}{\omega^2} \right. \nonumber \\ 
    & \left. - \frac{\omega_{\grad{T_s}}\left(\omega_{\kappa,s}+\omega_{\grad{B},s}\right)}{\omega^2} + 2b_s \frac{\omega_{\grad{T_s}}\omega_{\grad{B},s}}{\omega^2}\right) \biggr] \Biggr)
    \label{eq:nonres-ODE-limits}
\end{align}
where as a shorthand notation we introduced $\omega_{\grad{p_s}}=\omega_{\grad{n_s}}+\omega_{\grad{T_s}}$ corresponding to the diamagnetic frequency due to total pressure gradient, which may be written as $\omega_{\grad{p_s}}=\bm{k_\perp}\vdot \bm{v}_{\grad{p_s}}$ with $\bm{v}_{\grad{p_s}}$ the diamagnetic drift velocity that appears in the fluid description of the plasma \cite{Goldston1995}. Note in particular that the effect of the variation of the drift wells, i.e. those terms in \Cref{eq:geom_deriv_int_defs} proportional to $\gradpar{\omega_{ds}}$, has vanished altogether \Cref{eq:nonres-ODE-limits} as these are formally $\order{\delta^2}$ terms in the non-resonant limit of the asymptotic dispersion relation. \par 
Upon assuming the variation in the magnetic field strength is negligible\footnote{This approximation, though not used throughout this work, may be justified based on the neglect of trapped-ion effects, as it would physically correspond to the vanishing of the mirror force and the associated orbit modulation \cite{Romanelli1989IonTokamaks}.}, i.e. $\gradpar{\ln B}\approx 0$, and the geometry being limited to an $\hat{s}$-$\alpha$ tokamak, the asymptotic dispersion relation [\Cref{eq:asymp_disp_rel}] reduces to Eqns.~(1), and (21) from Ref.~\cite{Guo1989IonTransport,Coppi1990CandidatePlasmas},
respectively, with the trapped-electron term in the latter neglected, upon using \Cref{eq:nonres-ODE-limits} for the ODE coefficients. If, additionally to the aforementioned simplifications \Cref{eq:nonres-ODE-limits} is further expanded in the long-wavelength fluid limit ($b_s \sim \delta$) by using $\Gamma_0(b)\approx 1-b + 3b^2/4$ and $\Gamma_1(b)\approx b/2 - b^2/2$ and neglecting any emergent $\order{\delta^2}$ terms, then the resulting asymptotic dispersion relation reduces to Eqns.~(11), (1), (7) and (13) from Refs.~\cite{Biglari1998b,Cheng1981ElectrostaticTokamaks,Romanelli1989IonTokamaks,Horton1981ToroidalGradients}, respectively. 
Lastly, in the case of an arbitrary geometry, under the limiting conditions of the long-wavelength fluid-limit in conjunction with strong-$\grad{T_{\mathrm{i}}}$ drive, taken to be $\omega/\omega_{\grad{T_s}}\sim \omega_{\grad{n_s}}/\omega_{\grad{T_s}} \sim \delta \ll 1$, the lowest-order asymptotic dispersion relation further reduces to Eqn.~(15) from \cite{Plunk2014}.

\section{Comparison of Hermitian and non-Hermitian operator norms} \label{app:Herm_vs_nonHerm}
\begin{figure}[h]
    \centering
    \includegraphics[width=0.95\linewidth]{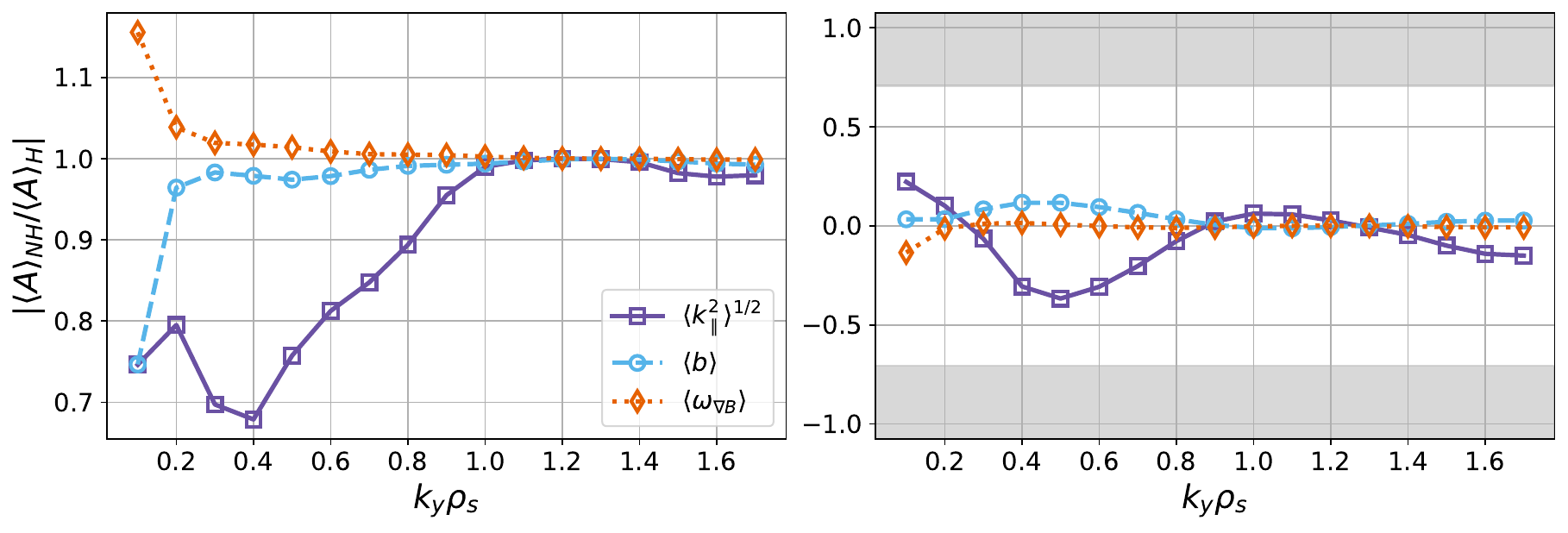}
    \caption{Impact of adopting non-Hermitian definitions for the operator norms of $k_\parallel$ (sold line, purple squares), $\omega_{ds}$ (dashed line, orange diamonds) and $b_s$ (doted line, blue circles) for CBC-like scenario, comparing the ratio of Hermitian to non-Hermitian norms (left) and ratio of imaginary component of the non-Hermitian norms with respect to its absolute value (right). The shaded area in the right plot corresponds to the region where $\abs{\Im[\expval{A}_{NH}]/\abs{\expval{A}_{NH}}}\geq1/\sqrt{2}$ and the non-Hermitian operator has a predominantly imaginary component.}
    \label{fig:CBC_operator_norms}
\end{figure}

\begin{figure}[t]
    \centering
    \includegraphics[width=0.95\textwidth]{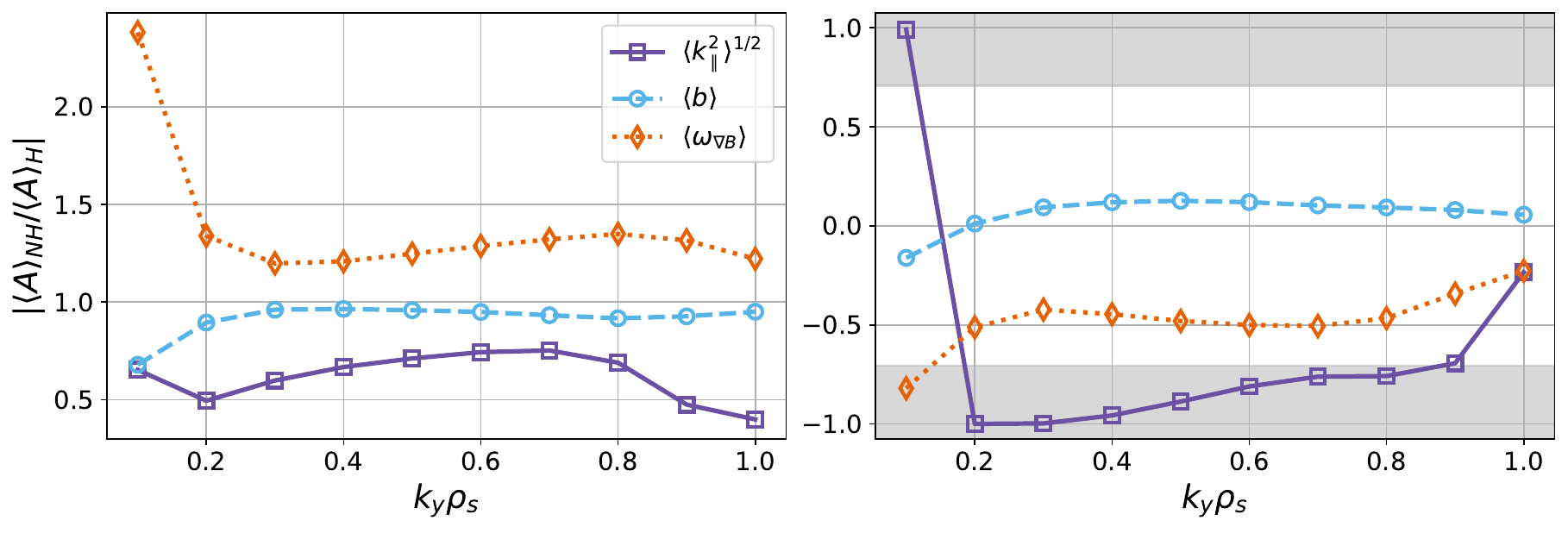}
    \caption{Impact of adopting non-Hermitian definitions for the operator norms of $k_\parallel$ (sold line, purple squares), $\omega_{ds}$ (dashed line, orange diamonds) and $b_s$ (doted line, blue circles) for the inward-shifted configuration of the LHD stellarator, comparing the ratio of Hermitian to non-Hermitian norms (left) and ratio of imaginary component of the non-Hermitian norms with respect to its absolute value (right). The shaded area in the right plot corresponds to the region where $\abs{\Im[\expval{A}_{NH}]/\abs{\expval{A}_{NH}}}\geq1/\sqrt{2}$ and the non-Hermitian operator has a predominantly imaginary component.}
    \label{fig:LHD_operator_norms}
\end{figure}

\begin{figure}[t]
    \centering
    \includegraphics[width=0.95\textwidth]{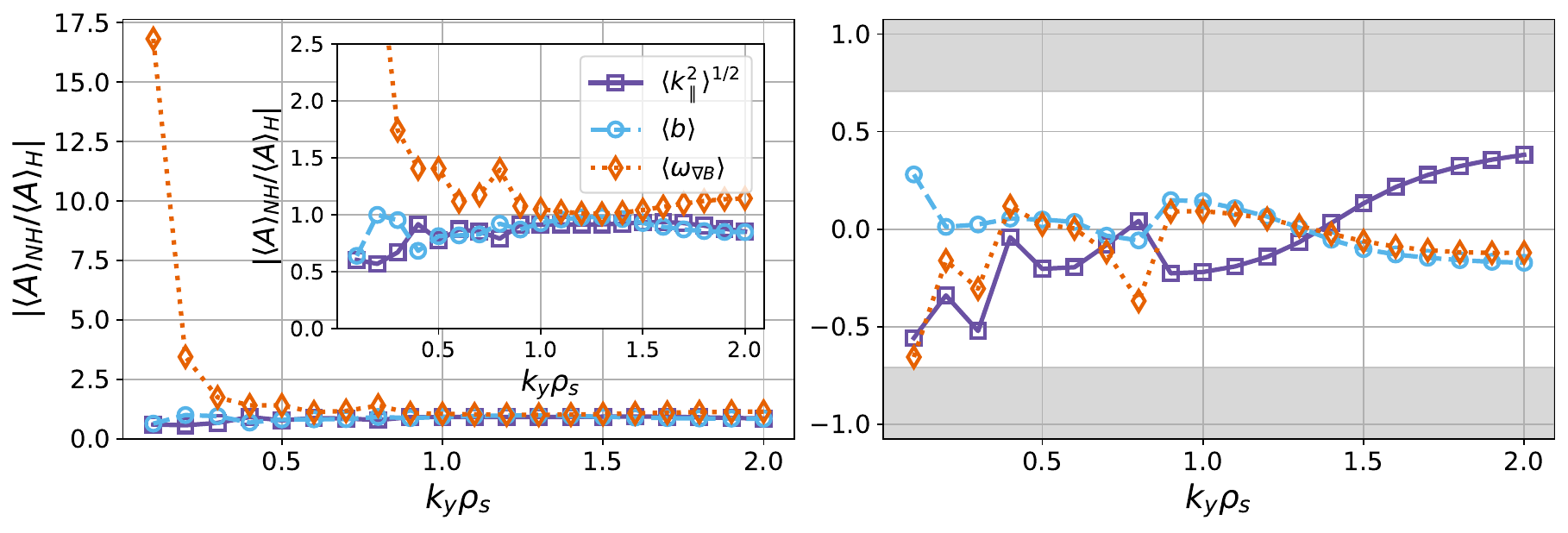}
    \caption{Impact of adopting non-Hermitian definitions for the operator norms of $k_\parallel$ (sold line, purple squares), $\omega_{ds}$ (dashed line, orange diamonds) and $b_s$ (doted line, blue circles) for the high-mirror configuration of the W7-X stellarator, comparing the ratio of Hermitian to non-Hermitian norms (left) and ratio of imaginary component of the non-Hermitian norms with respect to its absolute value (right). Note the large disparity in $\abs{\expval{\omega_{\grad{B}}}_{NH}/\expval{\omega_{\grad{B}}}_{H}}$ compared to other norms at low wavenumber in the left-plot, where the inset provides a detailed zoom around unit ratio to provide more insight. The shaded area in the right plot corresponds to the region where $\abs{\Im[\expval{A}_{NH}]/\abs{\expval{A}_{NH}}}\geq1/\sqrt{2}$ and the non-Hermitian operator has a predominantly imaginary component.}
    \label{fig:W7X_operator_norms}
\end{figure}

\begin{figure}
    \centering
    \includegraphics[width=0.95\textwidth]{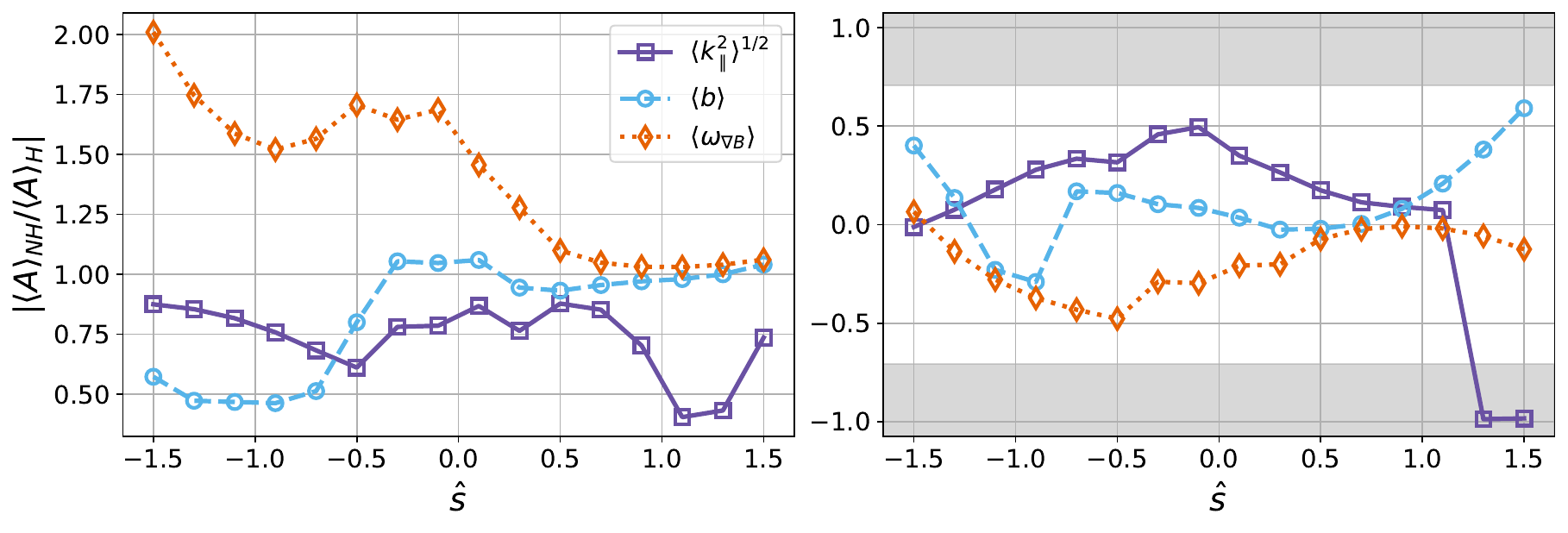}
    \caption{Impact of adopting non-Hermitian definitions for the operator norms of $k_\parallel$ (sold line, purple squares), $\omega_{ds}$ (dashed line, orange diamonds) and $b_s$ (doted line, blue circles) as a function of global shear in an $\hat{s}$-$\alpha$ tokamak, comparing the ratio of Hermitian to non-Hermitian norms (left) and ratio of imaginary component of the non-Hermitian norms with respect to its absolute value (right). The shaded area in the right plot corresponds to the region where $\abs{\Im[\expval{A}_{NH}]/\abs{\expval{A}_{NH}}}\geq1/\sqrt{2}$ and the non-Hermitian operator has a predominantly imaginary component.}
    \label{fig:salpha_operator_norms}
\end{figure}

To investigate the departure from Hermiticity of the operator norms \Cref{eq:kpar_SL_norm,eq:kperp_SL_norm,eq:omd_SL_norm} used in the semi-local dispersion relation \Cref{eq:semiloc_disp}, we investigate both the ratio of $\expval{A}_{NH}/\expval{A}_{H}$ between non-Hermitian norms (definitions as in main text) and Hermitian norms ($\hat{\phi}^2 \rightarrow |\hat{\phi}|^2$ and $(\gradpar{\hat{\phi}})^2 \rightarrow |\gradpar{\hat{\phi}}|^2$ in \Cref{eq:kpar_SL_norm,eq:kperp_SL_norm,eq:omd_SL_norm}), as well as the relative magnitude of the imaginary component of the non-Hermitian norm $\Im[\expval{A}_{NH}]/\abs{\expval{A}_{NH}}$. Consequently, the degree of non-Hermiticity follows from the departure of these two measures from unity and zero, respectively. These measures are examined in \Cref{fig:CBC_operator_norms,fig:LHD_operator_norms,fig:W7X_operator_norms} for the $k_y$ scans in the CBC-like tokamak, LHD, W7-X, respectively, and \Cref{fig:salpha_operator_norms} for the $\hat{s}$ scan of $\hat{s}$-$\alpha$ geometry. Note that in the aforementioned figures we consider $\sqrt{\langle k_\parallel^2 \rangle}$ as directly appears in the semi-local dispersion relation rather than \Cref{eq:kpar_SL_norm}, and we have considered 
\begin{align*}
    \expval{\omega_{ds}} = \int \left(\frac{\omega_{\grad{B}},s}{2} \frac{v_\perp^2}{v_{Ts}^2}+\omega_{\kappa,s} \frac{v_\parallel^2}{v_{Ts}^2}\right) \hat{\phi}^2 \dd{l} / \int \hat{\phi}^2 \dd{l} = \frac{v_\perp^2}{2v_{Ts}^2} \expval{\omega_{\grad{B},s}} + \frac{v_\parallel^2}{v_{Ts}^2} \expval{\omega_{\kappa,s}}
\end{align*}
by neglecting the orbit-modulation by field-strength in $v_\parallel,v_\perp$ (see \Cref{eq:orbmod}) -- as appropriate for a local dispersion model and consistent with the definition of $\expval{\omega_{ds}}$ from Refs.~\cite{Xie2016GlobalGradient,Hirose1995IonTokamaks} -- and report only on the norm for $\expval{\omega_{\grad{B},s}}$ since in the electrostatic limit $\bmog{\kappa}\rightarrow \grad_\perp {B}$, thus rendering the two components equal. \par
For the CBC-like tokamak case, the use of non-Hermitian norms hardly affects $\expval{b}, \expval{\omega_{\grad{B}}}$ from its Hermitian counterpart with the exception of $k_y \rho_s = 0.1$. The effective parallel wavenumber, however, is considerably affected resulting in smaller $|\langle k_\parallel^2 \rangle|^{1/2}$ for $k_y \rho_s < 1$ with a non-negligibly small imaginary component, though throughout the wavenumber scan $\langle k_\parallel^2 \rangle^{1/2}$ remains predominantly a real-number. \par

In the case of LHD, $\expval{b}$ remains almost completely real-valued and nearly equal to its Hermitian counterpart, however, unlike the tokamak case $\expval{\omega_{\grad{B}}}$ is consistently slightly larger than its Hermitian counterpart and also has a considerable imaginary component. With the exception of $k_y \rho_s = 0.1$, where the non-Hermitian $\expval{\omega_{ds}}$ is more than twice as large as its Hermitian counterpart, the imaginary component of the magnetic drift norm does, however, not exceed its real component, as indicated by the grey-shaded areas of the right-panel plots. The same does not hold for the effective parallel wavenumber, as $\langle k_\parallel^2 \rangle^{1/2}$ is found to be predominantly an imaginary number (with the exception of $k_y \rho_s = 1.0$), though like in the tokamak case its magnitude is smaller compared to its purely real-valued Hermitian counterpart. This stark contrast may be explained by the difference in high fidelity eigenmodes for the electrostatic potential shown in \Cref{fig:CBC_phi_comparison,fig:LHD_phi_comparison}; while both eigenmodes (with the exception of $k_y \rho_s = 0.1$ case in LHD) have roughly similar envelope shapes, the eigenmodes in LHD are characterised by an additional ripple in the $\hat{\phi}$ absent in the axisymmetric modes, with these qualitative features also being present in the smaller $\Im[\hat{\phi}/\hat{\phi}_{\mathrm{max}}]$ (not shown), thus leading to a qualitative different behaviour for $\gradpar{\hat{\phi}}$. \par
For the eigenmodes of W7-X geometry, the most striking feature is the significant difference between the Hermitian and non-Hermitian norms of $\expval{\omega_{\grad{B}}}$, with the latter being significantly larger at the lowest wavenumbers, before slowly becoming comparable to its Hermitian counterpart at $k_y \rho_{s}\approx 0.9$. Such a strong discrepancy might be expected if the use of non-Hermitian norm would introduce a significant imaginary component -- as it might in W7-X as the eigenmodes tend to be characterised by a significantly larger $\Im[\hat{\phi}]$ compared with the CBC-like tokamak and LHD, particularly at low wavenumber -- though at all times the drift norms satisfies $\abs{\Re[\expval{\omega_{\grad{B}}}]}>\abs{\Im[\expval{\omega_{\grad{B}}}]}$ (albeit only barely at $k_y \rho_s=0.1$), and by the same token we would then expect a significant difference between the norms of $\expval{b}$, which is mostly absent. Hence, a greater imaginary component of the norm resulting from larger $\Im[\hat{\phi}]$ cannot explain the observed discrepancies and must arise from a greater overlap of drift-well features with $\hat{\phi}^2 / \int \hat{\phi}^2 \dd{l}$ compared with $|\hat{\phi}|^2 / \int |\hat{\phi}|^2 \dd{l}$. The behaviour of the effective parallel wavenumber $\langle k_\parallel^2 \rangle^{1/2}$ is mostly of similar magnitude as its Hermitian counterpart for $k_y \rho_s > 0.3$, and has only a small imaginary component, much akin to what was observed for the CBC-like tokamak case. By contrast at lower wavenumbers the non-Hermitian norm $\langle k_\parallel^2 \rangle^{1/2}$ is roughly half of its Hermitian counterpart and has a more substantial imaginary component, though unlike in LHD, the real part of the effective wavenumber remains a substantial fraction of its total magnitude, as may be attributed to the fact that the eigenmodes lack definitive small-scale ripples on top of the envelope, as illustrated in \Cref{fig:W7X_eigenmodes}. \par
Finally, the $\hat{s}$ geometry scan in $\hat{s}$-$\alpha$ tokamak reveals a striking difference in behaviour of the different norms; a particular ``sweet spot'' exists near $0.3<\hat{s}<1.0$ where the Hermitian and non-Hermitian norms are nearly identical. At larger positive shear, the non-Hermitian FLR norm $\expval{b}$, despite remaining similar in magnitude to its Hermitian counterpart, slowly starts to gain a non-negligible imaginary part unlike what has been observed in other geometries. Likewise at significantly larger positive shear, the effective parallel wavenumber attains a significant imaginary component. Examining the corresponding eigenfunctions we observe that these findings coincide with a noticeably larger $\Im[\hat{\phi}]$ component for $\hat{s}>1$ compared to smaller positive shear. A second region with similar qualitative behaviour of the operator norms can be identified as $-0.5<\hat{s}<0.5$ where $\expval{b}$ remains nearly identical to its Hermitian counterpart, but $\expval{\omega_{\grad{B}}}$ slowly increases in magnitude while gaining a non-negligible imaginary component. In the same region the effective parallel wavenumber also gains in imaginary component, though does so while retaining an absolute value that is only marginally smaller than its Hermitian counterpart. Lastly for $\hat{s}<-0.5$, the imaginary component of both $\expval{\omega_{\grad{B}}}$ and $\langle k_\parallel^2 \rangle^{1/2}$ gradually decreases to zero, while their norms further diverge and converge to their Hermitian counterparts, respectively. Furthermore, in this same region we find $\expval{b}$ to suddenly be significantly smaller than its Hermitian counterpart, as has only been observed before in stellarator configurations at low wavenumber, despite the norm remaining largely real-valued. These findings may be explained from the qualitative difference in large $\hat{s}>0$ and large $\hat{s}<0$ eigenmodes. As shown in \Cref{fig:shear_scan_eigenmodes} the eigenmodes at large $\abs{\hat{s}}$ both tend to be strongly localised near the outboard mid-plane, but $\hat{s}>0$ modes are generally broader and more Gaussian in shape, while $\hat{s}<0$ modes are narrower but have a secondary preferential localisation near $\abs{z}\approx \pi$, with these secondary side bands gaining in prominence as $\hat{s}$ is further decreased. Consequently this asymmetry in the eigenmodes with shear, compared to the symmetry in the metric component $g^{yy}$ (see \Cref{fig:shat_geometry_effect}) related to FLR damping may explain the discrepancy in the behaviour in $\expval{b}$. Likewise, note that the shape of the drift wells $\mathcal{L}_y$ is also asymmetrically modified by $\hat{s}$, once more resulting in a better overlap of the shape of the drift wells with $\hat{\phi}^2/\int \hat{\phi}^2 \dd{l}$ compared to $|\phi|^2/\int |\hat{\phi}|^2\dd{l}$.

\section{Details on numerical implementation} \label{app:numerics_shenanigans}
\subsection{Mapping to Schrödinger-like eigenmode equation}
When adapted to the flux-tube coordinate system of \Cref{eq:FTcoo}, the asymptotic dispersion relation becomes an advection-diffusion-like equation in the field-aligned coordinate $z$
\begin{align}
    A(\omega,z) \dv[2]{\hat{\phi}}{z} + B(\omega,z) \dv{\hat{\phi}}{z} + C(\omega,z) \hat{\phi}(z)=0
    \label{eq:adv-diff-z}
\end{align}
albeit with slightly modified coefficients $A(z,\omega)=h_\parallel^{(2)}(\omega,z)/(\sqrt{g}(z) B(z))^2$ , $B(z,\omega)=h_\parallel^{(1)}(\omega,z)/(\sqrt{g}(z) B(z)) - h_\parallel^{(2)}(\omega,z)/(\sqrt{g}(z) B(z))^2 \pdv*{\ln[\sqrt{g}(z) B(z)]}{z}$ , $C(\omega,z)=\htorloc(\omega,z)-h_{\parallel}^{(0)}(\omega,z)$. Solving \Cref{eq:adv-diff-z} numerically using a finite difference discretisation may be challenging when the problem is advection dominated, as characterised my large magnitude of the effective Péclet number $P_e= B(\omega,z)\Delta z/A(\omega,z)$. As the advection coefficient $B(\omega,z)$ contains a $p=4$ term of the $\mathbb{I}_{n,m}^{p,q}$ integrals, $\abs{P_e}\gg 1$ may be feasible to occur near the zero-crossings of the magnetic drift without the use of appropriate weight functions $W(z;w)$ (see \Cref{sec:numerics-main}, and possibly elsewhere on the domain. \par
\begin{figure}[b!]
    \centering
        \includegraphics[width=0.95\textwidth]{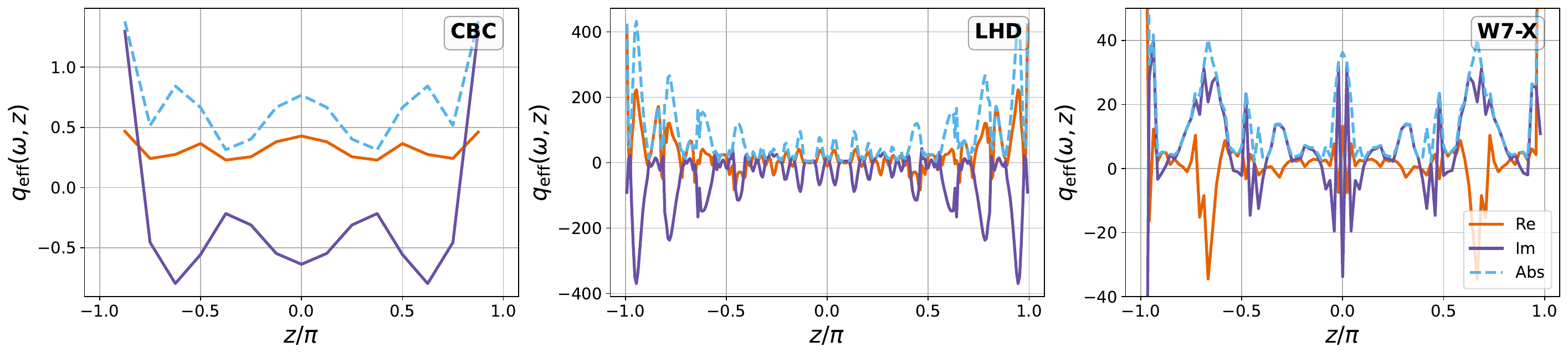}
    \caption{Variation of the effective potential $q_{\mathrm{eff}}(\omega,z)$ along the magnetic field for the flux-tube geometries of $\hat{s}$-$\alpha$ tokamak for CBC parameters (left), the inwards shifted configuration of the LHD stellarator (middle) and the high-mirror configuration of the W7-X stellatator (right). An ITG-compatible frequency of $\omega a/c_s=0.173+0.166i$ was considered at wavenumber of $k_y \rho_s = 0.3$, with gradients of $a/L_n = 2, a/L_T=4$. The effect of geometric variation has been included (i.e. $h_{\parallel}^{(1)}\neq h_\parallel^{(2)}\neq 0$) and weight functions $W(z,w)$ with half-widths $w=4,10,6$ grid points, respectively, are applied at the zero crossings of the magnetic drift.}
    \label{fig:eff_pots}
\end{figure}

Hence, to avoid such issues, we make use the integral transform
\begin{align}
    \hat{\phi}(z)=\psi(z) \exp(-\frac{1}{2}\int_{-\infty}^{z}\frac{B(\omega,z')}{A(\omega,z')}\dd{z'})
    \label{eq:phi-psi-transform}
\end{align}
which reduces \Cref{eq:adv-diff-z} to a Schrödinger-like eigenmode equation in $\psi(z)$
\begin{align}
    \dv[2]{\psi}{z} + \underbrace{\left(\frac{C(\omega,z)}{A(\omega,z)}-\frac{1}{2}\pdv{}{z}\left(\frac{B(\omega,z)}{A(\omega,z)}\right)-\frac{1}{4} \left(\frac{B(\omega,z)}{A(\omega,z)}\right)^2\right)}_{=q_{\mathrm{eff}}(\omega,z)} \psi(z) =0
    \label{eq:Schrodinger-psi-z}
\end{align}
with the transform \Cref{eq:phi-psi-transform} being valid provided that $A(\omega,z) \neq 0$ along the domain. Solving the Schrödinger-like eigenmode problem \Cref{eq:Schrodinger-psi-z} is significantly more numerically attractive than solving the advection-diffusion-like eigenmode problem \Cref{eq:adv-diff-z} as it decouples the discretisation of the derivatives from all the resonance physics, which is all lumped together into a single effective potential well function $q_{\mathrm{eff}}(\omega,z)$ the shape of which depends mostly on the magnetic geometry, compare \Cref{fig:eff_pots,fig:geometries}

Note that in the particular limit where geometric derivatives are neglected (viz. $h_{\parallel}^{(1)}=h_\parallel^{(0)}=0$), the original advection-diffusion form of the asymptotic dispersion relation \Cref{eq:adv-diff} also reduces to a Schrödinger-like eigenmode problem in the arc length coordinate $l$, or since the same approximation implies that $\pdv*{\left(\sqrt{g} B\right)}{z}=0$ and consequently $B(z,\omega)=0$, thus reducing \Cref{eq:adv-diff} directly to \Cref{eq:Schrodinger-psi-z} with a simplified potential $q_{\mathrm{eff}}(\omega,z)=C(\omega,z)/A(\omega,z)$. We remark that similar Schrödinger-like eigenmode problems to \Cref{eq:Schrodinger-psi-z} have been obtained for the ITG problem in the non-resonant limit for $\hat{s}$-$\alpha$ geometries, where $\sqrt{g}B=1/(q_0 R)$ is constant and $\gradpar{} \leftrightarrow \dv[]{z}$ are practically interchangeable, see e.g. Eqns.~(1), (7), (49), (18), and (9) in Refs.~\cite{Choi1980WeaklyModes,Guzdar1983Ion-temperature-gradientPlasmas,Horton1981ToroidalGradients,Biglari1998b,Mattor1991ToroidalTurbulence}, respectively, though in the aforementioned works the effect of geometry derivatives have been neglected.
In stellarator literature, noticeable appearance of a similar Schrödinger-like eigenmode problem appear in Eqns.~(1), (5), (2), (5) and (1) in Refs.~\cite{Bhattacharjee1983DriftStellarator,Dominguez1992,Kendl1999,Rafiq2001Ion-temperature-gradientGeometry,Bhattacharjee2026AndersonStellarators}, respectively, albeit employing a cold-fluid model for the ions.

\subsection{Discretisation schemes} \label{app:numerov_sec}
To solve \Cref{eq:Schrodinger-psi-z} numerically we estimate the second derivative with the 2nd-order accurate finite difference scheme
\begin{align*}
    \left.\dv[2]{\psi}{z}\right|_{z=z_j}\approx \frac{\psi_{j+1}-2\psi_{j}+\psi_{j-1}}{\Delta z^2} -\frac{\Delta z^2}{12}  \left.\dv[4]{\psi}{z}\right|_{z_j} - \frac{\Delta z^4}{360} \dv[6]{\psi}{z} + \order{\Delta z^6}
\end{align*}
where the field-aligned grid is taken to be uniform with grid spacing $\Delta z$. As a result of the original ODE \Cref{eq:Schrodinger-psi-z} the leading-order error term is not unknown and can subsequently also be estimated using the same 2nd-order finite differences, resulting in the following numerical scheme for solving \Cref{eq:Schrodinger-psi-z}
\begin{align}
    &\psi_{j+1} \left(1+\frac{\Delta z^2}{12} q_{\mathrm{eff}}(\omega,z_{j+1})\right) - 2 \psi_j \left(1 - \frac{5\Delta z^2}{12} q_{\mathrm{eff}}(\omega,z_j)\right) + \psi_{j-1} \left(1+\frac{\Delta z^2}{12} q_{\mathrm{eff}}(\omega,z_{j-1})\right) \nonumber\\
    &= \frac{\Delta z^6}{720} \left(5 \pdv[4]{q_{\mathrm{eff}}(\omega,z)\psi}{z}+2 \dv[6]{\psi}{z}\right).
    \label{eq:Numerov}
\end{align}
Upon assuming the error term to be vanishingly small \Cref{eq:Numerov} is known as the Numerov scheme \cite{Noumerov1924APerturbations,Numerov1927Note,Hairer1993SolvingI}, achieving 4th-order accuracy of the Schrödinger-like eigenmode equation with a compact finite-difference stencil. We provide explicit form of the error term to demonstrate that the scheme's accuracy is achieved not only if the solutions $\psi(z)$ is sufficiently slowly varying, but the same also needs to hold for the effective potential functions $q_{\mathrm{eff}}(\omega,z)$. This latter assumption tends to be violated when solving \Cref{eq:Schrodinger-psi-z} without the use of weight function $W(z;w)$, as the slow divergence in \Cref{eq:zero_drift_lim_integrals} can make the effective potential locally spiked near the zero crossings of the magnetic drift. The resonant integrals \Cref{eq:Inmq_anasolve} determining the shape of the $q_{\mathrm{eff}}(\omega)$ are calculated by bringing the derivative under the integral and analytically expanding $\pdv*[p-1]{\mathsf{g}_n(z(\omega,u_\perp)}{\omega}$, with the integrals subsequently evaluated via Gauss-Legendre quadrature by truncating the upper bound at $u_{\perp,\mathrm{max}}=5$. This has been observed to result in faster convergence when oscillatory integrands -- like the Bessel functions in our case -- are involved compared with conventional Gauss-Laguerre quadrature over the semi-infinite interval without truncation \cite{Trefethen2022ExactnessFormulas}. \par
The original eigenvalue problem \Cref{eq:NL_EVP} is thus solved instead as $\underline{M'}(\omega) \psi=0$ where $\underline{M'}(\omega)$ is the tri-diagional matrix encoding the Numerov scheme \Cref{eq:Numerov}. The matrix $\underline{M'}(\omega)$ is constructed on the internal domain $j=1,2,\cdots,N-1$, with Dirichlet boundary conditions $\psi_{0}=\psi_{N}=0$ applied at the edges of the domain at $z=\pm N_{\mathrm{ext}}\pi$, with $N_{\mathrm{ext}}$ the number of connections considered for the extended ballooning domain representation of the geometry generated by \Cref{eq:geometry-extend-formulae}. As the numerical evaluation of the determinant is ill-conditioned, the requirement $\det \underline{M'}(\omega^\star)=0$ is instead replaced by the equivalent requirement $\sigma_{\mathrm{min}}(\omega^\star)=0$ \cite{Datta2010NumericalApplications}, where $\sigma_{\mathrm{min}}$ is the smallest singular value of the matrix $\underline{M'}$, and the eigenvectors $\psi$ are taken as the corresponding right-singular vectors to $\sigma_{\mathrm{min}}(\omega^\star)$. To determine the eigenfrequencies, we evaluate $\sigma_{\mathrm{min}}(\omega)$ on a coarse-grained frequency grid with resolutions $\Delta\omega_R, \Delta\gamma$ to identify minima $\omega_{\mathrm{min}}$, and subsequently refine the minima by locally optimising $\sigma_{\mathrm{min}}(\omega)$ using the {\fontfamily{cmtt}\selectfont COBYQA} method \cite{rago_thesis} constraint to the region of $\omega_{\mathrm{min}} \pm (\Delta \omega_R + i\Delta\gamma )$ until $\sigma_{\mathrm{min}}(\omega) < 10^{-8}$ is achieved. Lastly, with the eigenfrequencies $\omega^\star$ and eigenvectors $\psi^\star$ determined, the corresponding eigenfunctions of the electrostatic potential solving the original advection-diffusion-like eigenmode problem are reconstructed by \Cref{eq:phi-psi-transform}. Lastly, as the goal of the asymptotic expansion is to permit non-local solutions, we require that the obtained eigenmodes have a width of at least $0.05 \pi$ along the field-aligned coordinate $z$, where this width is determined from the standard deviation in $z$ using $\abs{\phi}$ as a weight.

\subsection{Mapping onto symmetry states} \label{app:BallTear_separation}
Unstable eigenmode solutions to the Vlasov-Poisson system typically tend to display symmetry around the outboard mid-plane ($z=0$), resulting in what are coined as ballooning-parity eigenmodes ($\hat{\phi}(-z)=\hat{\phi}(z)$) and tearing-parity eigenmodes ($\hat{\phi}(-z)=-\hat{\phi}(z)$) \cite{Pueschel2019OnDevices,Ishizawa2015ElectromagneticPlasmas}. These symmetry properties of the admissible eigenmodes are essentially baked into the structure of the system matrix $\underline{M}(\omega)$. These symmetries are, however, not guaranteed to be numerically preserved by the null-vectors for $\hat{\phi}^\star=[\hat{\phi}_1,\hat{\phi}_2,\cdots,\hat{\phi}_{N-1}]^\top$ as these are solved on a coordinate basis. Emergence of asymmetric eigenmodes tend to be more prone to happen for eigenmodes localised at finite ballooning angle where $\hat{\phi}(z)\approx 0$ near the outboard mid-plane -- making the distinction between the two parity types harder to resolve -- and singular vectors tend to appear in pairs of asymmetric modes, as has also been seen in high-fidelity simulations particulalry at low magnetic shear \cite{McKinney2021Kinetic-ballooning-modeEquilibria}. Furthermore, a ballooning parity and tearing parity eigenmode may be characterised by similar eigenfrequencies, which may render it hard to resolve both solutions with the method used to determine the eigenfrequencies described in \Cref{app:numerov_sec}, potentially resulting in overlooked eigenmodes. \par
As this latter issue is found to be particularly limiting, we proceed to enforce these parity states on the eigenfunctions by solving \Cref{eq:Schrodinger-psi-z} on an orthonormal symmetry basis defined by
\begin{align}
    \bm{e}_{j}^{+} =& \begin{cases}
        \frac{1}{\sqrt{2}} \left(\bm{e}_j+\bm{e}_{N-j}\right) & j=1,2,\cdots,\frac{N}{2}-1 \\
        \bm{e}_{N/2} & j=\frac{N}{2}
    \end{cases} \nonumber \\
    \bm{e}_{j}^{-} =& \frac{1}{\sqrt{2}} \left(\bm{e}_j-\bm{e}_{N-j}\right) \quad j=1,2,\cdots,\frac{N}{2}-1
    \label{eq:coo_to_symmetry}
\end{align}
where $\bm{e}_j$ are the unit vectors of the coordinate basis,it is assumed that the grid of the field aligned coordinate is centred around $z=0$ and contains an odd number of grid points $N+1$. Consequently, the eigenmode vector in coordinate basis may be written as 
\begin{align}
    \hat{\phi} = \begin{pmatrix}
        Q_{\mathrm{BP}} & Q_{\mathrm{TP}} 
    \end{pmatrix}
    \begin{pmatrix}
        \hat{\phi}_{\mathrm{BP}} \\
        \hat{\phi}_{\mathrm{TP}}
    \end{pmatrix}
\end{align}
where $\psi_{\mathrm{BP}}, \psi_{\mathrm{TP}}$ are the eigenvectors of the ballooning and tearing-parity state, respectively, and $Q_{\mathrm{BP}}, Q_{\mathrm{TP}}$ are the matrices that map the transformation $\bm{e}_j^{\pm}\rightarrow \bm{e}_{j}$, i.e. the inverse of \Cref{eq:coo_to_symmetry}, and are given by
\begin{align}
    Q_{\mathrm{BP}} = \frac{1}{\sqrt{2}}
    \mqty[1 & 0 & 0 & \cdots & 0 \\
         0  & 1 & 0 & & \vdots \\
         \vdots &  & \ddots &  & \vdots \\
         \vdots &  &  & 1 &0 \\
         \vdots &  &  & 0 & \sqrt{2} \\
         \vdots &  &  & 1 & 0 \\
         \vdots &  & \iddots & & \vdots \\
         0 & 1 &  &  & \vdots \\
         1 & 0 & 0 & \cdots & 0 ],
         \quad 
         Q_{\mathrm{TP}} = \frac{1}{\sqrt{2}}
    \mqty[1 & 0  & \cdots & 0 \\
         0  & 1 &  & \vdots \\
         \vdots &  & \ddots & \vdots \\
         \vdots &   &  &1 \\
         \vdots &   &  & 0 \\
         \vdots &   &  & -1 \\
         \vdots & & \iddots & \vdots \\
         0 & -1 &   & \vdots \\
         -1 & 0 & \cdots & 0 ].
         \label{eq:Qtp_Qbp_def}
\end{align}
Note that the transformation \Cref{eq:coo_to_symmetry} is unitary, i.e. $\mqty(Q_{\mathrm{BP}} & Q_{\mathrm{TP}}) \mqty(Q_{\mathrm{BP}}^\top \\ Q_{\mathrm{TP}}^\top) = \underline{I}$ where $\underline{I}$ is the identity matrix. Performing this change of basis to the symmetry space then modifies the eigenmode problem $\underline{M}(\omega)\hat{\phi}=0$ to
\begin{align}
    \mqty(Q_{\mathrm{BP}}^\top \\ Q_{\mathrm{TP}}^\top) \ \underline{M}(\omega) \ \mqty(Q_{\mathrm{BP}} & Q_{\mathrm{TP}}) \mqty(\hat{\phi}_{\mathrm{BP}} \\ \hat{\phi}_{\mathrm{TP}}) \equiv \underline{\tilde{M}}(\omega) \mqty(\hat{\phi}_{\mathrm{BP}} \\ \hat{\phi}_{\mathrm{TP}}) = 0
    \label{eq:parity_EVP}
\end{align}
where the effective system matrix $\underline{\tilde{M}}$ takes on the block-structure form
\begin{align*}
    \underline{\tilde{M}}(\omega) = \mqty(Q_{\mathrm{BP}}^\top \underline{M}(\omega) Q_{\mathrm{BP}} & Q_{\mathrm{BP}}^\top \underline{M}(\omega) Q_{\mathrm{TP}} \\
    Q_{\mathrm{TP}}^\top \underline{M}(\omega) Q_{\mathrm{BP}} & Q_{\mathrm{TP}}^\top \underline{M}(\omega) Q_{\mathrm{TP}}) \equiv \mqty(\tilde{M}_{BB}(\omega) & \tilde{M}_{BT}(\omega) \\ \tilde{M}_{TB}(\omega) & \tilde{M}_{TT}(\omega)).
\end{align*}
The eigenmode problem of \Cref{eq:parity_EVP} is in general not independently solvable for $\hat{\phi}_{\mathrm{BP}},\hat{\phi}_{\mathrm{TP}}$ as the off-diagonal blocks $\tilde{M}_{BT}, \tilde{M}_{TB}$ result in coupling of the two parity states. However, in the special case where the system matrix $\underline{M'}\in\mathbb{C}^{N-1\times N-1}$ is centro-symmetric, i.e. if $M'_{ij}=M'_{N-i,N-j}, \ i,j=1,\cdots,N-1$, then it follows that $\tilde{M}_{BT}=\tilde{M}_{TB}=0$ and the even/odd parity eigenvectors can be determined each from their own eigenmode problem $\tilde{M}_{BB}(\omega) \hat{\phi}_{BP}=0, \ \tilde{M}_{TT}(\omega) \hat{\phi}_{TP}=0 $. \par 
The system matrix $\underline{M}$ characterising \Cref{eq:adv-diff-z} will be centrosymmetric if the ODE coefficients satisfy $A(\omega,-z)=A(\omega,z), B(\omega,-z)=-B(\omega,z), C(\omega,-z)=C(\omega,z)$. For stellarator symmetric flux-tubes it follows that the geometry-related terms in the dispersion relation satisfy \cite{Sugama2011MomentumPlasmas,Morren2026FastApproach}
\begin{equation}
    \begin{aligned}
        g^{xx}(-z) =&g^{xx}(z) \\
        g^{yy}(-z)=&g^{yy}(z) \\
        g^{xy}(-z) =& -g^{xy}(z) 
    \end{aligned}
    \quad \quad
    \begin{aligned}
        \sqrt{g}(-z) =& \sqrt{g}(z) \\
        B(-z) =& B(z) \\
        \mathcal{L}_{x}(-z)=&-\mathcal{L}_x(z)
    \end{aligned}
    \quad \quad
    \begin{aligned}
        \mathcal{L}_{y}(-z)=&\mathcal{L}_{y}(z) \\
        \mathcal{K}_{x}(-z)=&-\mathcal{K}_{x}(z) \\
        \mathcal{K}_{y}(-z)=&\mathcal{K}_{y}(z) \\
    \end{aligned}
\end{equation}
hence, as the only odd-parity elements of the geometry are the radial drift operators $\mathcal{L}_{x}(z)=\mathcal{K}_{x}(z)$ and we assume the eigenmodes to be characterised by $k_x=0$, it follows that the ODE coefficients have the above required properties\footnote{Recal \Cref{eq:semi-ana-integrals_expand} that $h_{\parallel}^{(1)}\propto \gradpar{\ln \mathrm{geo}}$ and $h^{(0)}_\parallel\propto \{\gradpar[2] \ln \mathrm{geo}, \gradpar{\ln \mathrm{geo}}^2\}$, thus becoming an odd and even-parity function, respectively due to the derivatives of the geometry.}, thus rendering the ballooning-parity eigenmode problem and tearing-parity eigenmode problem independent. Furthermore, under those conditions it also follows from \Cref{eq:phi-psi-transform} that the parity property of the eigenmodes $\hat{\phi}(z)$ is completely determined by the parity of $\psi(z)$, and that the effective potential satisfies $q_{\mathrm{eff}}(\omega,z)=q_{\mathrm{eff}}(\omega,-z)$. Consequently, the system matrix $\underline{M'}(\omega)$ characterising \Cref{eq:Schrodinger-psi-z} is also centro-symmetric, and hence we may exchange $\underline{M}(\omega)\rightarrow \underline{M'}(\omega), \hat{\phi}_{\mathrm{BP}}\rightarrow \psi_{\mathrm{BP}}, \hat{\phi}_{\mathrm{TP}}\rightarrow \psi_{\mathrm{TP}}$ to solve for the two independent parity states by solving for the even/odd symmetry solutions to the simpler Schrödinger-like eigenmode problem.

\section{Compliance filter} \label{app:weight_func_design}
\subsection{Properties of the weight function}
The main goal of the compliance weight function $W(z)$ is to locally suppress the divergent nature of the higher-order resonant integrals $\mathbb{I}_{n,m}^{p,q}$ near the zero-crossings of the magnetic drift, while allowing to maintain the physics associated with the spatial variation of the magnetic drift wells along the field line in regions of finite magnetic drift. To avoid the discontinuities inherent of the step function implied by this description, we consider a family of sigmoid curves $S_N$
\begin{align}
    S_{N}(x)  = 
        \begin{cases}
            0 & x\leq 0 \\
            x^{N+1} \sum_{k=0}^{N} \binom{N+k}{k} \binom{2N+1}{N-k} (-x)^k & 0<x<1 \\
            1 & x\geq 1
        \end{cases}
        \label{eq:Hermite_SN_filt}
\end{align}
where $\binom{l}{p}=l!/(p!(l-p)!)$ is the usual binomial coefficient, as obtained by Hermite interpolation which smoothly connects the suppressed state $W(z)=0$ to the saturated (i.e. unfiltered) state $W(z)=1$ on the interval $[0,1]$ while ensuring $C^{N}$ continuity at the clamps. \par 
\Cref{eq:Hermite_SN_filt} accurately describes the ramp-up of the weight function to the right of a zero crossing, and we will also need its complement capable of interpolating from unity to zero to the left of a zero crossing. As the functions $S_N(z)$ have the property of being point-symmetric around $(1/2,1/2)$ the equivalent ramp-down segment of the weight function is readily obtained by reflecting the $S_N(x)$ curve around this point, resulting in $S_N(1-x)$, which smoothly interpolate from unity to zero on the interval $x\in[0,1]$. \par
\begin{figure}
    \centering
    \includegraphics[width=0.95\textwidth]{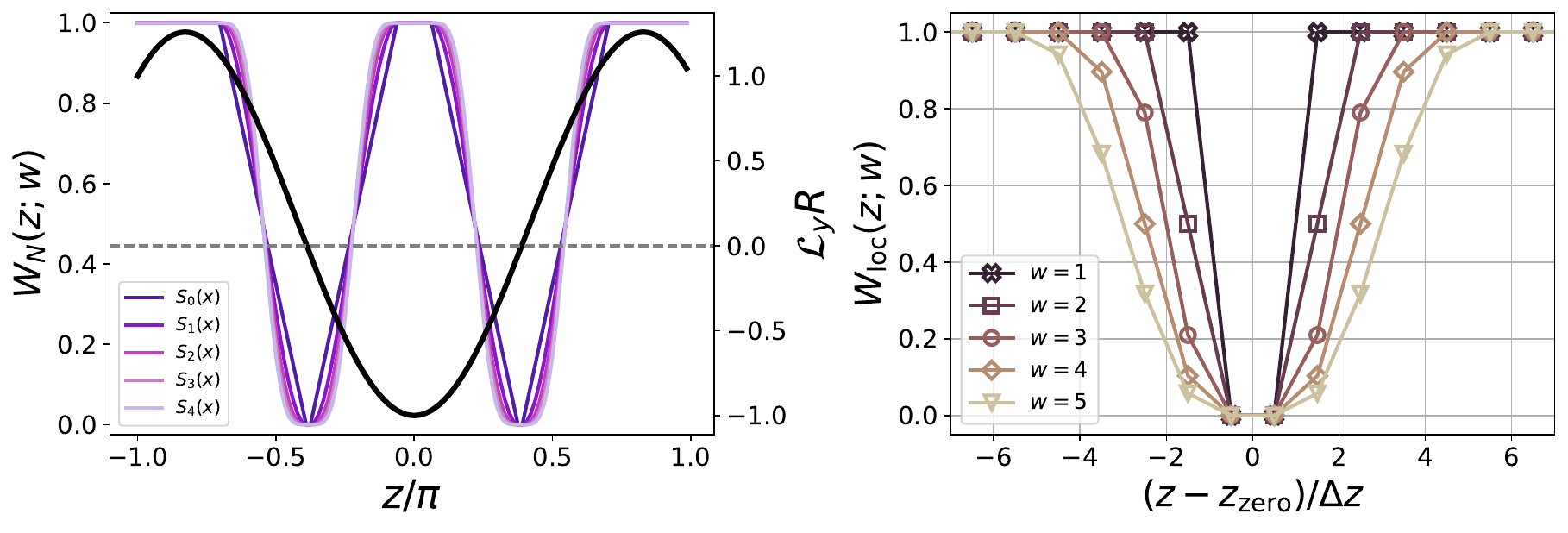}
    \caption{Effect of Hermite interpolant order $N$ (left) and width parameter $w$ (right) on the weight functions $W(z;w)$ obtained by combining the double $S$-shaped filter \Cref{eq:up-down-filter-discretised}. Higher-order filters are indicated by lighter pink/purple hues, and wider filters are indicated by lighter sepia hues. Also shown in black on the left is the binormal component of the curvature drift operator is shown for the case of an $s-\alpha$ geometry characterised by $\hat{s}=-0.3, \alpha_{\textrm{MHD}}=0$, whose intersections with the dotted dashed line identify the zero crossings. For the purpose of illustration, on the left we have used a resolution of $N_z=128$ grid points along the field line and set the filter half-width to $w=20$, though both should be considered unrealistically large for application. On the right, we consider a filter generated by second-order Hermite interpolation $S_2(x)$, locally zoomed-in around one of the zero crossings of the drift, whose location $\omega_d(z_\mathrm{zero})=0$ is for the purpose of visualisation taken to be exactly halfway the points gridpoints $z_{j}, z_{j+1}$ for which $\mathcal{L}_y$ changes sign.}
    \label{fig:weight_func_example}
\end{figure}

 
Consequently, combining this up and down ramp, the filter $W_{\mathrm{loc}}(z;w)$ which locally smoothly suppresses the divergent integrals for one zero crossing of the magnetic drift over a distance of $2w$ is given by
\begin{align}
    W_{\mathrm{loc}}(z;w) = S_N\left(\frac{z-z_\mathrm{right}}{w}\right) + S_N\left(\frac{z_{\mathrm{left}}-z}{w}\right)
    \label{eq:up-down-filter-ramp}
\end{align}
with $z_{\mathrm{left,right}}$ the points directly adjacent to the left and right of a sign change of the magnetic drift, or in the case of an exact zero $z_{\mathrm{left}}=z_{\mathrm{right}}$. When embedded on a grid with uniform spacing $\Delta z$, such that $z_k=k\Delta z$, \Cref{eq:up-down-filter-ramp} may be compactly implemented as
\begin{align}
    W_{\mathrm{loc}}(k;w)=S_N\left(\min\left(\frac{d(k)}{w},1\right)\right)
    \label{eq:up-down-filter-discretised}
\end{align}
where we have made use of the clamping property of \Cref{eq:Hermite_SN_filt} for $z<0$ and $z>1$, for simplicity in notation we let $w\rightarrow w\Delta z$, and introduced $d(k)=\min(|j-k|,|k-(j+1)|)$ as the nearest distance in number of grid points to the zero crossing under consideration, with $j,j+1$ the grid indices of the points adjacent to a sign change of the sampled magnetic drift, as determined by $\omega_d(z_{j+1})\omega_d(z_{j})<0$. The total weight function $W(z;w)$ applied to the integrals $\mathbb{I}_{n,m}^{p,q}$ is then determined as $W(z;w)=\prod_{\mathrm{zero \ crossings}} W_{\mathrm{loc}}(z;w)$, where the multiplication of local filters ensures $C^{N}$ continuity of each local filter is preserved in the global weight function, at the slight trade-off of potential enhanced suppression of the ODE coefficient if two subsequent zero crossings occur within less than $2w$ gridpoints. An example of the resulting weight function obtained from the multiplication of these double sigmoid curves in a tokamak geometry is shown in \Cref{fig:weight_func_example}, showing that the weight functions smoothly decrease from unity at finite magnetic drift to zero at the zero crossings of the drift.  \par

As this weight function is applied to the ODE coefficients of \Cref{eq:adv-diff}, artificial kinks in the behaviour of the ODE coefficients may arise when the weight function only ensuring $C^0$ continuity and are undesirable for obtaining smooth solutions for the eigenfunction. Hence, given the relatively rapid convergence of the weight function's shape with the interpolation order $N$ as exemplified in \Cref{fig:weight_func_example} combined with the fact that we demand $C^{2}$ continuity of the eigenfunctions to obtain solutions to \Cref{eq:adv-diff}, we opt to use the $S_2(z)$ function, whose $S$-shaped interpolation stage is described by $z^3 \left(10-15z + 6z^2\right)$. This special case of \Cref{eq:Hermite_SN_filt} is colloquially known as the ``smootherstep'' filter and has widespread use in computer graphics applications \cite{Perlin2002ImprovingNoise}. Additionally, due to the point symmetry of $S_N(z)$ function around $(1/2,1/2)$ the weight function's half-width parameter needs to be at minimum $w\geq3$ grid points in order to distinguish $S_2(x)$ from its lowest-order $S_0(x)$ counterpart providing only linear interpolation, and a classical step function discretely changing from $0$ to $1$ for $w=2$ and $w=1$, respectively, see \Cref{fig:weight_func_example}.

\subsection{Sensitivity tests} \label{app:sensitivity_study}
In order to find the appropriate filter width $w$ for each geometry, we perform a sensitivity test of the eigenfrequency on the filter width for each geometry. As the filter width is increased, the eigenfrequency obtained from the full dispersion model should converge to the eigenfrequency predictions from the hybrid model, where the kinetic integrals related to the parallel inhomogeneity of the magnetic drift are globally suppressed along the field line, corresponding to the $w\rightarrow \infty$ limit of the compliance filter. \par
\begin{figure}[ht]
    \centering
    \includegraphics[width=0.95\textwidth]{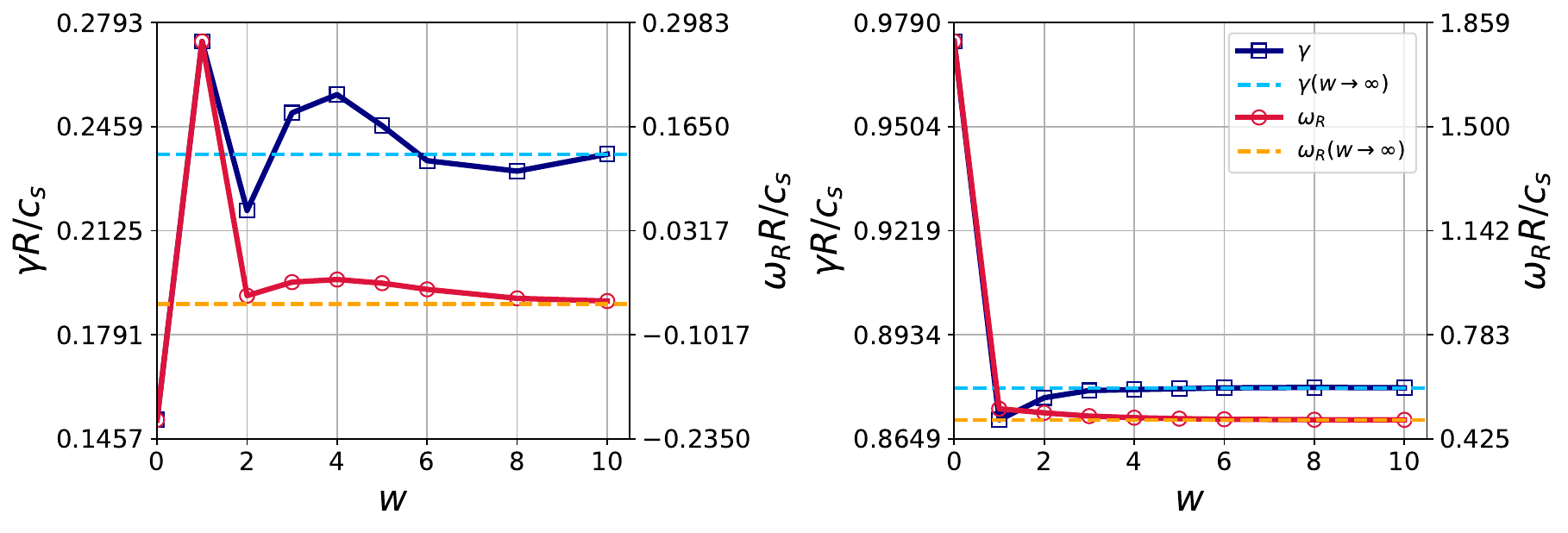}
    \caption{Effect of the compliance filter weight function width parameter on the growth rate (solid line, blue squares) and propagation frequency (solid line, red circles) of the most unstable eigenmode in the CBC-like tokamak geometry for $k_y \rho_s =0.1$ (left) and $k_y \rho_s = 1$ (right). Dashed cyan/orange lines indicate the growth rate/propagation frequency prediction from the hybrid model, respectively. }
    \label{fig:CBC_filter_sensitivity}
\end{figure}

\begin{figure}[ht]
    \centering
    \includegraphics[width=0.95\textwidth]{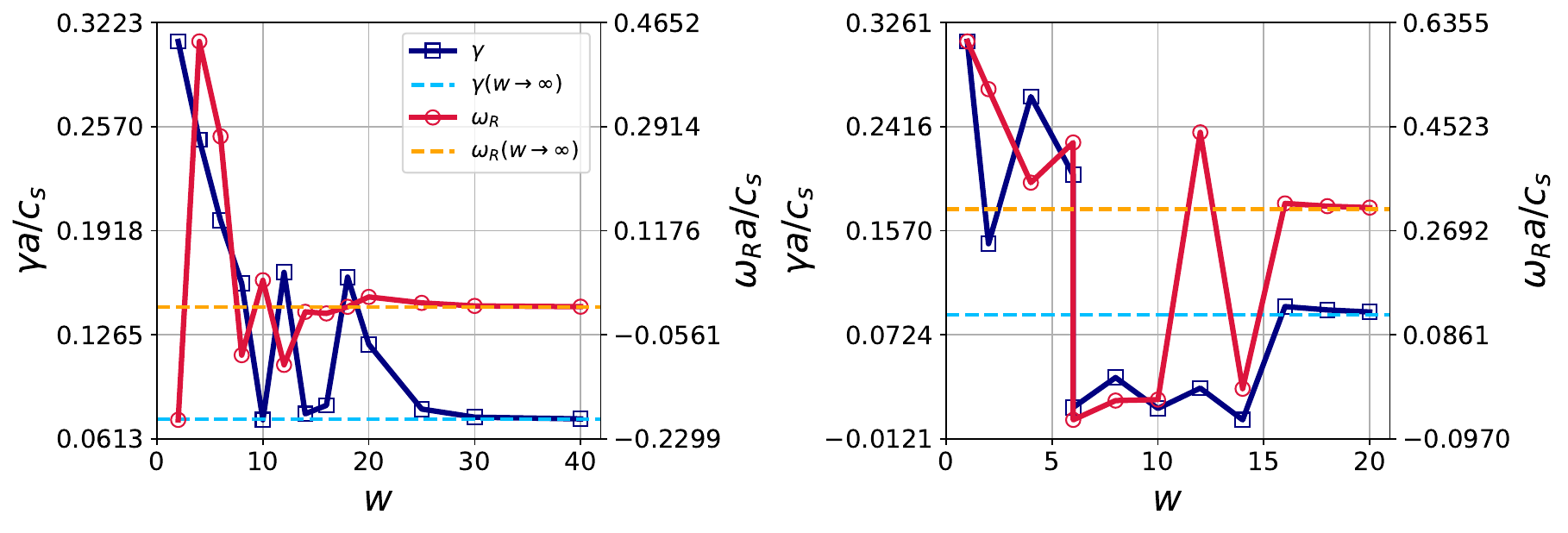}
    \caption{Effect of the compliance filter weight function width parameter on the growth rate (solid line, blue squares) and propagation frequency (solid line, red circles) of the most unstable eigenmode in the inward-shifted configuration of LHD for $k_y \rho_s =0.1$ (left) and $k_y \rho_s = 1$ (right). Dashed cyan/orange lines indicate the growth rate/propagation frequency prediction from the hybrid model, respectively. }
    \label{fig:LHD_filter_sensitivity}
\end{figure}

\begin{figure}[h!]
    \centering
    \includegraphics[width=0.95\textwidth]{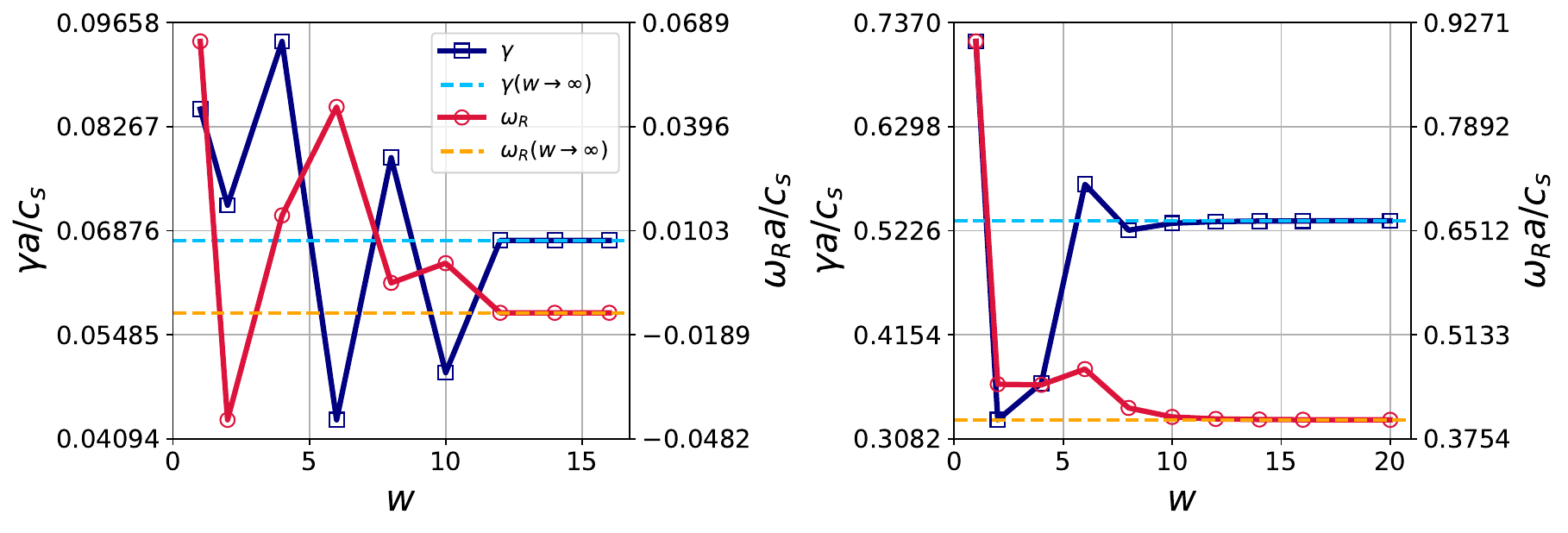}
    \caption{Effect of the compliance filter weight function width parameter on the growth rate (solid line, blue squares) and propagation frequency (solid line, red circles) of the most unstable eigenmode in the high-mirror configuration of W7-X for $k_y \rho_s =0.1$ (left) and $k_y \rho_s = 1$ (right). Dashed cyan/orange lines indicate the growth rate/propagation frequency prediction from the hybrid model, respectively. }
    \label{fig:W7X_filter_sensitivity}
\end{figure}

 
Results for the change in eigenfrequency of the most unstable eigenmode -- as determined including the correction for finite-transit-frequency damping given by \Cref{eq:freq_Landau_correction} -- at both $k_y \rho_s = 0.1$ and $k_y \rho_s = 1$ are shown in \Cref{fig:CBC_filter_sensitivity,fig:LHD_filter_sensitivity,fig:W7X_filter_sensitivity} for the high-shear tokamak, LHD and W7-X geometries, respectively. Note that only in the high-shear tokamak case it was possible to find eigenmode solutions satisfying $\det \underline{M}(\omega)=0$ without applying a weight function, with both stellarator cases not yielding a frequency for $w=0$ that could reduce the smallest singular value of the matrix below $10^{-6}$, two orders of magnitude above our solution threshold, and are hence omitted from the plots. Additionally, due their denser eigenmode spectra compared with the high-shear tokamak, the eigenfrequency for both stellarator cases is more sensitive to the choice of the filter width, as a change in $w$ generally both results in a small shift of the frequency spectrum combined with a perturbation of the corresponding eigenmode profiles, both of which affect the selection rule for the most unstable eigenmode, as evident by the zigzags in \Cref{fig:LHD_filter_sensitivity,fig:W7X_filter_sensitivity}. \par

In all geometries, however, the critical filter width $w_{\textrm{crit}}$ beyond which the solutions converge to the predictions of the hybrid model decreases with increasing wavenumber, as expected based on a drift-wave ordering for the frequency and the behaviour of the integrals near the zero crossings of the magnetic drift [\Cref{eq:zero_drift_lim_integrals}]. In general $w_{\textrm{crit}}$ increases as the number of grid points $N_z$ used to discretise the field-aligned $N_z$ is larger (compare \Cref{tab:crit_width_param} with \Cref{tab:GENE_res}). As a trade-off between retaining the physics related to the parallel inhomogeneity of the magnetic drift and circumventing the deleterious effect of the divergence of the associated integrals near the zero crossings, the optimum filter width $w$ should, ideally at each wavenumber, be smaller than $w_{\textrm{crit}}$ but not so small that the eigenfrequency strongly diverges from the hybrid model since, as the parallel drift inhomogeneity ought to be an asymptotically small effect. With this in mind, $w=4,10,6$ are used throughout the main text for the high-shear tokamak, LHD and W7-X, respectively.

\begin{table}[t]
\caption{Critical filter width parameter $w_{\textrm{crit}}$  beyond which the solutions of the full asymptotic dispersion model converge to the solutions of the hybrid asymptotic dispersion, which weight and suppress the kinetic integrals related to parallel inhomogeneity in the magnetic drift, respectively, at both low- and high binormal wavenumber high-shear tokamak, and LHD and W7-X stellarator geometries.}
\centering
\begin{tabular}{l | c c c c}
\hline
 & CBC-like & LHD & W7-X \\
\hline
$k_y \rho_{s}=0.1$ & 8 & 25 & 12 \\
$k_y \rho_{s} = 1$ & 3 & 16 & 8  \\
\hline
\end{tabular}
\label{tab:crit_width_param}
\end{table}

\section{Zero-drift simulations} \label{app:toroidal_vs_slab}
We perform an additional \textsc{Gene} where the magnetic drifts are artificially suppressed ($\mathcal{L}_x=\mathcal{L}_y=\mathcal{K}_x=\mathcal{K}_y=0$), leaving all other parameters equal. A similar type of numerical experiment has previously been considered by i.e. Ref.~\cite{Costello2023TheStellarators} to differentiate between toroidally-driven and slab-driven branches of the Universal Instability in stellarator geometries, and we thus expect this test to yield information about the nature of the ITG instabilities observed in the high-fidelity \textsc{Gene} simulations discussed in the main text. Thus, if curvature effects form the dominant physical mechanism for instability, we expect the resulting eigenfrequencies and eigenmodes to be substantially different between the ``Real geometry'' and the ``Zero-drift geometry'', whereas strong similarities would indicate that the instabilities significantly rely on parallel streaming. We remark, however, that the resulting geometries are purely synthetic and do not strictly constitute a valid MHD equilibrium. \par 
As the underlying physical mechanisms behind the instabilities have changed by modifying the flux tube geometry in which the eigenmodes are embedded, convergence tests have been performed to determine the appropriate resolutions to resolve these slab-like instabilities, resulting in the numerical resolutions shown in \Cref{tab:GENE_nodrift_res}. \par

\begin{table}[b!]
\caption{Converged numerical resolutions of high-fidelity \textsc{Gene} simulations with artificially suppressed magnetic drifts for flux-tube simulations of binormal wavenumber scan in the CBC-like tokamak geometry, LHD and W7-X stellarators, and the geometry scan for $\hat{s}-\mathrm{\alpha}$ tokamak at fixed $k_y \rho_s = 0.2$. $N_z, N_{v\parallel}, N_{\mu}$ refer to the number of grid points used for the field-aligned coordinate, parallel velocity grid and magnetic moment, respectively, while $N_{kx}$ refers to the number of Fourier modes considered in the radial envelope of the fluctuations for the twist-and-shift boundary condition \Cref{eq:twist-and-shift}.}
\centering
\begin{tabular}{l | c c c c}
\hline
 & CBC-like & LHD & W7-X & $\hat{s}\mathrm{-}\alpha$ \\
\hline
$N_z$ & 16 & 256 & 96 & 32 \\
$N_{kx}$ & 11 & 15 & 61  & 27 \\
$N_{v\parallel}$  & 32 & 36 & 24 & 24 \\
$N_{\mu}$ & 12 & 12 & 8 & 8\\
\hline
\end{tabular}
\label{tab:GENE_nodrift_res}
\end{table}

\begin{figure}[t]
    \centering
        \includegraphics[width=0.95\textwidth]{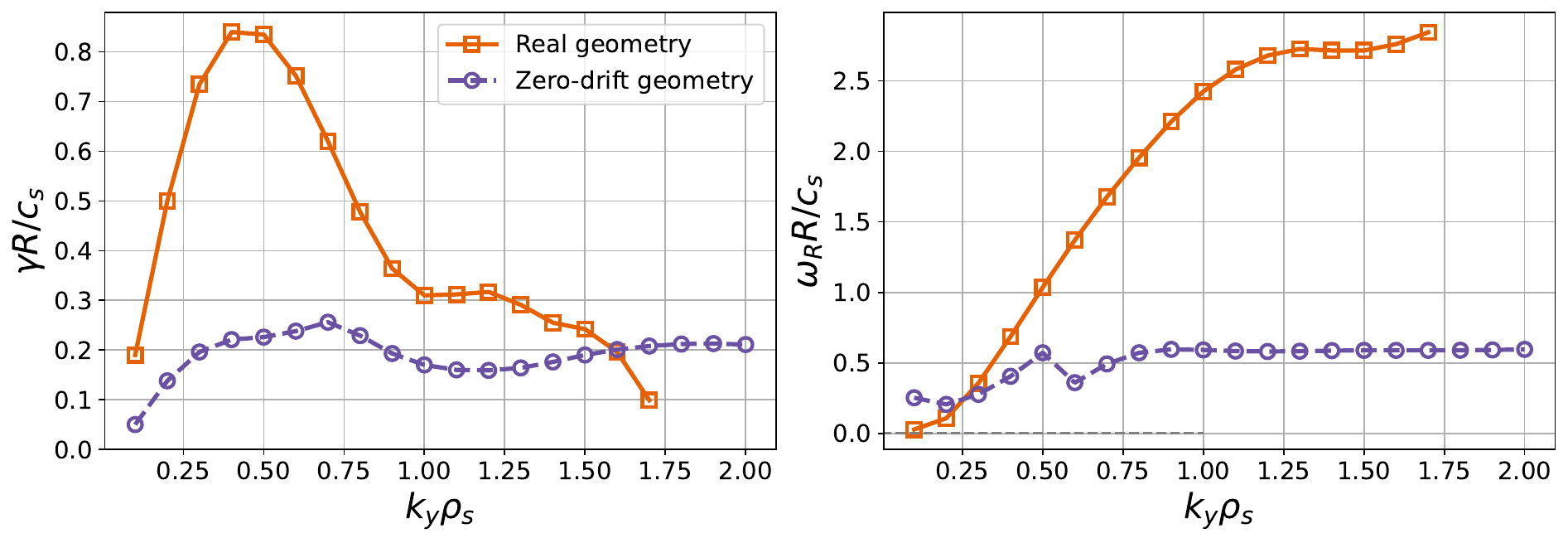}
        \includegraphics[width=0.95\textwidth]{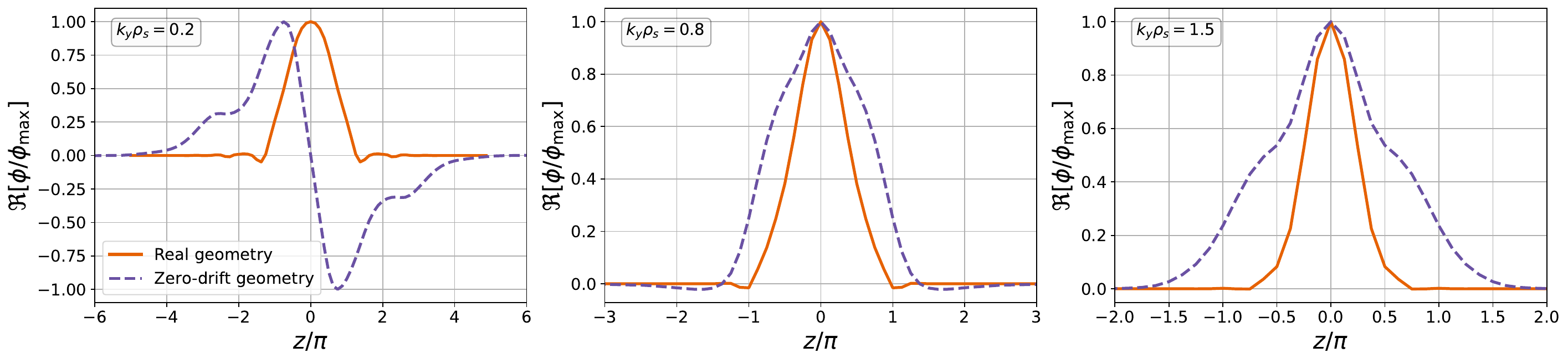}
    \caption{Effect of the presence of the magnetic drift on instabilities for the CBC-like scenario, showing the changes in eigenfrequency (top row) and eigenmodes (bottom row) when the real geometry (solid orange lines) is replaced by an artificial geometry with suppressed magnetic drifts (dashed purple lines).}
    \label{fig:CBC_drift_no_drift}
\end{figure}

\begin{figure}[t!]
    \centering
        \includegraphics[width=0.95\textwidth]{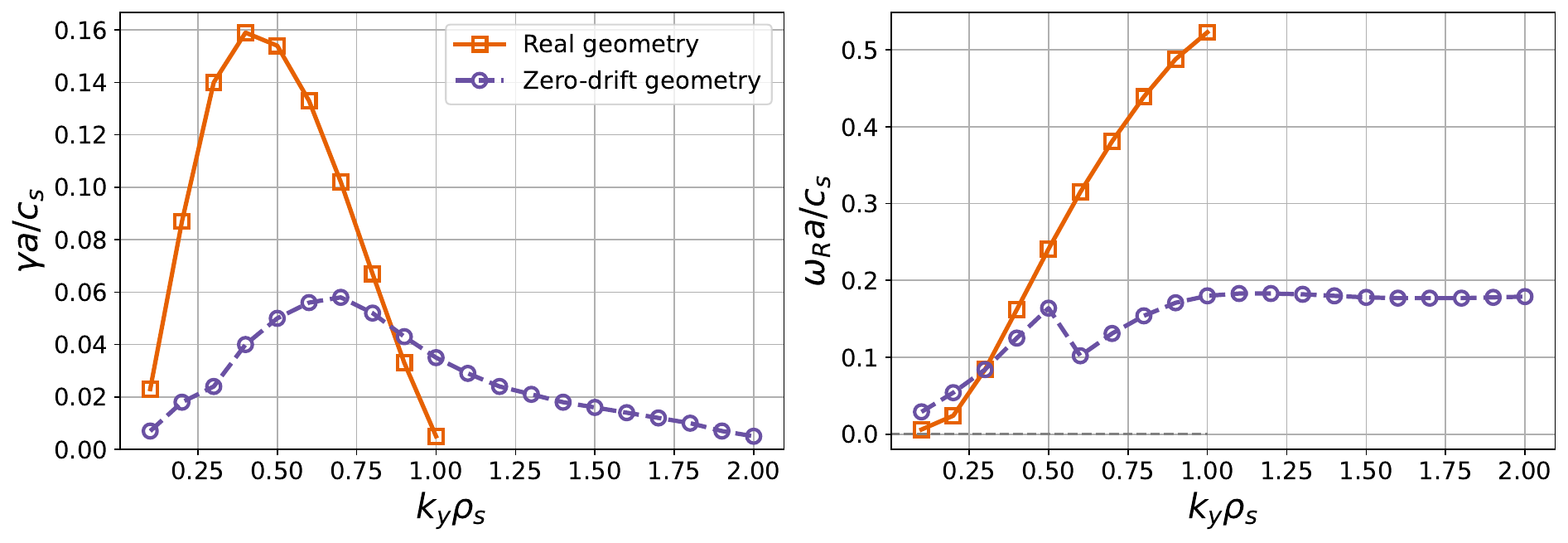}
        \includegraphics[width=0.95\textwidth]{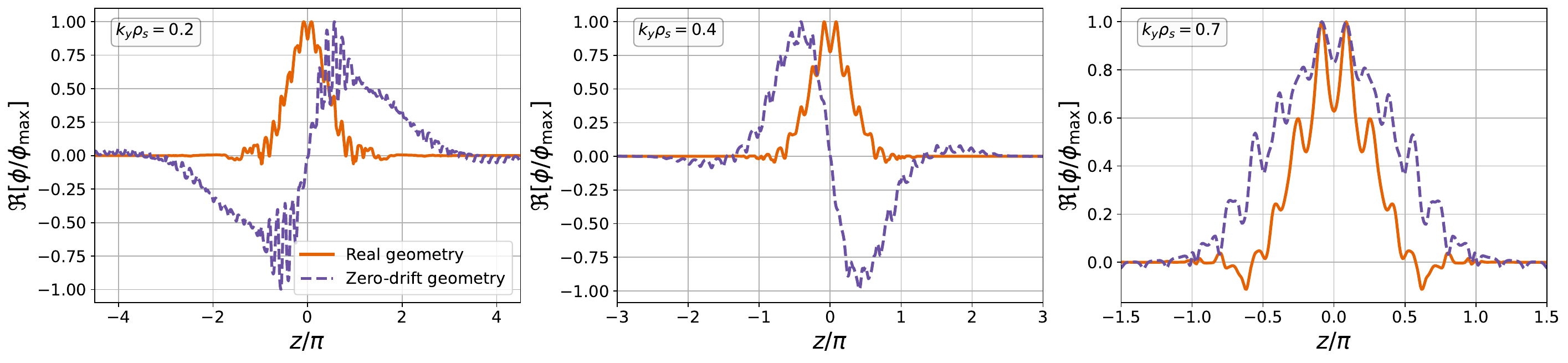}
    \caption{Effect of the presence of the magnetic drift on instabilities in the inward-shifted LHD configuration, showing the changes in eigenfrequency (top row) and eigenmodes (bottom row) when the real geometry (solid vermilion lines) is replaced by an artificial geometry with suppressed magnetic drifts (dashed blue lines).}
    \label{fig:LHD_drift_no_drift}
\end{figure}

\begin{figure}[t]
    \centering
        \includegraphics[width=0.95\textwidth]{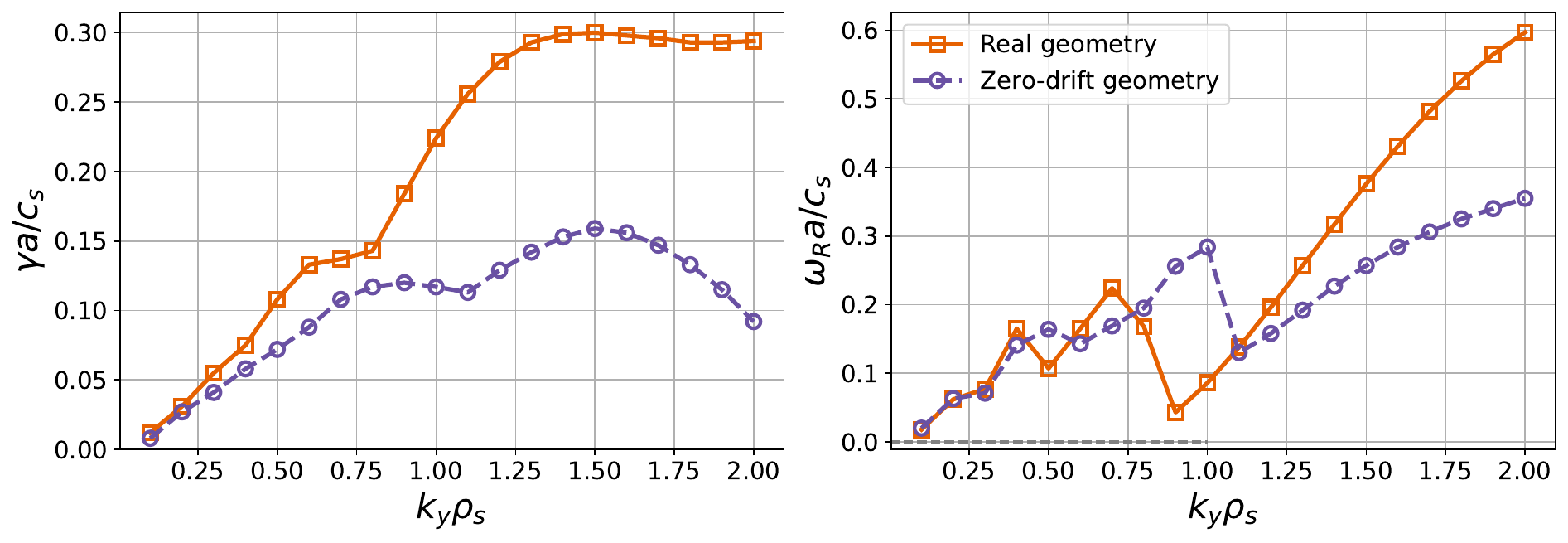}
        \includegraphics[width=0.95\textwidth]{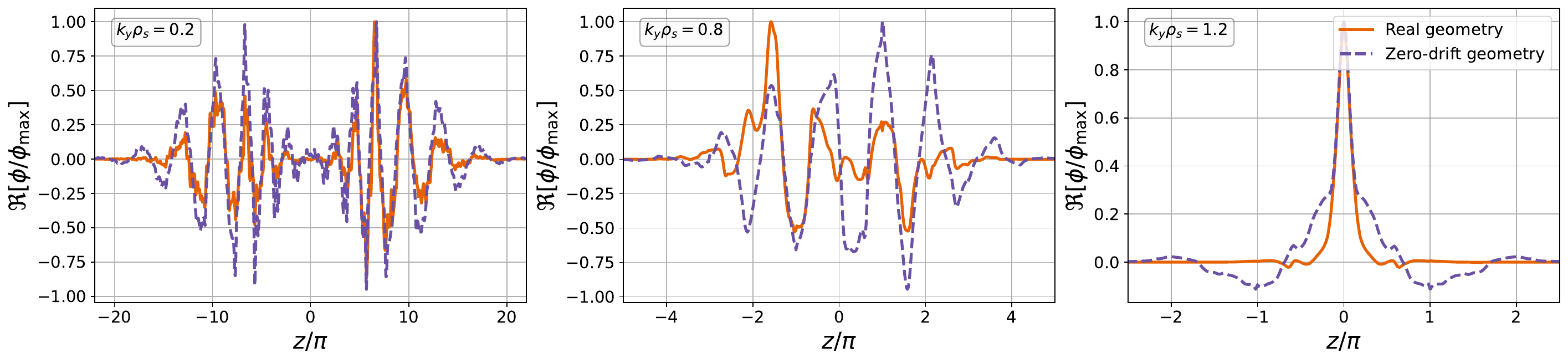}
    \caption{Effect of the presence of the magnetic drift on instabilities in the high-mirror configuration of W7-X, showing the changes in eigenfrequency (top row) and eigenmodes (bottom row) when the real geometry (solid vermilion lines) is replaced by an artificial geometry with suppressed magnetic drifts (dashed blue lines).}
    \label{fig:W7X_drift_no_drift}
\end{figure}

In \Cref{fig:CBC_drift_no_drift,fig:LHD_drift_no_drift,fig:W7X_drift_no_drift} we directly compare eigenfrequencies and electrostatic potential eigenmodes as a function of the binormal wavenumber in the realistic and artificial zero-drift geometries of the CBC-like tokamak, inward-shifted LHD and high-mirror W7-X configurations, respectively. In both the CBC-like tokamak and LHD cases, the eigenfrequency spectrum and eigenmodes are substantially different with the effects of magnetic drifts suppressed. While at low wavenumbers the propagation frequencies $\omega_R$ are roughly similar up to $k_y \rho_s \approx 0.5$, the growth rates in the artificial zero-drift geometries are roughly $2\mathrm{-}4$ times as large, indicating that at the considered gradients of $a/L_{n}=2, a/L_{T}=4$ the instabilities are driven by toroidal resonances. Furthermore, despite the similarity in $\omega_R$ the eigenmodes are distinctly different; being of the strongly localised ballooning-parity type for toroidal modes, while significantly broader tearing-parity type modes emerge for the slab-like modes driven by transit resonances. At higher wavenumbers another contrast appears where the slab-like instabilities persist up to shorter wavelengths beyond the point where the toroidally driven eigenmodes get stabilised -- particularly noticeable in LHD where the toroidally driven mode is stabilised beyond $k_y \rho_s > 1$ while the slab-like instability persists up to $k_y \rho_s =2$. These shorter wavelength slab-like eigenmodes are of the ballooning-parity type and thus more closely resemble their toroidal counterparts, although they are characterised by broader envelopes, indicating a weaker susceptibility to FLR damping. In W7-X, however, both eigenfrequency spectra and the eigenmodes are nearly identical up to $k_y \rho_s \approx 0.8$, with growth rates being marginally smaller in the artifical geometry with suppressed magnetic drifts. This may indicate that transit effects play a significant role in the establishment of instabilities in the long-to-intermediate wavelength range, with the presence of magnetic drift wells providing a modulating on the eigenmodes by providing additional instability drive, thus clearly falling outside of the realm of an asymptotic treatment for transit effects. At the intermediate-to-small wavelength regime, however, a clear distinction between the slab-like eigenmodes and finite-drift eigenmodes occur, with the latter characterised by growth rates that are roughly a factor of two larger and becoming strongly localised around the outboard mid-plane, thus being similar to the conventional toroidally-driven modes observed in CBC and LHD cases. We note, however, that these strong similarities between the instabilities in the finite-drift and (artificial) zero-drift geomtries of W7-X does seem to contrast with the predictions from the free-energy diagnostic shown in \Cref{fig:Free-energy}, which predict that the toroidal-drive is at least $2.5$ times stronger than the drive/damping from parallel streaming. \par

\begin{figure}[b!]
    \centering
        \includegraphics[width=0.95\textwidth]{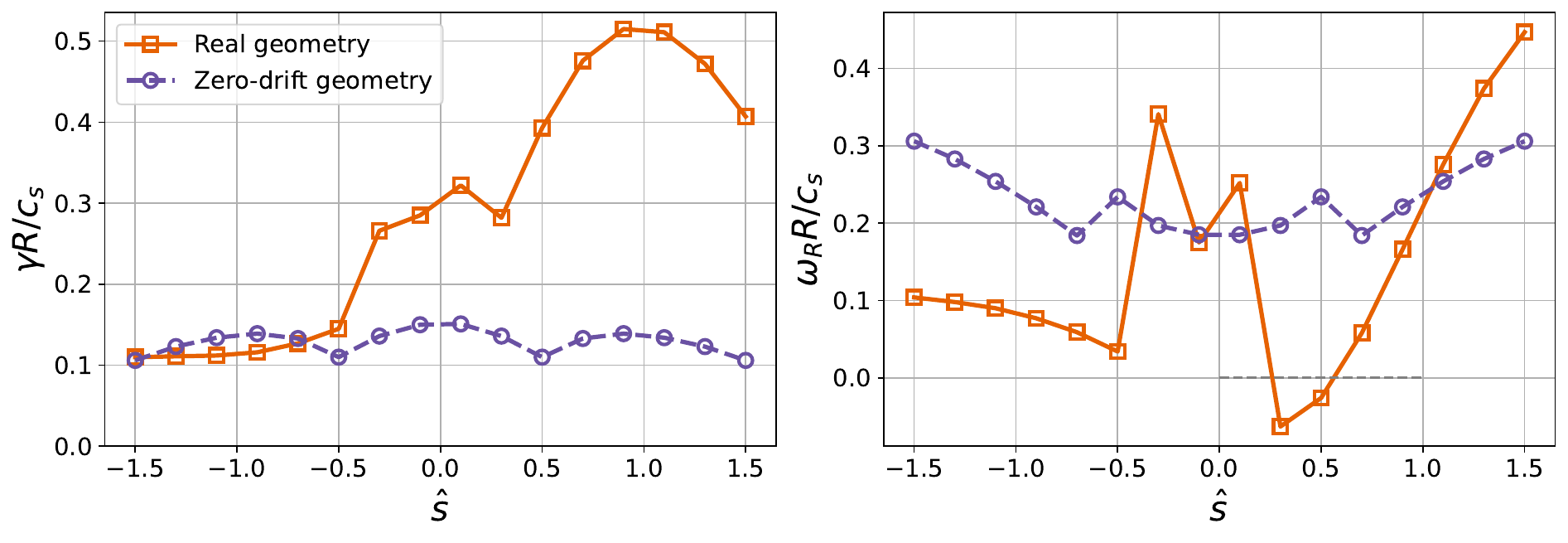}
        \includegraphics[width=0.95\textwidth]{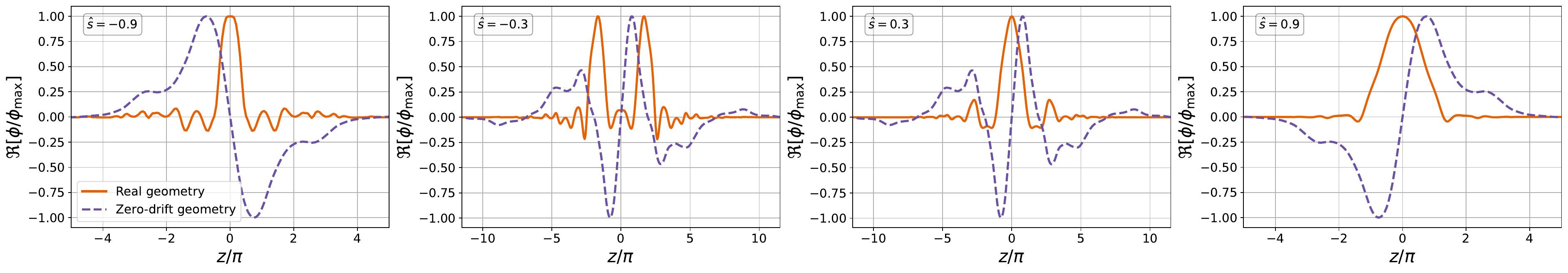}
    \caption{Effect of the presence of the magnetic drift on instabilities as the shear is varied in an $\hat{s}$-$\alpha$, showing the changes in eigenfrequency (top row) and eigenmodes (bottom row) when the real geometry (solid vermilion lines) is replaced by an artificial geometry with suppressed magnetic drifts (dashed blue lines).}
    \label{fig:shat_drift_no_drift}
\end{figure}

Additionally, in \Cref{fig:shat_drift_no_drift} the eigenfrequencies and eigenmodes are compared between finite-drift and zero-drift flux tubes of an $\hat{s}$-$\alpha$ tokamak as a function of (global) magnetic shear $\hat{s}$ at a fixed binormal wavenumber of $k_y \rho_s = 0.2$. Most notably, when the magnetic drifts are artificially suppressed both the eigenfrequency spectrum and the corresponding eigenmodes are symmetric\footnote{Accounting for the fact that $\hat{\phi}(z)$ and $-\hat{\phi}(z)$ are both solutions to the linear dispersion relation with the same $\omega$, and that the normalisation procedure of the displayed eigenfunctions essentially make the sign of a tearing-parity eigenmode arbitrary, as both $z_{\mathrm{max}}$ and $-z_{\mathrm{max}}$ have the same absolute magnitude of the eigenfunction} around $\hat{s}=0$, whereas a strong asymmetry exists for the instabilities in the finite-drift flux tubes. This asymmetry is a direct consequence of the effect of $\hat{s}$ on the shape of the drift wells, see \Cref{fig:shat_geometry_effect}, which strongly reduce both the depth and width of the bad-curvature region as $\hat{s}$ is reduced, thus significantly affecting toroidally-driven modes, whereas slab-like modes are only affected through shear via FLR damping, which for $k_x=0$ modes is symmetric in $\hat{s}$ (see also \Cref{fig:shat_geometry_effect}). Although the growth rates of the negative-shear instabilities are similar for $\hat{s}<-0.5$ between zero and finite-drift cases, their propagation frequencies differ by a factor of $2\mathrm{-}3$, and the corresponding eigenmodes are also vastly different; corresponding to a strongly localised ballooning-parity eigenmode for the finite-drift case, while a significantly broader tearing-parity eigenmode occurs for the zero-drift case. This clearly classifies the observed instabilities in the high-fidelity simulations of corresponding to the toroidal-branch of ITG modes, despite the near balance in the free-energy pumping rate by the curvature drift and parallel streaming terms.

\end{document}